\documentclass[fleqn,usenatbib]{mnras}

\usepackage{newtxtext,newtxmath}

\usepackage[T1]{fontenc}
\usepackage{ae,aecompl}

\usepackage{graphicx}	
\usepackage{amsmath}	
\usepackage{amssymb}	

\def\app#1#2{%
  \mathrel{%
    \setbox0=\hbox{$#1\sim$}%
    \setbox2=\hbox{%
      \rlap{\hbox{$#1\propto$}}%
      \lower1.1\ht0\box0%
    }%
    \raise0.25\ht2\box2%
  }%
}

\PassOptionsToPackage{pdfpagelabels=false}{hyperref}
\PassOptionsToPackage{nonamebreak}{natbib}   

\title[Conduction and cooling in superbubbles]{Evolution of supernovae-driven superbubbles with conduction and cooling}

\author[El-Badry et al.]{
Kareem El-Badry$^{1, 3}$\thanks{E-mail: kelbadry@berkeley.edu},
Eve C. Ostriker$^{2}$,
Chang-Goo Kim$^{2,3}$,
Eliot Quataert$^{1}$,
\newauthor
\,\,Daniel R. Weisz$^{1}$
\\
$^{1}$Department of Astronomy and Theoretical Astrophysics Center, University of California Berkeley, Berkeley, CA 94720\\
$^{2}$Department of Astrophysical Sciences, Princeton University, Princeton, NJ 08544\\
$^{3}$Center for Computational Astrophysics, Flatiron Institute, New York, NY 10010
}

\date{Submitted to MNRAS, February, 23, 2019}

\pubyear{2018}

\begin{document}
\label{firstpage}
\pagerange{\pageref{firstpage}--\pageref{lastpage}}
\maketitle

\begin{abstract}
We use spherically symmetric hydrodynamic simulations to study the dynamical evolution  and internal structure of superbubbles (SBs) driven by clustered supernovae (SNe), focusing  on the effects of thermal conduction and cooling in the interface between the hot bubble interior and cooled shell. Our simulations employ an effective diffusivity to account for turbulent mixing from nonlinear instabilities that are not captured in 1D. The conductive heat flux into the shell is balanced by a  combination of cooling in the interface and evaporation of shell gas into the bubble interior. This evaporation increases the density, and decreases the temperature, of the SB interior by more than an order of magnitude relative to simulations without conduction. However, most of the energy conducted into the interface is immediately lost to cooling, reducing the evaporative mass flux required to balance conduction. As a result, the evaporation rate is typically a factor of $\sim$\,3-30 lower than predicted by the classical similarity solution of Weaver et al. (1977), which neglects cooling. Blast waves from the first $\sim$30 SNe remain supersonic in the SB interior because reduced evaporation from the interface lowers the mass they sweep up in the hot interior. Updating the Weaver solution to include cooling, we construct a new analytic model to predict the cooling rate, evaporation rate, and temporal evolution of SBs. The cooling rate, and hence the hot gas mass, momentum, and energy delivered by SBs, is set by the ambient ISM density and the efficiency of nonlinear mixing at the bubble/shell interface. 
\end{abstract}




\section{Introduction}
\label{sec:intro}
Correlated supernovae (SNe) sweep out large cavities in the interstellar medium (ISM) around young star clusters. These ``superbubbles'' (SBs) are filled with hot gas and often survive for tens of Myr, with repeated SN explosions providing energy to keep their interiors pressurized. Hot gas within SBs is observed to emit X-rays, and recombination lines are observed to originate at the interface between hot SB interiors and cooled shells of swept-up ISM \citep[e.g.][]{Cash_1980, Dunne_2001, Cecil_2002, Townsley_2006, Jaskot_2011}. Resonant lines such as O VI, usually observed in absorption, trace warm-hot gas in the interface \citep[e.g.][]{Savage_2006, Bowen_2008, Li_2017}.
The effects of SBs can also be seen in the distributions of dust and gas in the Milky Way and external galaxies, where SBs that have broken out of galactic disks leave holes in the neutral gas distribution \citep[e.g.][]{Heiles_1979, Brinks_1986, Puche_1992, Weisz_2009, Bagetakos_2011} and dust filaments tracing chimneys that vent hot gas from the disk into the halo \citep[e.g.][]{Norman_1989, Strickland_2004}.  At much earlier stages, expanding SBs may be important to the destruction of the molecular clouds that host nascent super star clusters \citep[e.g.][]{Leroy_2018}.

Most stars form in clusters, so a large fraction of SNe explode inside SBs. SBs are therefore the dominant source of hot gas in galaxies and are thought to play a critical role in the launching of galactic winds \citep[e.g.][]{Chevalier_1985, Nath_2013, Sharma_2014, Vasiliev_2015, Kim_Ostriker_2017, Kim_2018, Fielding_2017, Fielding_2018}. The momentum injected into the ISM by SBs is a key driver of turbulence, which regulates star formation \citep[e.g.][]{Norman_1996, Elmegreen_2004, McKee_2007,Ostriker_2011,Kim_2013}.

When SNe explode in isolation, the energy is often  deposited in high-density gas that can rapidly cool. Radiative losses thus remove $\sim$90\% of the initial energy of isolated SNe \citep[e.g.][]{Cioffi_1988,Blondin_1998,Thornton_1998}, while at the same time producing an expanding shell with terminal momentum of order $10^5 M_\odot {\rm km s^{-1}}$ \citep[e.g.][]{Kim_2015,Iffrig_2015,Walch_2015,Martizzi_2015}. In contrast, when clustered SNe  explode within SBs, a larger fraction of the energy may be retained as internal hot gas, while simultaneously accelerating the surrounding warm/cold ISM. Once a bubble has been excavated by early SNe and/or stellar winds, the SNe that explode at later times deposit energy in a hot, low-density bubble interior. Continued SN explosions maintain a strong termination shock and keep SBs overpressurized long after the leading shock becomes radiative. Cooling is inefficient in the low-density hot gas in SB interiors, with most radiative losses occurring in the thin interfaces between hot and cool components of the SB.

At the outer edge of a SB, there is an abrupt transition between low-density, hot gas ($T\sim 10^{6-7}$\,K; \citealt{Townsley_2006}) and the shell of dense gas that has been swept up by the expanding shock and subsequently cooled. In this region, which is typically quite thin ($\lesssim$\,5\,pc; e.g. \citealt{Kregenow_2006, Sankrit_2007}), thermal conduction carries energy from the hot interior into the cool shell, and cool gas evaporates into the bubble. This interface, as well as the interfaces around cold clouds embedded in the bubble, is where most cooling losses occur \citep[e.g.][]{Li_2017}, since it is there that the temperature passes through the peak of the cooling curve. Cooling is aided by hydrodynamic instabilities in the interfaces that drive mixing of hot and cool gas \citep{Miles_2009, Sano_2012, Krause_2013, Inoue_2013, Fierlinger_2016}, and by the propagation into the shell of shocks that re-heat cooled gas. 

Many theoretical studies have investigated the structure and dynamics of SBs. \citet{Castor_1975} and \citet{Weaver_1977} developed an analytic similarity solution to predict the expansion history and internal structure of bubbles driven by an energy source with constant mechanical luminosity. Their model was designed to describe a bubble cleared out by stellar winds rather than SNe, but subsequent studies \citep[e.g.][]{McCray_1987, Maclow_1988} showed that, because the SNe rate predicted for a simple stellar population with a typical IMF is roughly constant over $\sim$40\,Myr, their solution (which we hereafter refer to as the ``Weaver solution'') can be straightforwardly adapted to describe SBs driven by discrete, repeated SNe. Of course, treatment of discrete SNe explosions as a continuous energy source becomes less accurate when the time interval between SNe is long. An assumption made in adapting the Weaver solution for SBs driven by SNe is that the blasts produced by  individual SNe become subsonic and share their energy fully with the hot bubble interior, rather than propagating as shocks to deposit energy in the shell.  

The Weaver solution has since been widely used to interpret observations of SBs in the Milky Way and nearby galaxies. Many subsequent studies have carried out one-, two-, and three-dimensional simulations of SBs driven by discrete SNe  \citep[e.g.][]{Tomisaka_1981, Tomisaka_1986, Silich_1996, Sharma_2014, Gentry_2017, Kim_2017, Yadav_2017,Vasiliev_2017,Fielding_2018, Gentry_2019}. These studies have broadly validated some of the predictions of the Weaver solution when its assumptions apply, particularly the predicted scaling relations of the SB expansion rate, energy, and momentum with the SN rate, ambient density, and SB age. They have also highlighted some effects that are not captured in the spherical similarity solution, such as the collision of shocks from discrete SNe with the cooled shell, the development of dynamical instabilities at the bubble/shell interface, the departure from spherical symmetry arising from interaction with the inhomogeneous ISM, the ablation over time of cold clouds left behind within the bubble, and the vertical extension and breakout of SBs from galactic disks. Most notably, the Weaver solution does not account for cooling losses in the shell/bubble interface, so simulations with significant radiative cooling generally produce less energetic SBs than predicted by the analytic model.   

Most previous numerical studies have not explicitly modeled thermal conduction. In the Weaver solution, conduction has no effect on the  bubble energetics and expansion rate  -- it only changes the structure of the SB interior, increasing its density and decreasing its temperature. Some numerical studies have 
used the evaporative mass flux into the bubble interior that is predicted by the Weaver solution as a ``subgrid'' model for effects of thermal conduction \citep[e.g.][]{Maclow_1988, Silich_1996,Keller_2014}. A few recent studies have tested the effects of explicitly including conduction as a source term in the energy equation in some of their simulations \citep{Sharma_2014, Gupta_2016, Yadav_2017}, but the effects of conduction on the shell/bubble interface and on the interior structure of SBs have yet to be examined in detail. Moreover, the prediction of \citet{Weaver_1977} for the evaporative mass flux from the shell into the bubble -- which is based on the assumption that the energy fluxes from conduction and evaporation exactly cancel -- has not been tested.  

In this paper, we present results from idealized, one-dimensional simulations of SBs in a uniform medium for a wide range of SB parameters. 
We extend previous studies by explicitly solving for both thermal conduction and cooling, treating turbulent mixing at the shell/bubble interface via a parameterized term in the energy equation.
We study in detail how the evaporative mass flux due to conduction varies with model parameters such as thermal conductivity, ambient density, SN rate, and nonlinear mixing efficiency.  We also identify transitions in the SB character as the bubble ages, demonstrating that the continuous energy injection approximation is only valid at late times (which we quantify). We develop a modified analytic solution for the SB expansion rate and interior properties that takes into account both conduction  and cooling in the shell, with the latter induced by small-scale mixing at the hot/cool interface. This analytic solution including cooling can be considered a successor to the no-cooling Weaver solution.  As we shall show, cooling not only slows SB expansion and reduces momentum injection, but also alters the structure of the conductive bubble/shell interface and reduces the evaporation rate. Throughout the paper, we compare our simulations and analytic models to the classical Weaver solution.  

The remainder of this paper is organized as follows. In Section~\ref{sec:predictions}, we review the classical theory of SB dynamics and the effects of conduction on the bubble interior in the absence of cooling. We summarize the  numerical methods and model parameters we use for our SB simulations in   Section~\ref{sec:numerical}. Section~\ref{sec:results} presents and analyzes our fiducial simulations, focusing on a single choice of SN rate and ambient density. Here we present radial profiles (Section~\ref{sec:profiles}), time-evolution of integrated SB properties (Section~\ref{sec:integrated_props}), and the energy balance of fluxes through the shell/bubble interface (Section~\ref{sec:fluxes}). Section~\ref{sec:pred_cooling_loss} develops an analytic model to predict the internal structure and evolution of  integrated properties of SBs with cooling. Here we compare the model predictions to a suite of simulations with a range of ambient densities, SNe rates, and nonlinear mixing efficiencies. Section~\ref{sec:discrete_shocks} considers the evolution of SBs at early times, when the effects of discrete SNe cannot be accurately modeled under the approximation of continuous energy injection. Here we also compare to the early-time dynamical evolution that would be driven by stellar winds. We summarize our results, compare to previous work, and discuss astronomical implications in Section~\ref{sec:discussion}. The Appendices provide further details about the numerical models used in the simulations. 

\section{Classical Theory}
\label{sec:predictions}

\subsection{Dynamical evolution}
\label{sec:radial_ev}

In this section, we summarize the classical theory of spherical SB evolution driven by continuous energy injection. 

After a short period of adiabatic expansion that is analogous to the Sedov solution for a single SN, high-density shocked gas at the leading edge of a SB cools radiatively, forming a thin, dense shell that is separated from the hot bubble interior by a contact discontinuity. The timescale for shell formation is set by the cooling time in the shocked gas, which is generally short ($\ll 0.1$\,Myr) compared to the lifetime of the SB \citep[e.g.][]{Kim_2017}.

After shell formation, the structural properties of a hot, pressurized bubble expanding into a homogeneous ISM are set by energy and force balance. \citet{Weaver_1977} derived a similarity solution for the evolution of such a bubble under the assumption that cooling losses within the bubble and inner interface are negligible (although the leading shock and outer layer of the shell are assumed to be radiative). Their solution assumes a continuous energy injection luminosity from stellar winds, but it can be adapted to describe a bubble driven by a series of SN explosions with fixed time-averaged power \citep[e.g.][]{McCray_1987,Maclow_1988, Koo_1992, Kim_2017}. 

Neglecting gravity,\footnote{For both the classical SB solution and the models developed in this work, the global force of gravity in countering shell expansion is negligible compared to pressure over the full range of SB parameters we consider.} the expansion of the pressure-driven spherical SB is described by a momentum equation:
\begin{align}
    \label{eq:mom_bubble}
    \frac{{\rm d}}{{\rm d}t}\left(\frac{4\pi}{3}R^{3}\rho_{0}\frac{{\rm d}R}{{\rm d}t}\right)=4\pi R^{2}P.
\end{align}
Here $R$ and $P$ represent the radius and internal pressure of the SB, and $\rho_0$ is the density of the ambient medium. It is assumed that (a) the SB radial momentum is dominated by the cooled shell, $p_{{\rm rad}}\approx M_{{\rm shell}}v_{{\rm shell}}$, where $v_{{\rm shell}}={\rm d}R/{\rm d}t$ is the shell velocity, and the shell contains most of the mass that has been swept up, so $M_{{\rm shell}}\approx\left(4\pi/3\right)R^{3}\rho_{0}$; (b) the external pressure is negligible; and (c) the SB interior is isobaric due to the short sound crossing time within the bubble. 

Assuming that the internal energy of hot gas in the SB interior is reduced by adiabatic expansion with negligible cooling losses, the internal energy of the SB follows
\begin{align}
    \label{eq:energy_bubble}
    \frac{{\rm d}}{{\rm d}t}\left(\frac{4\pi}{3}R^{3}\frac{P}{\gamma-1}\right)=\dot{E}_{{\rm in}}-4\pi R^{2}\frac{{\rm d}R}{{\rm d}t}P,
\end{align}
where $\dot{E}_{\rm in} = E_{\rm SN}/\Delta t_{\rm SNe}$ is the time-averaged energy input rate from SNe, $E_{{\rm SN}}\approx10^{51}\,{\rm erg}$ is the mechanical energy injected by one SN, $\Delta t_{\rm SNe}$ is the time interval between SNe explosions, and $\gamma=5/3$ is the ratio of specific heats.

Equations~(\ref{eq:mom_bubble}) and~(\ref{eq:energy_bubble}) can be directly solved for $R(t)$ and $P(t)$, yielding 
\begin{align}
    \label{eq:weaver_R_t}
    R\left(t\right)&=\left(\frac{125}{154\pi}\right)^{1/5}\dot{E}_{{\rm in}}^{1/5}\rho_{0}^{-1/5}t^{3/5}\\
    \label{eq:weaver_P_t}
    P\left(t\right)&=\frac{5}{22\pi}\left(\frac{125}{154\pi}\right)^{-3/5}\dot{E}_{{\rm in}}^{2/5}\rho_{0}^{3/5}t^{-4/5}.
\end{align} 
This solution is only valid up to the time when the pressure in Equation~(\ref{eq:weaver_P_t}) becomes comparable to that of the ambient ISM.

The total energy of the SB is the sum of the thermal energy of the hot interior and the kinetic energy of the dense shell:
\begin{align}
    E_{{\rm SB}}&=E_{{\rm kin}}+E_{{\rm th}}, \nonumber\\
    \label{eq:Etot_defn2}
    &=\frac{1}{2}M_{{\rm shell}}v_{{\rm shell}}^{2}+\frac{4\pi}{3}R^{3}\frac{P}{\gamma -1} .
\end{align}
Substituting Equations~(\ref{eq:weaver_R_t}) and~(\ref{eq:weaver_P_t}) into Equation~(\ref{eq:Etot_defn2}) yields 
\begin{align}\label{eq:Enfracs}
E_{{\rm kin}}=\frac{15}{77}\dot{E}_{{\rm in}}t;\qquad E_{{\rm th}}= \frac{5}{11}\dot{E}_{{\rm in}}t,
\end{align}
 so that 
\begin{align}
    \label{eq:Etot_Weaver}
    E_{{\rm SB}}\left(t\right)=\frac{50}{77}\dot{E}_{{\rm in}}t.
\end{align} 
That is, $50/77 \approx 65\%$ of the energy input from SNe is retained in the SB, and $27/77\approx 35\%$ is lost. 
The energy loss represents the portion of the $P\,{\rm d}V$ work done by the expanding bubble that goes into heating the ISM gas at the leading shock where it is swept up by the expanding shell. 
While the kinetic energy of the shocked ISM gas is retained, it is assumed that the post-shock thermal energy is quickly radiated away in the exterior layer of the shell:

\begin{align} \label{eq:cooling_shocked_ism} \dot{E}_{{\rm cooling,\,shocked\,ism}}=\frac{27}{77}\dot{E}_{{\rm in}}.\end{align}
The relative proportions of thermal to kinetic energy in the classical SB solution are
\begin{align}
    \label{eq:Ekin_Weaver}
    E_{{\rm kin}}
    =\frac{3}{10}E_{{\rm SB}};\qquad  E_{{\rm th}}= \frac{7}{10}E_{{\rm SB}},
\end{align} which is very similar to the energy proportions for the Sedov solution for a single non-radiative supernova blast ($E_{{\rm kin}}=0.28E_{{\rm SN}};\,E_{{\rm th}}=0.72E_{{\rm SN}}$). 

To summarize, the classical Weaver solution assumes that there is negligible cooling in the bubble interior and shell/bubble interface. As a result, the dynamics and structural properties of the SB (i.e., radius, pressure, momentum, total energy, and the ratio of thermal to kinetic energy) are simply set by force balance and energy conservation, after accounting for radiative losses in the leading shock. Thermal conduction has no effect on these properties; 
at a given pressure, conduction controls the density and temperature of the bubble interior (Section \ref{sec:analytic_interior}).

In  Sections \ref{sec:sb_dyn} and  \ref{sec:cooling_eff} we provide a re-analysis of the predicted dynamical evolution allowing for cooling.   

\subsection{Bubble interior}
\label{sec:analytic_interior}
\citet{Weaver_1977} also proposed a similarity solution to describe the interior structure of a wind-driven bubble. In their solution, radial profiles of temperature, density, and velocity are determined by simultaneously solving the energy conservation and continuity equations within the bubble, under the assumptions that cooling losses in the bubble and interface are negligible and the bubble expansion is described by Equations~(\ref{eq:weaver_R_t})-(\ref{eq:weaver_P_t}). In this model, the conductive energy flux from the bubble into the shell is exactly balanced by the evaporation of cool gas from the shell into the bubble, such that the shell does not gain or lose thermal energy.

In fact, even without making assumptions regarding the detailed interior structure of the SB, the evaporation rate of the shell into the bubble can be estimated using the approach adopted by \citet{Cowie_1977} to compute the evaporation rate of a cold cloud embedded in a hot medium.  
The total conductive heat flux through a surface at radius $r$ is $-4\pi r^{2}\kappa\left({\rm d}T/{\rm d}r\right)$, where $\kappa$ and ${\rm d}T/{\rm d}r$ are the local thermal conductivity and temperature gradient, respectively.  The inward mass flux in the frame of the expanding shell is $\dot{m}=4\pi r^{2}\left(v_{\rm shell}-v\right)\rho$, where $\rho$ and $v$ are the local density and flow velocity at the radius $r$ through which $\dot{m}$ is measured. In the absence of cooling, energy balance requires 
\begin{align}
\label{eq:energy_balance}
4\pi r^{2}(v_{\rm shell}-v)({\cal E}+P)\approx\frac{5}{2}\dot{m}c_{s,{\rm \,iso}}^{2} =-4\pi r^2\kappa\frac{{\rm d}T}{{\rm d}r},
\end{align}
where ${\cal E}$ is the energy density, which can be approximated as $P/(\gamma-1)$ under the assumption that thermal energy dominates in the SB interior. Here we have defined the isothermal sound speed $c_{s,{\rm \,iso}} = \sqrt{k_{\rm B}T/\mu m_{\rm p}}$, where $\mu$ is the mean molecular weight, $k_{\rm B}$ is Boltzmann's constant, and $m_{\rm p}$ is the proton mass. The mass flux required to balance the conductive heat flux is then 
\begin{align}
\label{eq:mdot}
\dot{m}=\frac{-8\pi \mu m_{\rm p} r^2\kappa\left({\rm d}T/{\rm d}r\right)}{5k_{B}T}.
\end{align}

The mass flux profile $\dot{m}(r, t)$ is thus fully specified when the temperature profile and thermal conductivity are known. Taking $\kappa = CT^{5/2}$ for the isotropic thermal conductivity (\citealt{Spitzer_1962}; ``$C$'' is a constant) and considering the ``no cooling'' limit, a local expansion of the temperature profile near the interface \citep[e.g.][]{Cowie_1977} can be used to show that 
$r{\rm d}T/{\rm d}r\to\left(2/5\right)T_{\rm int}$,
where $T_{{\rm int}}$ is the asymptotic interior temperature inward from the conductive interface.

The evaporative mass flux at the shell/bubble interface in the no-cooling limit thus becomes 
\begin{align}
\label{eq:mdot_cowie_mckee}
\dot{m}_{\rm nc}=\frac{16\pi\mu m_{\rm p} R C T_{\rm  int}^{5/2}}{25k_{B}},
\end{align} 
where $R$ is the SB radius.
This expression is identical to that derived by \citet{Cowie_1977} for the evaporation rate of a cold cloud in a hot medium.

Applying Equation~(\ref{eq:mdot_cowie_mckee}) to compute the evaporation rate into the SB interior requires knowledge of $T_{\rm int}$.  A simple estimate can be obtained by assuming a uniform interior with density $\rho=P/(kT_{\rm int}/\mu m_{\rm p})=\left(\int\dot{m}\,{\rm d}t\right)/\left(4\pi R^{3}/3\right)$ and combining Equation~(\ref{eq:mdot_cowie_mckee}) with Equations~(\ref{eq:weaver_R_t}) and~(\ref{eq:weaver_P_t}). 
We denote the resulting value as  $T_{\rm int} = T_{\rm hot,nc}$ to make clear that it depends on the no-cooling assumption:
\begin{align}
\label{eq:Tclass}
T_{{\rm hot,\,nc}}(t)&=\left(\frac{205}{84}\frac{R^{2}P}{Ct}\right)^{2/7}\propto\rho_{0}^{2/35}\dot{E}_{{\rm in}}^{8/35}C^{-2/7}t^{-6/35} .
\end{align}
Substituting into Equation~(\ref{eq:mdot_cowie_mckee}) and integrating yields
\begin{align}
    \label{eq:Mhot_class}
M_{{\rm hot,\,nc}}(t)&=\left(\frac{112\pi}{205}\right)\left(\frac{205}{84}\right)^{5/7}\frac{\mu m_{p}}{k_{b}}R^{3}P^{5/7}\left(\frac{tC}{R^{2}}\right)^{2/7}\\ &\propto\rho_{0}^{-2/35}\dot{E}_{{\rm in}}^{27/35}C^{2/7}t^{41/35}.\nonumber
\end{align}
Of particular note is the weak dependence of $T_{\rm hot,\,nc}$ on time, the input energy rate, and the ambient density. With a constant energy input rate, $\dot M_{\rm hot,nc}$ has the same weak scalings as $T_{\rm hot,\,nc}^{-1}$.  

\citet{Weaver_1977} did not assume a constant-density interior, but instead calculated the interior structure by numerically integrating the equations of continuity and energy conservation in the SB interior under the assumption of self-similarity. This yielded predictions similar to Equations (\ref{eq:Tclass}) and (\ref{eq:Mhot_class}), except that our temperature and mass normalizations are respectively 22\% and 46\% lower.   

Several previous studies \citep[e.g.][]{Maclow_1988,  Silich_1996, Keller_2014} have used Equation~(\ref{eq:mdot_cowie_mckee}) as a subgrid model to approximate the effects of thermal conduction in simulations of SBs. We note, however, that Equations~(\ref{eq:mdot}) and~(\ref{eq:mdot_cowie_mckee}) are {\it not} expected to hold if there is significant cooling in the shell/bubble interface. Cooling will remove some of the energy conducted into the shell, reducing the amount of evaporation needed to balance conduction. We provide an analysis of the effects of cooling on the evaporation rate and temperature in Section \ref{sec:mhot_modified_pred}. 

\section{Numerical simulations}
\label{sec:numerical}

We simulate a 1D domain in spherical coordinates using the \texttt{Athena++} code\footnote{\url{https://princetonuniversity.github.io/athena/}} \citep{Stone_2019}, a new version of \texttt{Athena} \citep{Stone_2008, Stone_2010}. We use reflecting and outflow boundary conditions, respectively, at the inner and outer boundaries. The background medium is characterized by a mean density $\rho_0$ and pressure $P_0$; we adopt $P_{0}=3000\,k_{B}\,{\rm K\,cm^{-3}}\times(\rho_{0}/(\mu_{\rm H}\,m_{{\rm p}}{\rm cm^{-3}}))$ as a fiducial value. We assume a mean molecular weight $\mu =0.62$, appropriate for ionized gas,\footnote{Note that the mean molecular weight $\mu$ should in principle vary throughout the simulation domain, depending on the ionization state of the gas. For simplicity, we fix $\mu = 0.62$.} where $n = \rho/(\mu m_{\rm p})$, and $\mu_{\rm H} = 1.4$, with $n_{\rm H} = \rho/(\mu_{\rm H}m_{\rm p})$.

We solve the equations of hydrodynamics  on a 1D grid with uniform linear cell resolution $\Delta r$. Thermal conduction contributes a flux term to the energy equation:
\begin{align}
\frac{\partial {\cal E}}{\partial t}+ \nabla\cdot\left(\left({\cal E}+P\right){\bf v}\right)=\nabla\cdot\left(\kappa\nabla T\right)-
n\left[n\Lambda\left(T\right)-\Gamma\left(T\right)\right]
\end{align}
where $\kappa$ is thermal conductivity and $\cal{E}$ is the energy density. 
The gas temperature is $T=P/(nk_{B})$, and $\Lambda \equiv (\mu/\mu_{\rm H})^2 \Lambda_{\rm H}$ and $\Gamma \equiv (\mu/\mu_{\rm H}) \Gamma_{\rm H}$ are cooling and heating rate  coefficients. Subscripts of ``H'' denote the coefficients with respect to the hydrogen number density.

We use an operator-split implementation of cooling and heating adapted from \citet{Fielding_2018}. The cooling curve $\Lambda_{\rm H}(T)$ is a 10-segment piecewise power law fit to  the cooling curve from \citet{Oppenheimer_2013} for $T>10^4 \rm K$ and the one from \citet{Koyama_2002} at lower temperatures; see Appendix~\ref{sec:cooling_curve} for details. We adopt an absolute temperature floor of 100 K. We modify the cooling curve from \citet{Fielding_2018} slightly to drop off steeply over $100 < T/{\rm K} < 500$. This modification prevents the temperature in the shell from reaching the floor, which can lead to numerical problems, but it has little effect on the dynamics of the SB. 

We use a constant, spatially uniform photoelectric heating rate of $\Gamma_{\rm H}=n_{{\rm H,0}}\Lambda\left(P_{0}\mu m_{p}/(k_{B}\rho_{0})\right)$ for $T<10^4\,\rm K$; here $n_{{\rm H,0}}$ is the hydrogen number density of the ambient medium. The scaling of $\Gamma_{\rm H}$ with $n_{\rm H,0}$ crudely accounts for the increased photoelectric heating rate when $n_{\rm H,0}$ and the star formation rate are higher. The practical consequence of adopting this heating rate is that heating and cooling exactly balance in the ambient medium. At $T>10^4$\,K, we take $\Gamma_{\rm H} = 0$, since we expect dust grains to be destroyed at higher temperatures. 

\subsection{Thermal conduction}
\label{sec:thermal_cond}

Conduction in an ionized gas is primarily mediated by electrons. The thermal conductivity for a hydrogen plasma is \citep{Spitzer_1962}:
\begin{align}
\label{eq:spitzer}
\kappa_S=\frac{1.70\times10^{11}T_{7}^{5/2}}{1+0.029\ln\left(T_{7}n_{\rm e,-2}^{-1/2}\right)}\,{\rm erg\,s^{-1}cm^{-1}K^{-1}},
\end{align}
where $T_7 \equiv T/(10^{7}\,\rm K)$ and $n_{\rm e,-2} \equiv n_{\rm e}/(10^{-2}\,\rm cm^{-3})$, with $n_{\rm e}$ the number density of free electrons. We take $n_{\rm e} = 1.2 n_{\rm H}$, appropriate for an ionized gas with solar metallicity. Equation~(\ref{eq:spitzer}) includes a correction for the self-consistent electric field produced by the diffusing electrons. 

Because the electron density only enters Equation~(\ref{eq:spitzer}) in the logarithm, the density-dependence of Spitzer conductivity is very weak. To a good approximation, $\kappa$ can thus be modeled as being a function of temperature only, $\kappa_S \approx C\times (T/{\rm K})^{5/2}$, with $C= 6\times 10^{-7}\,{\rm erg\,s^{-1}cm^{-1}K^{-7/2}}$. Our simulations, however, use the full expression from Equation~(\ref{eq:spitzer}).

Equation~(\ref{eq:spitzer}) does not hold at low temperatures, where the gas becomes neutral. At low temperatures,  we take the conductivity for neutral atomic collisions from \citet{Parker_1953}: 
\begin{align}
\kappa_{P}=2.5\times10^{5}T_{4}^{1/2}\,{\rm erg\,s^{-1}\,cm^{-1}\,K^{-1}},
\end{align}
where $T_4 \equiv T/(10^4\,\rm K)$. The Spitzer and Parker conductivities are equal at $T=6.6\times 10^4$\,K, so we transition between the two forms at this temperature. The effects of including Parker conductivity at low temperatures are very small, since values of $\kappa_{P}$ in the shell are orders of magnitude lower than $\kappa_S$ in the bubble and interface. 

In the classical limit in which many collisions occur over the scale length of the temperature gradient, the conductive heat flux is 
\begin{align}
q=-\kappa \nabla T. 
\label{eq:q_cond}
\end{align}
However, Equation~(\ref{eq:q_cond}) overestimates the true heat flux when the electron mean free path is large compared to the temperature scale height, $T/|\nabla T|$. When the temperature gradient is very steep, the diffusive energy flux predicted by Equation~(\ref{eq:q_cond}) exceeds the energy flux that can actually be carried by electrons, which is $q_{{\rm max}} \sim (3/2)\rho c_{s,\,{\rm iso}}^{3}$ \citep{Parker_1963, Cowie_1977}.\footnote{The saturated heat flux is $q_{{\rm max}}\sim n_{e}k_{{\rm B}}T_{e}v_{e}$, where $T_{e}$ and $v_{e}$ represent the temperature and thermal speed of the electrons. The value of $v_{e}$ depends on the electric field set up by the electron current and distribution function of the electrons. \citet{Cowie_1977} calculated $q_{{\rm sat}}\approx 5\phi\rho c_{s,\,{\rm iso}}^{3}$, where $\phi$ is an uncertain dimensionless constant of order unity. Laser fusion experiments suggest $\phi\approx0.3$ \citep{Max_1980, Balbus_1982}, leading to our adopted $q_{{\rm sat}}\approx \frac{3}{2}\rho c_{s,\,{\rm iso}}^{3}$. Measurements of the heat flux in the solar wind suggest slightly larger values \citep[e.g.][]{Bale_2013}.} Following \citet{Balbus_1982}, we use a conductive flux that smoothly interpolates between the classical and saturated values: 
\begin{align}
\label{eq:q_sat}
\frac{1}{|q|}=\frac{1}{\left|-\kappa\nabla T\right|}+\frac{1}{\left(3/2\right)\rho c_{s,\,{\rm iso}}^{3}}.
\end{align}
This expression recovers Equation~(\ref{eq:q_cond}) in the  limit where $\left|-\kappa\nabla T\right|\ll\left(3/2\right)\rho c_{s,\,{\rm iso}}^{3}$, and it asymptotes to $q=(3/2)\rho c_{s,\,{\rm iso}}^{3}$ in the opposite limit. Saturation is implemented via an effective conductivity, $\kappa_{\rm eff}$, that includes the effects of saturation:
\begin{align}
\label{eq:kappa_eff}
\kappa_{{\rm eff}}^{-1}=\kappa^{-1}+\left|\nabla T\right|/((3/2)\rho c_{s,\,{\rm iso}}^{3}).
\end{align}
The actual conductive heat flux is then $q=-\kappa_{\rm eff}\nabla T$.

We examine the effects of saturation of conduction in detail in Appendix~\ref{sec:saturation}. Accounting for saturation slightly decreases the average conductive energy flux from the bubble into the shell, and thus also decreases the evaporative mass flux into the bubble. At early times, the instantaneous heat flux immediately after the arrival of a shock from a new SNe at the shell is more than an order of magnitude larger when saturation is not included than when it is. However, in a time-averaged sense, the effects of saturation on integrated bubble properties are relatively minor, because at late times the magnitude of the classical conductive flux, $|\kappa \nabla T|$, is comparable to $(3/2)\rho c_{s,\,{\rm iso}}^3$. For example, including saturation decreases the hot gas mass produced per SN at late times by $\sim 15\%$ for our default SB parameters (Figure~\ref{fig:saturation}). 

Our simulations become prohibitively expensive when $\kappa$ is very large. For computational tractability, we impose a ceiling on $\kappa$ (see Appendix~\ref{sec:kappa_max}). We ensure that the ceiling is sufficiently high that it has negligible effects on our results.

\subsection{Mixing due to nonlinear instabilities}
\label{sec:mix}
In 3D simulations (and in real SBs and SN remnants), the thickness of the shell/bubble interface in which cooling occurs is enhanced by nonlinear development of Kelvin-Helmholtz (KH), Rayleigh-Taylor (RT), Richtmyer-Meshkov (RM) and related instabilities  \citep[e.g.][]{Vishniac_1989, Blondin_1998, Bucciantini_2004, Folini_2006, Ntormousi_2011, Michaut_2012, Sano_2012, Pittard_2013, Badjin_2016}. Such instabilities and the resulting turbulent mixing do not occur in 1D simulations, leading the interface region to become extremely thin and diminishing cooling losses (see Appendix~\ref{sec:res_test} and \citealt{Gentry_2017}), but their effects can be approximated through subgrid models \citep[e.g.][]{Duffell_2016}.

To model consequences of turbulence that are not directly captured in one-dimensional simulations, we include an effective diffusivity in the energy equation to represent 
nonlinear mixing. Because we wish to represent the mixing of material of different temperature phases due to instability, we implement the model through the conduction operator with a temperature-independent conductivity, $\kappa_{\rm mix}$. The actual conductivity, $\kappa$, in each simulation zone is set to the larger of $\kappa_{\rm mix}$ and thermal conductivity ($\kappa_S$ or $\kappa_{P}$, with appropriate modifications for saturation of conduction) described in Section~\ref{sec:thermal_cond}.

The energy diffusion associated with mixing may be characterized by a length scale and effective nonlinear velocity amplitude. We adopt a parameterization 
\begin{align}
\label{eq:kappa_mix}
\kappa_{{\rm mix}}=\frac{(\lambda\delta v)\times \rho k_{{\rm B}}}{\mu m_{\rm p}}.
\end{align}
 We treat the product $(\lambda \delta v)$ as an adjustable, spatially uniform constant that parameterizes the efficiency of mixing. The value of $(\lambda \delta v)$ will be different for each unstable mode, but we expect mixing to be dominated by fluctuations of the largest-scale modes. 
 
We estimate a possible range of values of  $(\lambda \delta v)$ as follows. 
Instabilities driven by shear, such as the KH and RM instabilities, have a linear growth rate $\omega\sim k v_{{\rm rel}}(\rho_{{\rm high}}/\rho_{\rm low})^{-1/2}$,
where $\rho_{\rm high}$ and $\rho_{\rm low}$ are the densities of the high- and low-density media (in this case, those of the shell and the bubble interior), $v_{\rm rel}$ is the tangential velocity difference across the corrugated shell-bubble interface, and $k = 2\pi/\lambda$ is the wavenumber.  The most effective mixing is likely to be caused by modes with growth rate comparable to the SNe rate, $1/\Delta t_{\rm SNe}$, since modes that grow more slowly will be suppressed by the arrival of blast waves from subsequent supernovae, and modes that grow more quickly are associated with smaller spatial scales and a thinner nonlinear mixing layer. Setting the growth rate equal to $1/\Delta t_{\rm SNe}$ yields a wavelength $\lambda \sim \Delta t_{\rm SNe}v_{\rm rel}(\rho_{{\rm high}}/\rho_{\rm low})^{-1/2}$.
We can multiply by the typical nonlinear velocity difference for shear-driven instabilities,
$\delta v \sim \omega \lambda \sim v_{{\rm rel}}(\rho_{{\rm high}}/\rho_{{\rm low}})^{-1/2}$,
yielding 
\begin{align}
\label{eq:lambda_dv}
\lambda\delta v& \sim \frac{\Delta t_{\rm SNe}v_{{\rm rel}}^{2}}{\rho_{{\rm high}}/\rho_{{\rm low}}}\\&=0.1\,{\rm pc\, km\,s^{-1}}\left(\frac{v_{{\rm rel}}}{10\,{\rm km\,s^{-1}}}\right)^{2}\left(\frac{\rho_{{\rm high}}/\rho_{{\rm low}}}{100}\right)^{-1}\left(\frac{\Delta t_{{\rm SNe}}}{0.1\,{\rm Myr}}\right).
\end{align}
Note that in the above, we focus on parametric scalings and omit factors of $2 \pi$.
For SBs expanding in the multiphase ISM, corrugation and tangential velocities may develop as a result of the shock overrunning cold, dense clouds. With hot gas flowing in to the interface at velocities of several hundred $\rm km\,s^{-1}$, depending on the level of corrugation in the front, the tangential velocity difference could be $10\lesssim v_{\rm rel}/(\rm km\,s^{-1}) \lesssim 300$.  Our simulations with conduction typically have $\rho_{\rm high}/\rho_{\rm low} \sim 100$ at the bubble/shell interface, so for $\Delta t_{\rm SNe} = 0.1$\,Myr, a plausible range of $\lambda \delta v$ is $0.1\lesssim\lambda\delta v/({\rm pc\,km\,s^{-1}})\lesssim10$. We also note that the value of $\lambda \delta v$ could in principle change over time, since at late times (a) individual SNe blast waves no longer arrive at the interface as discrete shocks, (b) the dominant instability may change, and (c) a significant fraction of the mixing and cooling may occur at the edges of cool clouds within the SB interior, not at the shell-interior interface.

We emphasize that the above is intended solely to provide a sense of the plausible level and parametric scalings for  the coefficient in mixing-driving energy exchange, and should not be taken as more than a conceptual guide. 
The ``correct'' value of $\lambda \delta v$, as well as any parameter dependencies and time dependencies, is presently quite uncertain. 
Although instabilities do aid mixing in 3D simulations of SBs, they are likely only marginally resolved in existing global simulations, and/or may be seeded by grid-scale numerical noise rather than density and velocity perturbations that are representative of the real  turbulent, multiphase ISM.  
To constrain the mixing-driven diffusivity, focused high-resolution three-dimensional simulations of the interface that allow for a realistic inhomogeneous ambient medium to seed structure are essential. As we shall show, the net cooling rate is related to $\lambda \delta v$, so a possible path forward in future work is to use the cooling rates from 3D simulations as a calibration point and adjust $\lambda \delta v$ to match their energetics. We will explore such calibration further in future work.

Given the uncertainty in the mixing rate, we compare simulations with a range of values of $\lambda \delta v$, representing different efficiencies of nonlinear mixing, throughout this paper. We find that the cooling rate in the simulations is not fully converged at the resolution we can achieve for $\lambda \delta v < 1\,\rm pc\,km\,s^{-1}$, so in most cases we vary $\lambda \delta v$ between 1 and 10 pc\,km\,s$^{-1}$. For these values of $\lambda \delta v$, $\kappa_{S}\gg\kappa_{{\rm mix}}$ in the bubble interior, and $\kappa_{{\rm mix}}\gg\kappa_{S}$ in the cool shell. $\kappa_{{\rm mix}}$ begins to dominate over $\kappa_{S}$ in the region of the interface where the temperature falls below $\approx2\times10^{5}\,{\rm K}$ and $n_{{\rm H}}$ exceeds $\approx0.2\,{\rm cm^{-3}.}$ We show results for a wider range of $\lambda \delta v$ in Appendix~\ref{sec:subgrid} and compare the numerical convergence of simulations with and without mixing in Appendix~\ref{sec:res_test}. 

\subsection{Energy injection}
\label{sec:inject}
In our simulations, SNe explode with a uniform time spacing $\Delta t_{\rm SNe}$. When a SN occurs, we inject $E_{\rm SN}=10^{51}$\,erg of kinetic energy near
the inner boundary of the simulation domain. This energy is added to an ejecta mass of $M_{\rm ej} = 10\,M_{\odot}$, approximately representing the IMF-averaged total mass of SNe ejecta and swept-up circumstellar material. Within the injection region, we use a linear ejecta radial velocity profile, such that the radial velocity is 0 in the innermost cell and increases linearly to the outer edge of the injection region. For a given injection region size, the normalization of the velocity profile is set such that the total kinetic energy added per SN is $E_{\rm SN}$. 
Our fiducial simulations only include energy injection from SNe, which are expected to dominate the feedback energy budget over the lifetime of a young cluster. We consider the effects of stellar winds at early times in Section~\ref{sec:winds}.

We use a default injection region of $2 < r/{\rm pc} < 4$, with a reflecting inner boundary at $r=2$\,pc. We experimented with a range of injection region sizes, radial velocity profiles, and ejecta temperatures, and with injecting thermal rather than kinetic energy. In all cases, we find that although the properties of the bubble interior near the center of the SB are sensitive to the injection scheme, the structure of the SB near the shell/bubble interface and the integrated properties of the SB (i.e. radius, momentum, energy, and hot gas mass) are not. 

\subsection{Resolution}
\label{sec:res_intro}
We discuss convergence with resolution in Appendix~\ref{sec:res_test} but briefly summarize the results here. For 1D simulations without conduction or  mixing, convergence is difficult to achieve because cooling losses decrease with increasing resolution. In this case, most of the cooling losses within the SB occur in the thin interface between the cool shell and hot interior, where the temperature falls from $T\sim 10^6\,{\rm K}$ to $T\sim 10^3\,{\rm K}$, passing through the peak of the cooling curve at $T\sim 10^5 \rm K$. Without some diffusive process in the problem, at late times nothing in the problem sets a physical scale for the interface.\footnote{At early times, shocks from individual SNe propagate into the shell; this is discussed in Section~\ref{sec:discrete_shocks}.} As a result, the region in which significant late-time cooling occurs becomes narrower and narrower with increasing resolution, such that cooling losses become negligible at very high resolution. In the real ISM, the transition from high to low temperature occurs over a macroscopic scale due to a combination of thermal conduction and physical mixing. To represent the latter in 1D simulations, we apply an effective diffusion model. As a result of including explicit terms for conduction and mixing, our simulations converge at relatively low resolution.

Because the simulations with conduction are computationally more expensive (see Appendix~\ref{sec:kappa_max}), we compare simulations with and without conduction at different resolution levels. The fiducial simulations we show with conduction and mixing have $\Delta r = 0.024$\,pc and are well converged. Those without conduction have a higher resolution of $\Delta r = 0.0016$\,pc. Even at this resolution, the no-conduction case is not completely converged, but we show in Appendix~\ref{sec:res_test} that the integrated bubble quantities in the no-conduction simulations asymptotically approach the predictions of the classical SB theory without cooling (Section~\ref{sec:predictions}). Except where explicitly noted, all the simulations are converged at the resolution we present. 

\subsection{Definitions}
\label{sec:definitions}
We use the following definitions for different components of the SB in analyzing the simulations. We define the ``bubble'' component to include all gas with $T>10^5$\,K or $|v_{r}|>1\,\rm km\,s^{-1}$. This effectively selects all gas that has been perturbed; we define the ambient medium to include all gas that is not part of the bubble. We define the ``hot'' component of the bubble as all cells with $T > 10^5$\,K, and the ``shell'' component as all gas with $|v_{r}|>1\,\rm km\,s^{-1}$ and $T<2\times10^4$\,K. 

For each component, we calculate the integrated mass, radial momentum, and kinetic, thermal, and total energy by summing over the relevant densities for all cells within that component, multiplied by each cell's volume. We define the average temperature of gas in the hot component in terms of its total thermal energy:
\begin{align}
T_{{\rm hot}}\equiv \frac{2}{3}\frac{\mu m_{p}}{k_{b}}\frac{E_{{\rm hot,\,th}}}{M_{{\rm hot}}}.
\end{align}
We define $R_{\rm bubble}$ as the outer edge of the outermost cell in the bubble component and $R_{\rm shell}$ as the mass-weighted mean radius of cells in the shell component. The expansion velocity of the shell is $v_{{\rm shell}}={\rm d}R_{{\rm shell}}/{\rm d}t$.

We often normalize the SB energy, radial momentum, and hot gas mass by the number of SNe that have exploded prior to a given time. This is denoted with a caret; e.g.,
\begin{align}
\hat{E}_{{\rm SB}}\left(t\right)\equiv\frac{E_{{\rm SB}}\left(t\right)}{N_{{\rm SNe}}\left(t\right)}=\frac{E_{{\rm SB}}\left(t\right)}{\left\lfloor t/\Delta t_{{\rm SNe}}\right\rfloor +1},
\end{align}
where $\left\lfloor x \right\rfloor$ denotes the floor function of $x$.

\section{Fiducial Simulation Results}
\label{sec:results}
We present our simulation results in three parts. 

First, in this section, we examine in detail the dynamics and internal structure of a single ``fiducial'' simulation: one with ambient density $n_{\rm H,0}=1\,\rm cm^{-3}$ (comparable to the mean ISM density in the solar neighborhood) and a SNe interval $\Delta t_{\rm SNe}= 0.1\,\rm Myr$. Assuming a SNe rate of 1 SN per 100\,M$_{\odot}$ of stellar mass formed, and a constant average SNe rate over $\sim$\,40 Myr, similar to the predictions of stellar population synthesis models for a \citet{Kroupa_2001} IMF \citep[e.g.][]{Leitherer_1999}, the fiducial SNe rate of $\Delta t_{\rm SNe}= 0.1\,\rm Myr$ corresponds to a relatively large star cluster with $M_{\rm star}\sim 4\times 10^4\,M_{\odot}$. For our fiducial model we adopt $\lambda \delta v = 1\,\rm pc\,km\,s^{-1}$ when mixing is turned on.

Then, in Sections~\ref{sec:pred_cooling_loss} and~\ref{sec:discrete_shocks}, we present results for simulations with a range of SNe rate, ambient density, and nonlinear mixing efficiency. 

\subsection{Radial profiles}
\label{sec:profiles}

\begin{figure}
\includegraphics[width=\columnwidth]{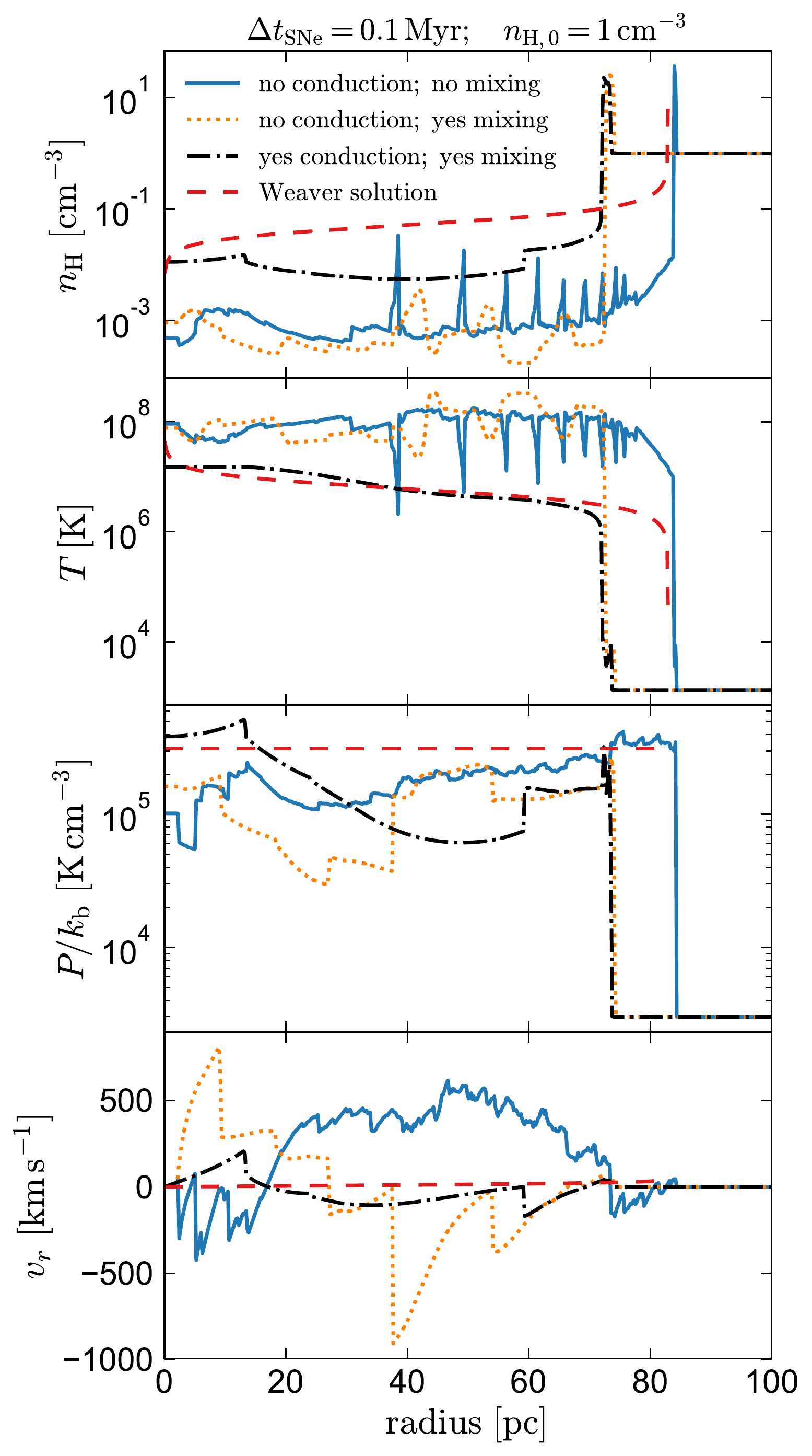}
\caption{Radial profiles of hydrogen number density, temperature, pressure, and radial velocity, for simulation snapshots at $t=1$\,Myr. 
We compare results for a simulation with no conduction or nonlinear mixing (blue), one with mixing but no conduction (gold), and one with both conduction and mixing (black). All models adopt fiducial values for the SN interval and ambient density, i.e.  $\Delta t_{\rm SNe}= 0.1\,\rm Myr$ and  $n_{\rm H,0}=1\,\rm cm^{-3}$.  }We also show (red) the analytic solution for continuous energy injection from \citet{Weaver_1977}, which includes conduction but neglects radiative cooling. In the SB interior, the simulation with conduction has higher density and lower temperature than those without conduction, but its density and pressure are {\it lower} than predicted by the Weaver solution.
\label{fig:profiles}
\end{figure}

Figure~\ref{fig:profiles} shows radial profiles of density, temperature, pressure, and radial velocity for simulations with our fiducial problem setup at $t=1$ Myr after SNe commence. The fiducial version (black) includes both conduction and nonlinear mixing. We also compare to a simulation without conduction or mixing (blue), and to one with only mixing but no conduction (gold).\footnote{We do not show a simulation with conduction and no mixing, because in this case the cooling rate is not converged at the resolution level our simulations can currently achieve, and the asymptotic behavior at high resolution is uncertain (see Appendix~\ref{sec:res_test}).}  

In the simulation without conduction or mixing, there is a large degree of substructure in the radial profiles of all quantities. Such substructure is absent in simulations of a single SN explosion; it arises from the interference of the forward and reverse shocks generated by different SNe within the SB. Even at late times, contact discontinuities caused by discrete SNe can be seen in the density and temperature profiles. These structures would very likely not survive in the interior of real superbubbles, as they would be smoothed out by both instabilities and conduction. 

We note that in the model without conduction but with mixing turned on, the sharpest density and temperature spikes are smoothed out.  While some turbulent mixing is expected to occur in SB interiors, in the real case the parameters characterizing mixing would presumably differ from those at the outer reaches of SBs, unlike our simple prescription.  Even though the detailed interior structure is unrealistic for both this case and the no-mixing, no-conduction case, the important physical point is that the mean interior density and temperature in these cases are similar.

For all quantities, conduction smooths out much of the structure in the bubble interior, erasing the memory of individual SNe. The density and temperature in the bubble interior are significantly higher and lower, respectively, in the simulation with conduction than in those without it, because conduction precipitates an evaporative mass flux from the shell into the bubble, diluting the hot gas in the interior.

The simulation that includes nonlinear mixing but not thermal conduction has a hot, low-density interior similar to the one without conduction or mixing. However, both it and the simulation with conduction and mixing produce a smaller SB at fixed time than the simulation with no conduction or mixing. As we shall show, the size and energy of the SB are set primarily by mixing, which determines the efficiency of cooling losses. Once the cooling rate is set, conduction sets the density and temperature of the SB interior. We note that the internal temperature in the interior is insensitive to cooling in the shell, however.  

Figure~\ref{fig:profiles} includes (in red) the profiles predicted by the Weaver similarity solution. The detailed internal structure of all simulations is different from the Weaver prediction, but much of this disagreement is a result of the discreteness of SNe explosions in the simulations; particularly at early times, the radial profiles in the bubble interior fluctuate significantly on short timescales whenever a new SN explodes. 

As expected, the Weaver solution predicts a higher-density and lower-temperature interior than is produced by the simulations without conduction. The temperature in the fiducial simulation with conduction and mixing is very close to the Weaver solution, but the density is lower.  As we will discuss in Section~\ref{sec:fluxes}, this is primarily a consequence of cooling losses in the shell/bubble interface, which decrease the evaporative mass flux into the bubble. 
The Weaver solution does not account for cooling within the bubble interior or bubble/shell interface, so the bubble size in the simulation without conduction or mixing, which also has very small cooling losses, is similar to that predicted by the Weaver solution. 

\subsection{Integrated SB properties}
\label{sec:integrated_props}

\begin{figure*}
 \includegraphics[width=\textwidth]{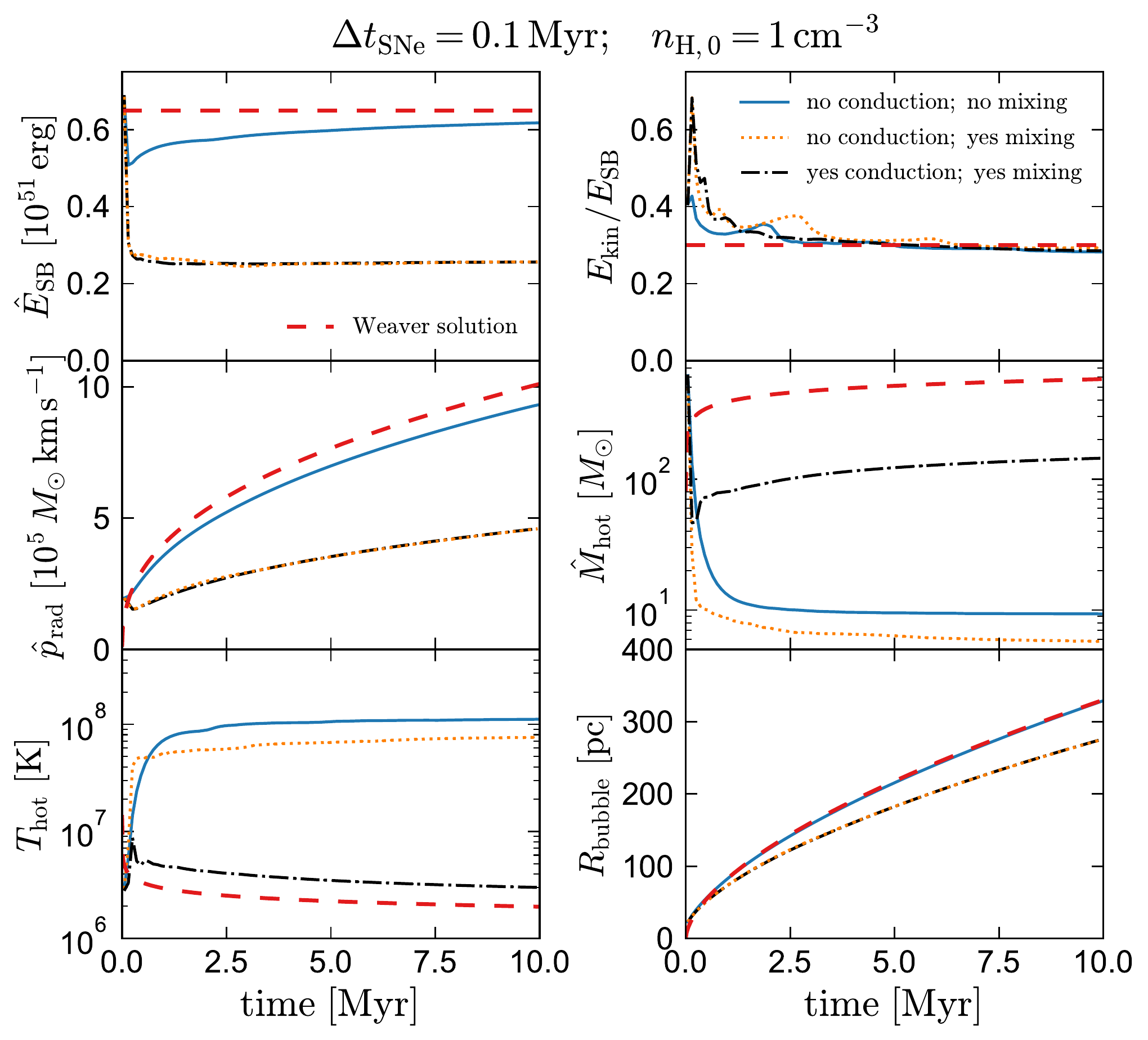}
\caption{Integrated SB properties as a function of time, averaged over 0.1\,Myr. As in Figure \ref{fig:profiles},  we compare results for a simulation without conduction or nonlinear mixing (blue), one with mixing but no conduction (gold), the fiducial simulation with both conduction and mixing (black), and the \citet{Weaver_1977} analytic solution for continuous energy injection and conduction but no cooling (red). The total bubble energy, radial momentum, and hot gas mass are normalized by the number of SNe that have exploded prior to a given time. Due to increased cooling, the simulations that include mixing have significantly lower SB energy and momentum than the simulation that does not. 
However, the simulation that includes conduction and the Weaver solution both produce much more hot gas in the bubble interior than those that do not, and have much lower temperatures. That is, mixing sets the cooling losses, but conduction sets the interior temperature and hot gas mass.}
\label{fig:integrated_bubble_properties}
\end{figure*}

We investigate the temporal evolution of the fiducial SB model in Figure~\ref{fig:integrated_bubble_properties}, which compares bubble-integrated properties for the same set of simulations and analytic Weaver model shown in Figure~\ref{fig:profiles}.

The total bubble energy and radial momentum are a factor of $\sim$2 lower in the simulations with mixing than in the one without it. Comparing the two simulations that include mixing, the radius, total energy, and momentum are essentially identical; i.e., once mixing is included, thermal conduction has a negligible effect on the cooling efficiency and dynamics of the SB. However, the difference between the simulations with and without conduction becomes clear when comparing the mass and temperature of hot gas in the bubble interior. In the simulations without conduction, $\lesssim 10 M_{\odot}$ of hot gas are added to the bubble per SN; this can be explained entirely by the ejecta mass that is added at each SN injection. With conduction, $M_{\rm hot}$ increases by $\sim 150\,M_{\odot}$ per SN at late times, primarily as a result of evaporation into the bubble from the shell. 

The dynamics of the simulation {\it without} conduction or mixing (blue) are in good agreement with the predictions of the Weaver similarity solution. The internal energy, size evolution, and ratio of thermal to kinetic energy in the Weaver solution are calculated without any consideration of the effects of conduction on the bubble interior: they follow from the assumptions of force balance and adiabatic expansion (Section \ref{sec:radial_ev}). At high resolution, cooling losses in the interface become very small in the simulation without conduction or mixing (Appendix~\ref{sec:res_test}), leading to similar SB energetics to what is predicted by the Weaver solution. 
Indeed, in the limit where cooling within the bubble and interface vanishes (or in the limit of an infinitesimally thin interface; see \citealt{Gentry_2017}), the energy, size, and momentum evolution should reproduce the Weaver solution almost exactly, modulo any modifications that result from the discreteness of SNe in the limit of large $\Delta t_{\rm SNe}$. 

In terms of $M_{\rm hot}$ and $T_{\rm hot}$, the results from the  simulation with conduction and mixing are in better agreement with the Weaver solution  than the simulations without conduction, but still have slightly higher temperature and a factor $\sim$\,4
less hot gas mass than the Weaver solution predicts. As we will show in Section~\ref{sec:fluxes}, both deviations are the result of cooling in the interface. 

The ratio of kinetic to thermal energy at late times is similar in all simulations and is comparable to the value $E_{\rm kin}/E_{\rm SB} = 3/10$ predicted by the Weaver solution. For a pressure-driven shell, this quantity depends only on the bubble's size evolution $R(t)$; the classical power law of $R\left(t\right)\sim t^{3/5}$ that is predicted for a constant energy input rate leads to $E_{\rm kin}/E_{\rm SB} = 3/10$. As we show in Section~\ref{sec:pred_cooling_loss}, including cooling in the interface does not change this prediction.

\subsection{Conduction, evaporation, and cooling in the interface}
\label{sec:fluxes}

\begin{figure}
\includegraphics[width=\columnwidth]{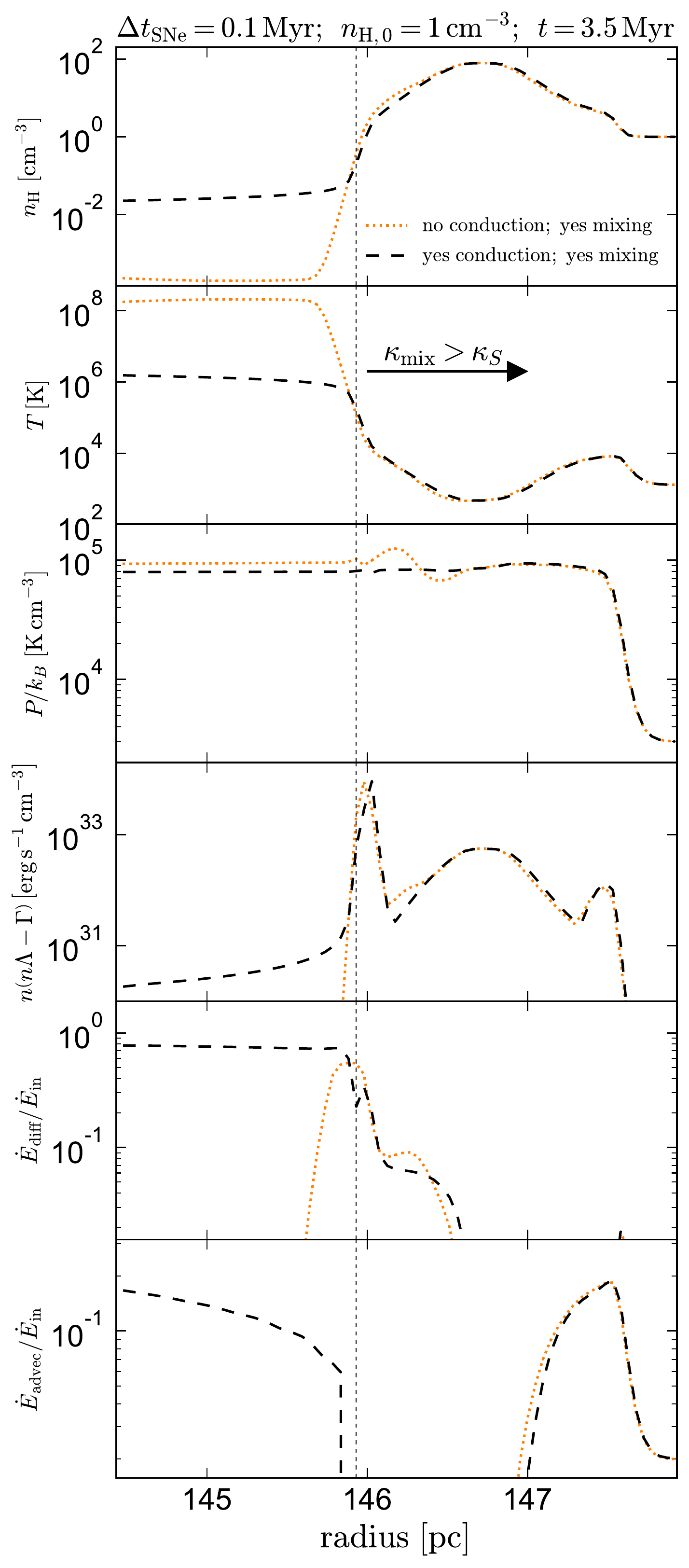}
\caption{Profiles of density, temperature, pressure, net cooling, diffusive energy flux (outward; including both Spitzer conduction and mixing terms), and advective flux (inward) in the shell/bubble interface. We compare the fiducial simulation that includes conduction and mixing (black) to one that includes mixing but no conduction (gold). The dashed vertical line shows where, in the simulation with conduction, the effective conductivity due to mixing (Equation~\ref{eq:kappa_mix}) begins to dominate over the Spitzer conductivity (Equation~\ref{eq:spitzer}).
In the bubble interior, the profiles of the two cases differ. In the simulation with thermal conduction, evaporation of material from the shell creates a mass flux inwards (lowest panel), which raises the density and lowers the temperature (top two panels) in the interior.}
\label{fig:interface}
\end{figure}

We now examine the interface between the cool shell and hot bubble interior. Figure~\ref{fig:interface} shows a zoom-in of the shell at $t=3.5$\,Myr, once shocks from discrete SNe have subsided and the interface has settled into quasi steady-state. We compare the fiducial simulation with conduction and nonlinear mixing (black) to the version with mixing but no conduction (gold). Within the shell and cooling region, the two simulations are very similar. In both cases, the interface between the bubble and shell is narrow, with the transition from $T\sim 10^{6}$\,K and $n_{\rm H}\sim 10^{-2}\,\rm cm^{-3}$ in the hot interior to $T\sim 10^{2}$\,K and $n_{\rm H}\sim 10^{2}\,\rm cm^{-3}$ in the shell occurring over a scale of $<1$\,pc. 
Most of the cooling occurs in a narrow  region in the interface on the inner edge of the shell, where the temperature passes through $T \sim 10^{4-5}$\,K. Some additional cooling occurs in the outer edge of the shell, where shocked ISM being swept up by the expanding bubble is heated above the temperature of the ambient ISM. 

In the simulation with conduction, the diffusive energy flux (the sum of conductive and mixing terms; 5th panel of Figure~\ref{fig:interface}) is nearly flat  interior to the shell's inner edge. Conduction sets the shape of the temperature profile near the edge of the bubble such that the conductive heat flux, $q \sim -r^2T^{5/2} {\rm d}T/{\rm d}r$, is approximately constant with radius, at least over scales that are small compared to the size of the SB. Most of the conductive energy flux is deposited directly in the shell/bubble interface; i.e., the conductive flux drops rapidly in the region where cooling losses are significant and goes to 0 deep in the shell. 

The dotted vertical line delimits the region to the left where (Spitzer) conduction dominates the diffusive energy flux, and to the right where mixing dominates the diffusive flux.  For both simulations, $\dot E_{\rm diff}$ is similar in the high-density, low-temperature regime, where mixing dominates over Spitzer conduction and cooling becomes strong.
In the simulation without conduction, the heat flux becomes negligible in the bubble interior, where $\kappa_{\rm mix}$ is small. 

The advective energy flux from evaporation of the shell into the bubble (bottom panel) is significantly smaller than the diffusive flux at late times; as we discuss below, this is a consequence of radiative losses in the interface. In the simulation without conduction, the advective flux is small and often negative (i.e. mass and energy are advected into the shell from the interior). For the simulation snapshot shown at $t=3.5$\,Myr, the advective energy flux in the simulation with conduction rises gradually toward the SB interior. However, the magnitude and shape of the advective flux profile varies with time as sound waves from individual SNe arrive at the interface.

We now compare the relative contributions of conduction, evaporation, and cooling in the interface region. We calculate the energy fluxes as follows. At each snapshot, we define the outer edge of the hot interior as the outermost zone with $T> 10^5\,\rm K$. We then measure the fluxes through a surface {\it 2 pc interior to this zone}. Because the temperature and pressure gradients in the shell/bubble interface are steep and the fluxes vary rapidly there, measuring the conductive and advective fluxes through a surface slightly interior to the bubble leads to more stable results. We note that in this region, the conductive flux is much larger than the flux from mixing.

The total energy fluxes are:
\begin{align}
\label{eq:dot_E_cond}
\dot{E}_{\rm diff}&=-4\pi r^{2}\kappa\frac{{\rm d}T}{{\rm d}r}\\ 
\label{eq:dot_E_advec}
\dot{E}_{\rm advec}&=4\pi r^{2}\left(v_{\rm shell}-v\right)\left({\cal E}+P\right)\approx  \frac{5}{2} \dot m c_{\rm s,iso}^2,
\end{align}
representing the energy diffusion (conduction plus mixing terms) into the shell and the advection out of the shell, respectively. 
Note that the sign of the velocity difference in $\dot{E}_{\rm advec}$ is such that an inward flow relative to the expanding shell (evaporation) is a positive advection flux.  Correspondingly, a bulk flow that carries energy into the shell (condensation) is a negative advection flux.  

Our results are not sensitive to the particular choice of measuring energy fluxes through a surface 2 pc interior to the shell boundary. As Figure~\ref{fig:interface} shows, the energy fluxes (particularly the advection term) are not strictly constant in radius at any particular time, but ``ripples'' from individual SNe explosions cancel out in a time-averaged sense, such that the time-averaged fluxes are consistent through any surface reasonably close to the shell boundary.

In calculating cooling losses, we wish to separate the cooling of shocked ISM exterior to the contact discontinuity (i.e., on the outer edge of the shell) from cooling in the interface between the hot interior and the cool shell. Only cooling in the interface on the inside of the shell, which removes energy that was carried in by conduction, is relevant for the energy balance between conduction and evaporation. Cooling in the outer edge of the shell primarily removes thermal energy from the $P{\rm d}V$ work done by the expanding bubble, so it should be discounted in calculations of the energy flux through the interface. 

Gas can in principle mix through the shell, and some cooling does occur within the shell, so it is not possible to cleanly separate cooling of shocked ISM on the shell exterior from cooling in the interface. However, because the temperature in the shell at late times is always well below the peak of the cooling curve, the fraction of cooling that occurs within the shell is sub-dominant. We define the boundary between cooling in the interface and cooling of shocked ISM on the shell exterior at the first local minimum in the temperature profile within the shell component. We thus define the cooling in the bubble/shell interface as: 
\begin{equation}       
\label{eq:L_int}
{\cal L}_{\rm int}\equiv \int_0^{r(T=T_{\rm min,\,shell})} n\left[n\Lambda\left(T\right)-\Gamma\left(T\right)\right] 4 \pi r^2 {\rm d}r.  
\end{equation}
$\mathcal{L}_{\rm int}$ in principle includes some cooling within the bubble interior but Figure~\ref{fig:interface} shows that this is negligible compared to the cooling in the interface.

\begin{figure}
\includegraphics[width=\columnwidth]{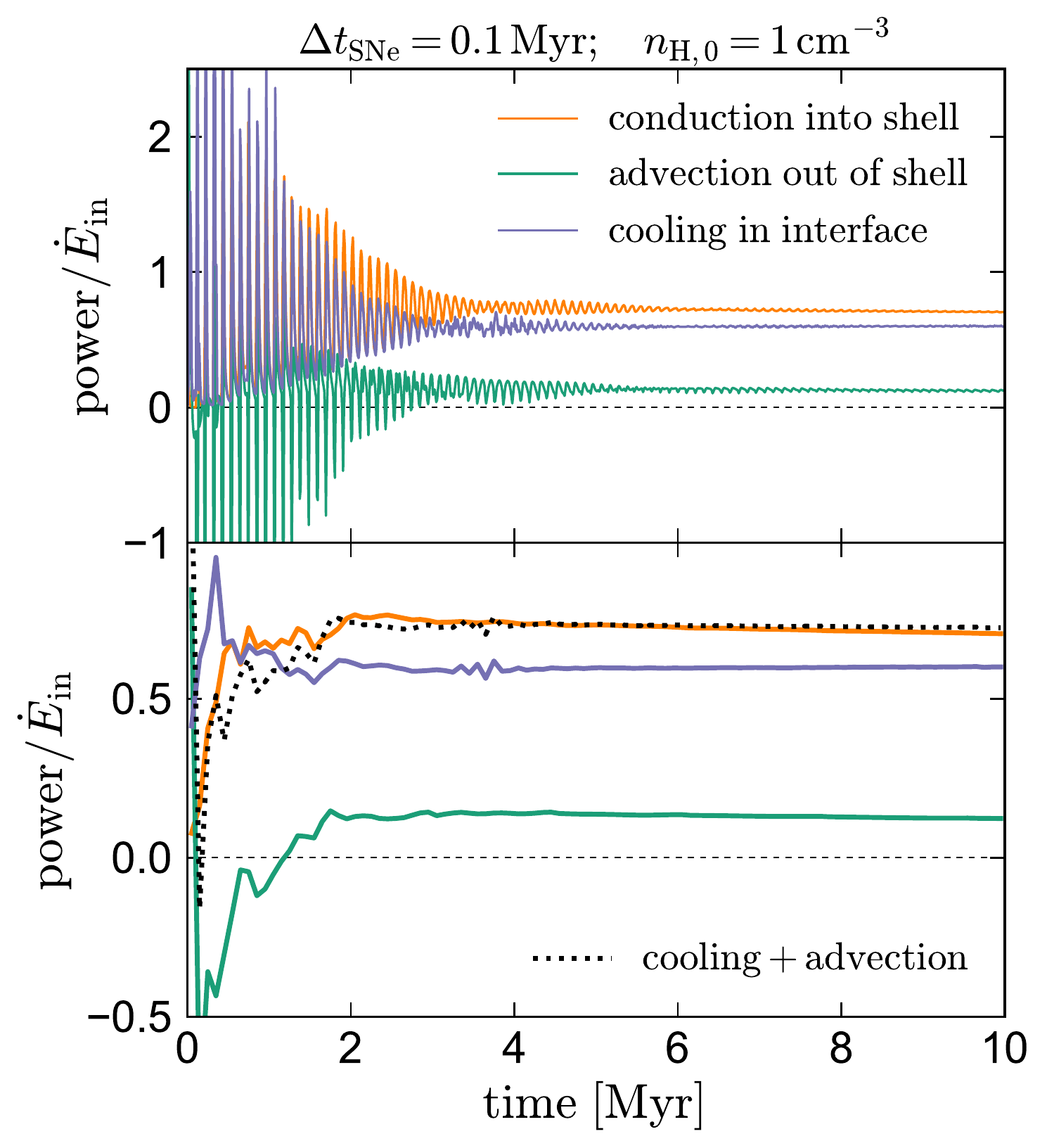}
\caption{Energy fluxes through the shell/bubble interface as a function of time. We compare the total conductive flux from the bubble to the shell (gold; Equation~\ref{eq:dot_E_cond}), the total advective flux from evaporation of the shell into the bubble (green;  Equation~\ref{eq:dot_E_advec}), and the net energy lost to cooling in the interface (purple; Equation~\ref{eq:L_int}). We average each term over 0.01 Myr (top) and 0.1 Myr (bottom). 
Cooling removes the majority of the energy conducted into the shell/bubble interface.
Thermal conduction is balanced by the sum of cooling and advection.}
\label{fig:energy_fluxes}
\end{figure}

Figure~\ref{fig:energy_fluxes} shows the energy flux through the interface as a function of time for the fiducial simulation with conduction and mixing. We compare the conductive energy flux into the shell, the advective energy flux out of the shell due to evaporation, and the cooling losses in the shell/interior interface. All quantities are normalized by the total time-averaged energy input rate from SNe. The top and bottom panels show the same fluxes, with the smoothing timescale longer in the bottom panel. 

The classical Weaver model, which neglects cooling in the interface, predicts that energy fluxes from conduction and advection should exactly cancel. If this were true, the gold and green lines in Figure~\ref{fig:energy_fluxes} would be equal. In fact this prediction does not hold true for the simulation: at late times, the conductive heat flux into the shell is a factor $\sim$5 larger than the advective energy flux from shell evaporation. Cooling losses are quite significant, with $\sim$4 times more energy radiated away from the interface than is evaporated from the shell. The relative importance of cooling and advection is relatively constant; i.e., the cooling efficiency does not vary significantly with time. The dotted line in the bottom panel of Figure~\ref{fig:energy_fluxes} shows the total interface energy losses from the combination of evaporation and cooling. This tracks the total energy conducted into the interface quite closely. This means that the basic picture of energy balance in the shell does hold, once cooling in the interface is accounted for. 

\begin{figure}
\includegraphics[width=\columnwidth]{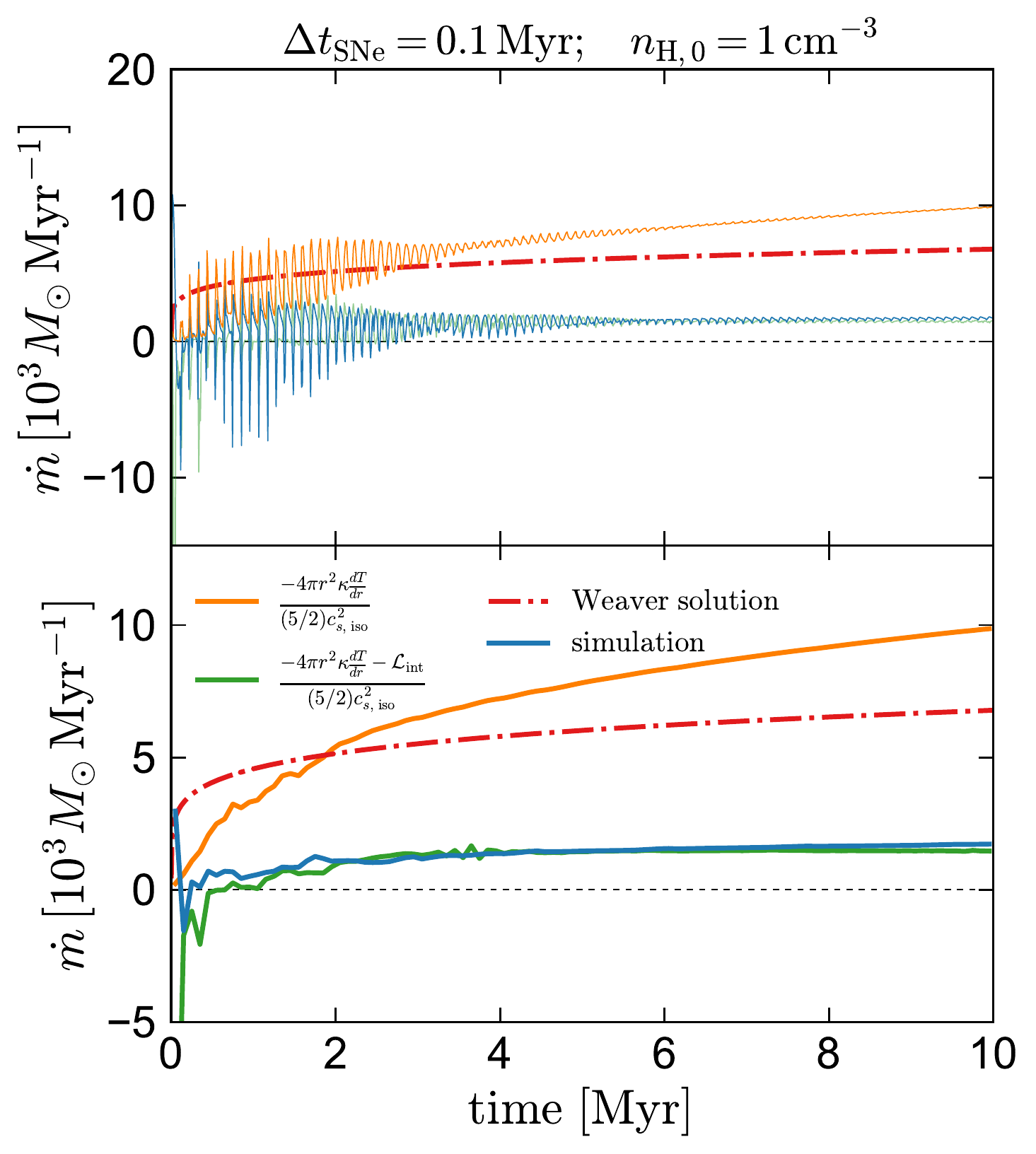}
\caption{Mass flux from the shell into the bubble. We average over 0.01 Myr (top) and 0.1 Myr (bottom). We compare the true mass flux in the simulation (blue), the value predicted by the \citet{Weaver_1977} model with 
conduction but no cooling (red), the mass flux expected if advection and conduction exactly cancel and cooling is ignored (gold), and the mass flux expected if advection and conduction exactly cancel and cooling is accounted for (green). The mass flux in the simulation is consistent with an evaporative flow carrying away all the energy that is conducted in and not lost to cooling. }
\label{fig:mass_fluxes}
\end{figure}

We next consider the {\it mass} flux into the bubble interior from evaporation. In the classical model without cooling, this is predicted by Equation~(\ref{eq:mdot}), but cooling in the shell/bubble interface significantly reduces the evaporation rate. In fact, if cooling is sufficiently strong, evaporation can be reversed and replaced with condensation of hot gas onto the shell \citep[e.g.][]{McKee_1977}.

Figure~\ref{fig:mass_fluxes} shows the mass flux from the shell into the bubble. The blue line shows the true mass flux measured in the simulation, $\dot{m}=4\pi r^2 (v_{\rm shell} - v)\rho$, while the gold line shows the mass flux predicted if advection and conduction exactly cancel and cooling in the interface is negligible (Equation~\ref{eq:mdot}), as is assumed in classical models. The true time-averaged mass flux is a factor of $\sim$5 lower than this prediction at late times, because cooling rather than advection removes most of the energy from the interface. 

When cooling in the shell-bubble interface is important, a term $\mathcal{L}_{\rm int}$ must be added to the left-hand side of Equation~(\ref{eq:energy_balance}), representing the losses from cooling in the interface. In this case, less evaporation from the shell is required for energy balance, because some of the energy conducted into the shell is lost to cooling. The revised mass flux is then 
\begin{align}
\label{eq:revised_mdot}
\dot{m}=\frac{-4\pi r^2\kappa\left({\rm d}T/{\rm d}r\right)-\mathcal{L}_{\rm int}}{\left(5/2\right)c_{s,{\rm \,iso}}^{2}}.
\end{align}
 
The green line in Figure~\ref{fig:mass_fluxes} shows the mass loss predicted from energy balance in the shell after cooling is accounted for (Equation~\ref{eq:revised_mdot}). This matches the true mass flux within 10\% at late times; any residual disagreement can likely be attributed to ambiguity in the separation of cooling in the interface from cooling of shocked ISM on the outside of the shell, which can lead the total cooling in the interface to be slightly over- or under-estimated.  The hot gas mass in the bubble can thus be understood purely as a consequence of the evaporation required to carry away ``surplus'' conduction energy that radiation is unable to remove. 

The red line in Figure~\ref{fig:mass_fluxes} shows the mass flux rate predicted by the Weaver solution (Equation~\ref{eq:mdot_cowie_mckee}). 
It is not identical to the gold line because the shape and normalization of the temperature profile inside the simulated SB differ somewhat from the predictions of the Weaver solution (see Figure~\ref{fig:profiles} and~\ref{fig:integrated_bubble_properties}). The higher temperature in the simulation is itself a result of cooling, which leads to less mass in the bubble interior. 

Figures~\ref{fig:energy_fluxes} and~\ref{fig:mass_fluxes} also show that at early times ($t\lesssim 3$\,Myr for the fiducial simulation parameters), all the energy and mass fluxes fluctuate significantly on a timescale comparable to $\Delta t_{\rm SNe}$. In particular, the mass and energy advection fluxes become negative for a brief period once every $\Delta t_{\rm SNe}$, when the blast wave from a new SNe reaches the shell. During these periods,  hot gas is being pushed {\it into} the shell instead of flowing out of it. While some minor fluctuations persist at later times, the advective fluxes in the bubble remain positive after $t\approx 3$\,Myr. As we will show in Section~\ref{sec:discrete_shocks}, the time after which blast waves from individual SNe cease pushing mass into the shell corresponds to the time when they become subsonic before reaching the shell.

\section{Modified solution with cooling}
\label{sec:pred_cooling_loss}

\subsection{Dynamics}
\label{sec:sb_dyn}
Our simulations show that at late times, the fraction of the total energy input that is lost to cooling is approximately constant in time, varying primarily with the ambient ISM density. We now provide a simple analytic argument to predict the bubble dynamics in the presence of cooling. 

We define the fraction of the energy input that is lost to cooling in the interface to be $\theta$, such that the cooling rate in the interface is $\mathcal{L}_{{\rm int}}\equiv \theta\dot{E}_{{\rm in}}$. The SB momentum equation (Equation~\ref{eq:mom_bubble}) is unchanged from the case without cooling, but in the internal energy equation (Equation~\ref{eq:energy_bubble}),   $\dot{E}_{\rm in}$ is replaced with $(1-\theta)\dot{E}_{\rm in}$.   
All the dynamical predictions of classical SB theory (Section~\ref{sec:radial_ev}; Equations \ref{eq:weaver_R_t}-\ref{eq:Ekin_Weaver}) still hold, after the substitution $\dot{E}_{\rm in}\to (1-\theta)\dot{E}_{\rm in}$.

We can then write the radius in terms of dimensional quantities as
\begin{multline}
\label{eq:weaver_R_t_cooling} R(t)=83\,{\rm pc}\times \left(1-\theta\right)^{1/5}\left(\frac{E_{{\rm SN}}}{10^{51}\:{\rm erg}}\right)^{1/5}\left(\frac{\Delta t_{{\rm SNe}}}{0.1\,{\rm Myr}}\right)^{-1/5}\times \\
\left(\frac{\rho_{0}}{1.4\,m_{p}\,{\rm cm^{-3}}}\right)^{-1/5}\left(\frac{t}{1\,{\rm Myr}}\right)^{3/5} \end{multline}
The radial evolution from Equation~(\ref{eq:weaver_R_t_cooling}) can also be translated to a prediction for the radial momentum per supernovae, $\hat{p}_{{\rm rad}}=\left(4\pi R^{3}\rho_{0}/3\right)\left({\rm d}R/{\rm d}t\right)/\left(t/\Delta t_{{\rm SNe}}\right)$: 
\begin{multline}
\label{eq:prad_cooling}
\hat{p}_{{\rm rad}}(t)=4\times10^{5}\,M_{\odot}\,{\rm km\,s^{-1}} \left(1-\theta\right)^{4/5}\left(\frac{E_{{\rm SN}}}{10^{51}\:{\rm erg}}\right)^{4/5} \times \\ \left(\frac{\Delta t_{{\rm SNe}}}{0.1\,{\rm Myr}}\right)^{1/5} \left(\frac{\rho_{0}}{1.4\,m_{p}\,{\rm cm^{-3}}}\right)^{1/5}\left(\frac{t}{1\,{\rm Myr}}\right)^{2/5}.
\end{multline}

\begin{figure*}
    \includegraphics[width=\textwidth]{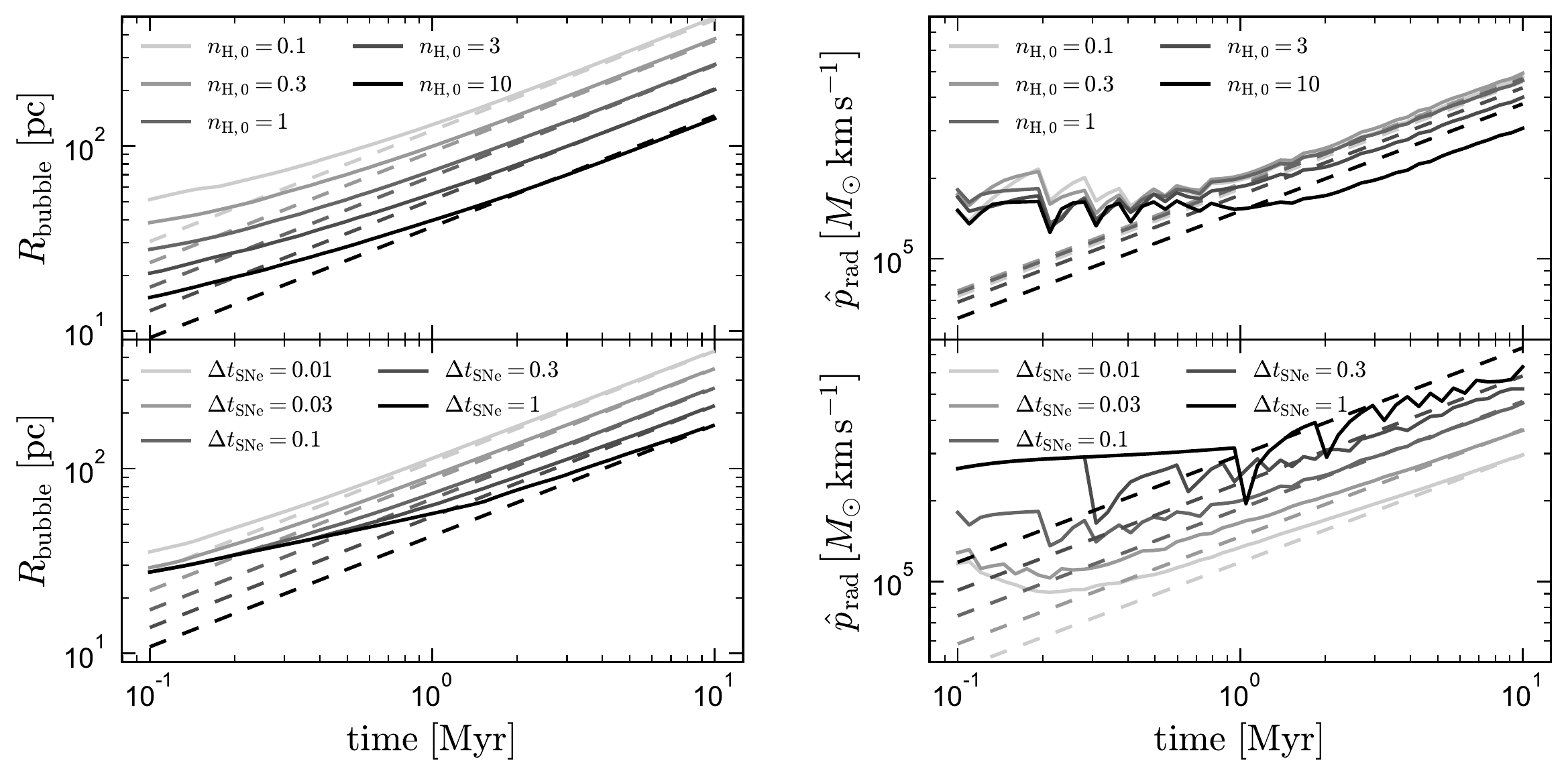}
    \caption{SB radius (left) and radial momentum per SN (right) as a function of time for simulations with different ambient densities (top) and different SN rates (bottom). All models in the upper panels adopt $\Delta t_{\rm SNe} =0.1$ and lower panels adopt $n_{\rm H,0}=1$. Dashed lines show the prediction of our modified solution with cooling (Equation~(\ref{eq:weaver_R_t_cooling}) for $R(t)$; Equation~(\ref{eq:prad_cooling}) for $\hat{p}_{\rm rad}(t)$), with $\theta$ computed from Equations~(\ref{eq:cooling_eth_ratio_norm}) and~(\ref{eq:theta}), using $A_{\rm mix}=3.5$. The simulations are generally in good agreement with the analytic prediction at late times. Deviations from the analytic prediction at early times are due to the discreteness of individual SNe and modified dynamics during the free and adiabatic expansion stages. }
    \label{fig:radius_evolution}
\end{figure*}

In Figure~\ref{fig:radius_evolution}, we compare the evolution of $R(t)$ and $\hat p_{\rm rad}(t)$ predicted by Equations~(\ref{eq:weaver_R_t_cooling}) and (\ref{eq:prad_cooling}) to a suite of simulations with a range of ambient densities (top panel) and a range of SNe rates (bottom panel). Computing the analytic prediction requires an estimate of  $\theta$ consistent with the input parameters for each simulation; we calculate this as described in the next section. We shall show that (at late times) $\theta$ is expected to depend on ambient density $\rho_0$ but not on $\Delta t_{\rm SNe}$. At late times, the agreement between the simulations and analytic prediction is good. At early times, the simulated SBs are larger and grow more slowly with time than predicted by Equation~(\ref{eq:weaver_R_t_cooling}).

The top panels of Figure~\ref{fig:radius_evolution} show that the time at which the simulations reach the behavior predicted by Equations~(\ref{eq:weaver_R_t_cooling}) and (\ref{eq:prad_cooling}) does not vary with ambient density (and thus, it does not vary with $\theta$): in all cases, the simulations converge to the predicted behavior after $1-2$ Myr, corresponding to $10-20$ SNe. The bottom panels show that the time to reach the predicted radial evolution does vary with $\Delta t_{\rm SNe}$: simulations with lower SNe rates reach the analytically predicted behavior at later times. 

For simulations with a low ambient density and/or high SN rate, the SB radius can reach several hundred pc by the end of the simulation. We run all simulations for 10 Myr to facilitate straightforward comparison. However, we note that our idealized simulations with a uniform ambient medium become less realistic  as the bubble grows large ($R\gg 100$\,pc), since real galactic disks are vertically stratified and SBs become elongated and break out once the bubble radius becomes comparable to the disk scale height \citep{Maclow_1988, Maclow_1989, Kim_2017, Kim_2018,Fielding_2018}.

For the set of simulations at fixed $\Delta t_{\rm SNe}$, the predicted and actual values of $\hat p_{\rm rad}$ are quite insensitive to density.  This is true in both early stages dominated by discrete SN events, and in later stages that approach the cooling-modified pressure-driven bubble solution. While Equation~(\ref{eq:prad_cooling}) includes an explicit positive dependence of $\hat p_{\rm rad}$ on $\rho_0$, this is more than compensated by the decrease of $1-\theta$ with $\rho_0$. Given the dependence of $\theta$ on $\rho_0$ that we derive in Section~\ref{sec:cooling_eff}, we expect $\hat{p}_{{\rm rad}}\propto \rho_{0}^{-1/5}$ at $n_{\rm H,0}\gg 1\,\rm cm^{-3}$. For the set of simulations at fixed $n_{\rm H,0}$, the increase of $\hat{p}_{\rm rad}$  with $\Delta t_{\rm SNe}$ is only due to the explicit dependence on $\Delta t_{\rm SNe}$ in Equation~(\ref{eq:prad_cooling}), as $\theta$ is expected to be independent of $\Delta t_{\rm SNe}$.  We note that agreement between the prediction and simulation also occurs later at larger $\Delta t_{\rm SNe}$.

\subsection{Predicting the cooling efficiency }
\label{sec:cooling_eff}
The interface cooling efficiency, $\theta =  {\cal L}_{\rm int}/\dot E_{\rm in}$, is expected to depend on the ambient density and on the efficiency of nonlinear mixing as parameterized by $\lambda \delta v$.
To estimate $\theta$, we assume that  cooling occurs only in a nonlinear mixing layer, where the conductivity is given by Equation~(\ref{eq:kappa_mix}), and that the pressure is constant within the interface.\footnote{We have verified in our simulations that once equilibrium is reached at late times, pressure is always constant (within $\sim$\,5\%; see e.g. Figure~\ref{fig:interface}) through the interface and that most cooling losses ($\gtrsim 95$\,\%) indeed occur in the nonlinear mixing layer.} 
The evaporative energy flux becomes negligible in the cooling layer, where the temperature and enthalpy of the gas are low, so the advection term in the energy equation may be neglected (see Figure~\ref{fig:interface}). For simplicity, we neglect photoelectric heating so the volumetric net cooling rate is $n^2 \Lambda(T)$.

We introduce a local coordinate $\xi = r/R$, where $R$ is the shell radius.  
Energy flux conservation then requires ${\rm d}\left(-\kappa_{\rm mix}{\rm d}T/{\rm d}r\right)/{\rm d}r=-n^{2}\Lambda$, or 
\begin{align}
\label{eq:cooling_equal_cond}
\frac{P}{R^2}(\lambda \delta v) \frac{{\rm d^2}\ln T }{{\rm d}\xi^2} =
\frac{P^{2}}{k_{{\rm B}}^{2}}\frac{\Lambda\left(T\right)}{T^{2}}.
\end{align}
Note that in the above, we have only worked to lowest order in $\xi$, neglecting curvature of the front since the cooling/mixing layer is very thin.

Based on the sharply peaked cooling profile shown in Figure \ref{fig:interface}, we adopt the {\it ansatz} that most of the cooling occurs at a characteristic temperature $T_{\rm pk}$, so that we can approximate $\Lambda(T)/T^2 \approx  \Lambda(T_{\rm pk})/T_{\rm pk}^2$. From dimensional analysis, we assume that in the mixing region, the curvature of the temperature profile follows $d^2\ln T/d \xi^2 \approx \alpha/(\Delta \xi)^2$ within a localized layer of thickness $\Delta \xi$, where $\alpha$ is a dimensionless constant. 
We can then write 
\begin{align}
\label{eq:delta_xi}
\Delta\xi=\left[\frac{\alpha\lambda\delta vk_{B}^{2}T_{{\rm pk}}^{2}}{R^{2}P\Lambda(T_{{\rm pk}})}\right]^{1/2}.
\end{align}

\begin{figure}
\includegraphics[width=\columnwidth]{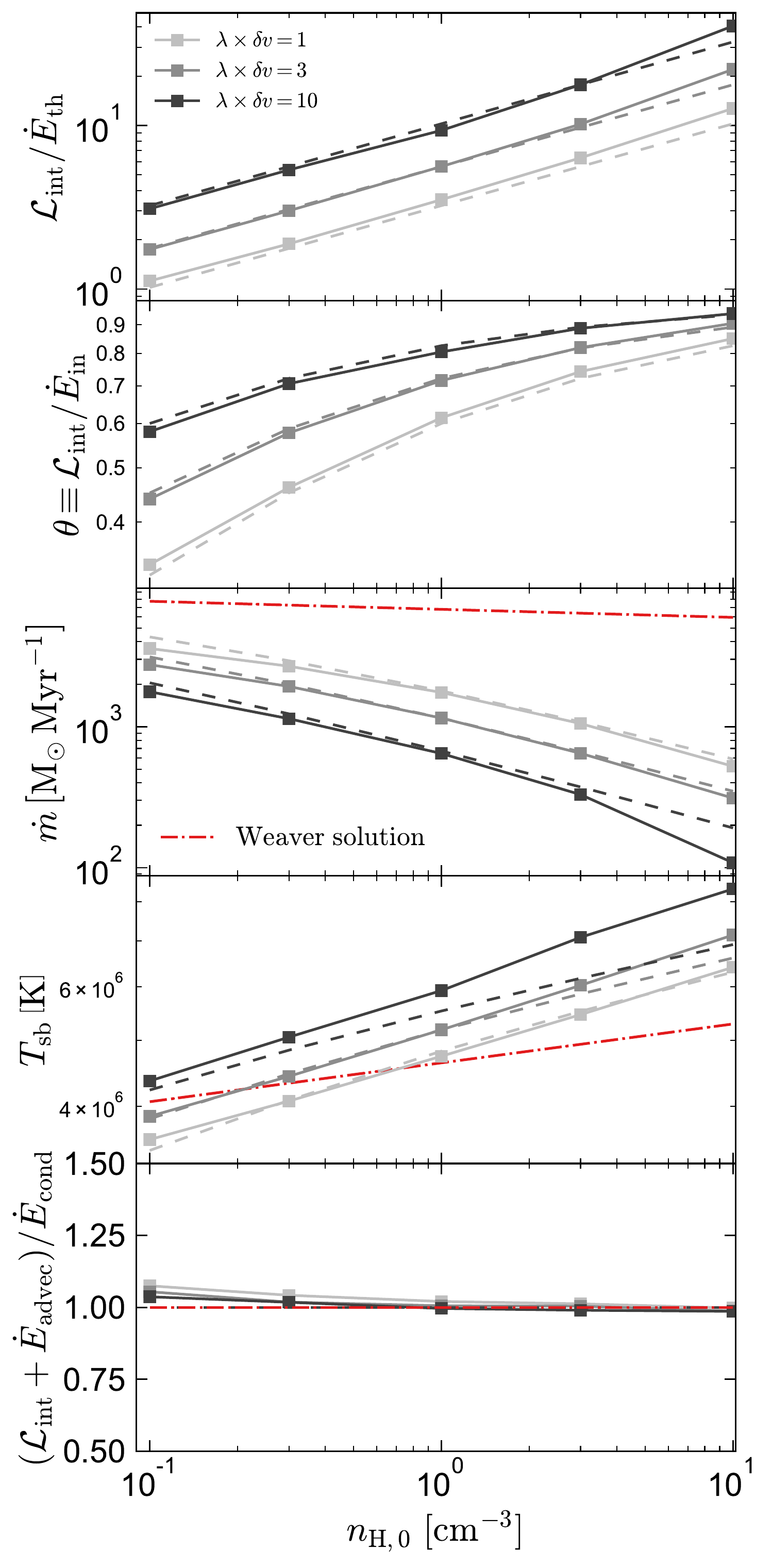}
\caption{
Parameter survey of SB simulations. Each square corresponds to a single simulation; we vary the ambient density ($n_{\rm H,0}$) and nonlinear mixing efficiency ($\lambda \delta v$) between simulations, with fixed $\Delta t_{\rm SNe} = 0.1$\,Myr. All results are measured at $t=10$\,Myr.   Dashed lines show the prediction from our calibrated analytic model (Sections~\ref{sec:cooling_eff} and \ref{sec:mhot_modified_pred}).  Red lines show predictions from the Weaver solution (no cooling).  
{\bf Panel 1}: Ratio of the cooling rate in the interface to the rate of thermal energy increase in the interior (Equation~\ref{eq:cooling_eth_ratio_norm}, with $A_{\rm mix}=3.5$).  
{\bf Panel 2}: Cooling efficiency (Equation~\ref{eq:theta}).
In the classical Weaver solution, there is no cooling, so ${\cal L}_{\rm int}=0$. 
{\bf Panel 3}: Mass evaporation rate from the shell into the bubble (Equation~\ref{eq:mdot_scaling}). Less evaporation is needed to balance conduction when cooling is efficient. 
{\bf Panel 4}: Temperature of the SB interior (Equation~\ref{eq:T_sb}).  This increases slightly at high density because the reduction in the evaporated mass exceeds the reduction in the thermal energy due to cooling.  
{\bf Panel 5}: Ratio of the combined energy losses from the interface (due to advection and cooling) to the total energy conducted into the shell. Energy flux balance predicts this ratio to be $\sim$\,1 in all cases.}
\label{fig:sne_rate_and_density}
\end{figure}

The total cooling rate in the interface is 
\begin{align}
\mathcal{L}_{\rm int}&=4\pi \int n^2\Lambda(T)\, r^2{\rm d}r
\approx 4 \pi R^3 P^2 \frac{\Lambda(T_{\rm pk})}{k_{\rm B}^2 T_{\rm pk}^{2}}\Delta \xi.
\end{align}
Finally, we substitute for $\Delta \xi$ from Equation~(\ref{eq:delta_xi}) and use Equations~(\ref{eq:weaver_R_t}), (\ref{eq:weaver_P_t}),  and~(\ref{eq:Enfracs}) for $R$, $P$, and $\dot{E}_{\rm th}$ with the modification $\dot{E}_{\rm in} \to (1-\theta)\dot{E}_{\rm in}$ . With the net cooling ${\cal L}_{\rm int} \equiv \theta \dot E_{\rm in}$, we obtain 
\begin{align}\label{eq:full_cooling_pred}
\frac{{\cal L}_{\rm int}}{\dot E_{\rm th}}&\equiv\frac{11}{5}\frac{\theta}{ 1-\theta}\\
&\approx  \frac{2\sqrt{7}}{5}    \left(\lambda\delta v\right)^{1/2}\rho_{0}^{1/2}\left[\frac{\alpha \Lambda\left(T_{\rm pk}\right)}{k_{\rm B}^2 T_{\rm pk}^{2}}\right]^{1/2}.
\end{align}
Because $T_{\rm pk}$ depends primarily on the adopted cooling function (and does not vary significantly with $\lambda \delta v$ or $\rho_0$; see Appendix~\ref{sec:cooling_curve}), we thus expect 
\begin{align}
\label{eq:cooling_eth_ratio_norm}
\frac{\mathcal{L}_{{\rm int}}}{\dot{E}_{{\rm th}}}\sim A_{\rm mix}\times\left(\frac{\lambda\delta v}{1\,{\rm pc\,km\,s^{-1}}}\right)^{1/2}\left(\frac{n_{{\rm H},0}}{1\,{\rm cm}^{-3}}\right)^{1/2},
\end{align} where $A_{\rm mix}$ is a dimensionless constant. If we adopt $\alpha=1$ and $T_{\rm pk}=2\times 10^4$\,K (in rough agreement with our simulations; see Appendix~\ref{sec:cooling_curve}), we obtain $A_{\rm mix}\approx 1.7$. 
From our simulations, we can directly measure the ratio $\mathcal{L}_{{\rm int}}/\dot{E}_{{\rm th}}$, and  we find $A_{\rm mix}\approx 3.5$ to provide a good fit to the numerical results (Figure~\ref{fig:sne_rate_and_density}). 

The cooling efficiency, $\theta$, can be expressed in terms of the ratio of cooling luminosity to the rate of thermal energy increase as 
\begin{align}
\label{eq:theta}
\theta=\frac{\mathcal{L}_{{\rm int}}/\dot{E}_{{\rm th}}}{11/5+\mathcal{L}_{{\rm int}}/\dot{E}_{{\rm th}}},
\end{align}
where the dependence of the quantity $\mathcal{L}_{{\rm int}}/\dot{E}_{{\rm th}}$ on physical parameters is predicted by Equation~(\ref{eq:cooling_eth_ratio_norm}).

Finally, we note that the cooling efficiency is independent of time (i.e. ${\cal L}_{\rm int}\propto t$)  because the time dependence of the width of the mixing region exactly compensates for that of $n^2 \Lambda(T) R^2$. 

\subsection{Predicting the mass flux rate}
\label{sec:mhot_modified_pred}
With the cooling rate specified, we now compute the mass flux rate into the bubble interior in the presence of cooling. This quantity does not follow immediately from energy balance, because the conductive heat flux into the interface also changes when cooling becomes important. 

Following \citet{Cowie_1977}, we begin by setting energy fluxes out of the shell, which result from the combination of evaporation and cooling, equal to the heat flux into the shell from conduction: 
\begin{align}
    \label{eq:full_energy_balance}
    \frac{5}{2}\dot{m}c_{\rm iso}^{2}+\mathcal{L}_{{\rm int}}=-4\pi r^{2}CT^{5/2}\frac{{\rm d}T}{{\rm d}r}.
\end{align}
Following \citet{Weaver_1977} (see also Section \ref{sec:analytic_interior}), we define a temperature scale 
\begin{align}
T_s \equiv   \left(\frac{R^{2}P}{Ct}\right)^{2/7} \propto (1-\theta)^{8/35} \rho_0^{2/35} \dot E_{\rm in}^{8/35} C^{-2/7} t^{-6/35}   
\end{align}
for the SB interior that depends on the coefficient $C$ in the conductivity (see Section  \ref{sec:thermal_cond}).
This has the same parameter and time dependencies as the ``no-cooling'' expression in Equation~(\ref{eq:Tclass}), with an additional factor $(1-\theta)^{8/35}$ due to cooling. We can now write the temperature profile in terms of a dimensionless function $\tau$ as 
\begin{align}\label{eq:taudef}
T(r)=\tau(\xi) T_s;  
\end{align}
we use $\xi=r/R$ for the dimensionless radial coordinate as before.  Dividing Equation~(\ref{eq:full_energy_balance}) by $4\pi R^3 P /t=2\dot E_{\rm th}$, we can write 
\begin{align}
    \label{eq:energy_balance_dimensionless}
    \frac{2}{5} w \tau +\frac{1}{2}\frac{\mathcal{L}_{{\rm int}}}{\dot{E}_{{\rm th}} }=-\xi^{2}\tau^{5/2}\frac{{\rm d}\tau}{{\rm d}\xi}.
\end{align}
In Equation~(\ref{eq:energy_balance_dimensionless}) we have introduced the parameter $w\equiv \dot{m}/\dot{m}_{\rm nc}$ to characterize the hot gas evaporation rate relative to the classical mass flux rate $\dot{m}_{\rm nc}$ predicted in the absence of cooling, where we substitute $T_s$ for $T_{\rm int}$ in Equation~(\ref{eq:mdot_cowie_mckee}) for $\dot m_{\rm nc}$. 

Equation~(\ref{eq:energy_balance_dimensionless}) is an ordinary differential equation for $\tau$ as a function of $\xi$ with parameters $w$ and ${\cal L}_{\rm int}/\dot E_{\rm th}$; this determines the behavior of $\tau\left(\xi\right)$ near the shell/interior interface. Physically, we know that $\tau\to0$ while $|d\tau/d \xi| \gg 1$ when $\xi\to1$, so that the first term on the left-hand side of Equation~(\ref{eq:energy_balance_dimensionless}) may be neglected.  We thus have $\tau^{5/2}d\tau/d\xi = (2/7) d \tau^{7/2}/d\xi \approx  - (1/2)\mathcal{L}_{\rm int}/\dot{E}_{\rm th} $.  The definition in  Equation~(\ref{eq:full_cooling_pred}) allows us to write
\begin{align}
    \label{eq:tau72}
\frac{{\rm d}}{{\rm d}\xi}\left(\tau^{7/2}\right)	&\approx-\frac{77}{20}\frac{\theta}{1-\theta}.
\end{align}
Using the $\tau(1)=0$ boundary condition, we find that the limiting behavior of $\tau\left(\xi\right)$ near $\xi\to1$ is 
\begin{align}
    \label{eq:tau_lim}
    \tau\left(\xi\right)=\left[\frac{77}{20}\frac{\theta}{1-\theta}\left(1-\xi\right)\right]^{2/7}.
\end{align}
This asymptotic behavior is  different from that in the classical case with no cooling, where the 2nd term on the left-hand side of Equation~(\ref{eq:energy_balance_dimensionless}) vanishes, yielding $\tau\left(\xi\right)\sim\left(1-\xi\right)^{2/5}$ \citep{Weaver_1977, Cowie_1977}. 

\begin{figure*}
    \includegraphics[width=\textwidth]{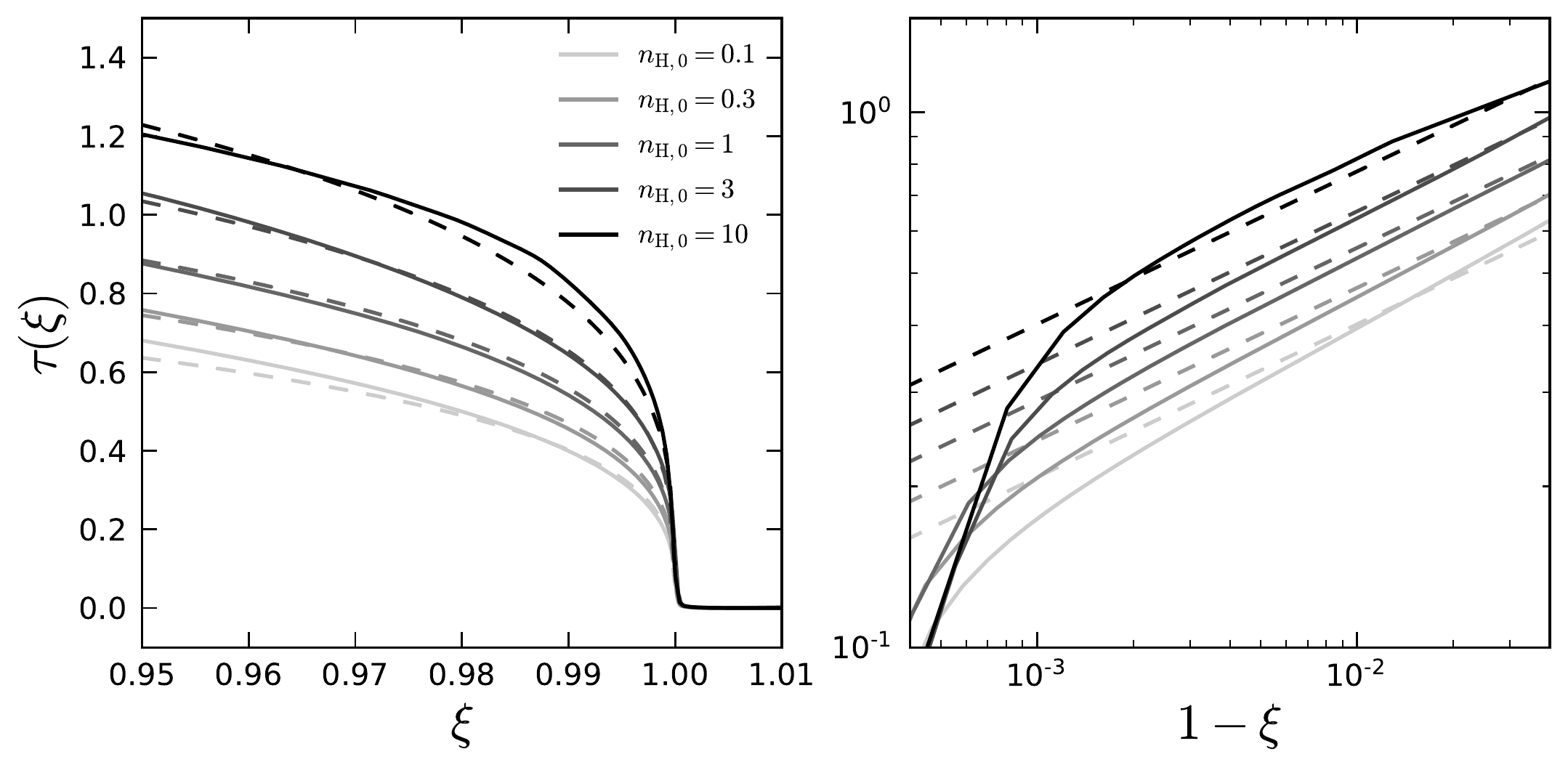}
    \caption{{\bf Left}: Profiles of $\tau = T/T_s$ as a function of $\xi = r/R$ for simulations with different ambient densities at $t = 10$\,Myr. All simulations have $\lambda \delta v =1\,{\rm pc\,km\,s^{-1}}$ and $\Delta t_{\rm SNe} = 0.1$\,Myr. Dashed lines show the prediction of Equation~(\ref{eq:tau_lim}) scaled by a constant factor of 1.24, which is calibrated to match the simulations and primarily accounts for saturation of conduction. {\bf Right}: Same profiles plotted on a logarithmic scale. The simulations follow the predicted $\tau \sim (1-\xi)^{2/7}$ behavior in the region where cooling is negligible and the conductivity follows the Spitzer formula ($\xi \lesssim 0.99$). In the nonlinear mixing region ($\xi \to 1$), cooling losses become important, and the temperature profile steepens.}
    \label{fig:tau_profiles}
\end{figure*}

In Figure~\ref{fig:tau_profiles}, we show the temperature profiles for simulations with five different ambient densities at late times ($t\approx 10$\,Myr). The solid lines show the actual profiles from the simulation, while the dashed lines show the prediction of Equation~(\ref{eq:tau_lim}). In calculating the analytic prediction, we use Equations~(\ref{eq:cooling_eth_ratio_norm}) and~(\ref{eq:theta}) to predict $\theta$, assuming $A_{\rm mix}=3.5$. We also multiply the prediction of Equation~(\ref{eq:tau_lim}) by a constant factor of 1.24 in all cases. As we describe in Appendix~\ref{sec:saturation}, saturation of conduction in the simulations, which was not accounted for in the calculations above, slightly reduces the mass flux in the simulations and increases the temperature relative to the analytic prediction. Once this constant multiple is included, the scaling with $\theta$ and spatial dependence on $\xi$ predicted by Equation~(\ref{eq:tau_lim}) is in good agreement with the simulations. 

The scalings of $\tau$ and $T_s$ with $\theta$ and other parameters of the problem can be combined to predict the temperature in the interior of the SB. From Equation~(\ref{eq:tau_lim}), we can parameterize the temperature profile near $\xi\to1$ as $T=T_{{\rm sb}}\left(1-\xi\right)^{2/7}$, implying a predicted interior temperature $T_{{\rm sb}}\sim 1.8 [\theta/(1-\theta)]^{2/7} T_s$ (including the factor 1.24 to account for saturated conduction). 
As a scaling relation, we have
\begin{multline}
    \label{eq:T_sb}
    T_{{\rm sb}}	=T_{0}\frac{\theta^{2/7}}{\left(1-\theta\right)^{2/35}}\left(\frac{\Delta t_{{\rm SNe}}}{0.1\,{\rm Myr}}\right)^{-8/35} \times \\ \, \left(\frac{\rho_{0}}{1.4\,m_{p}\,{\rm cm}^{-3}}\right)^{2/35}\left(\frac{t}{1\,{\rm Myr}}\right)^{-6/35}\left(\frac{C}{6\times10^{-7}\,{\rm cgs}}\right)^{-2/7},
\end{multline}
where $T_{0}\approx7.7\times10^{6}\,{\rm K}$ is a calibrated constant. The Spitzer conductivity coefficient $C$ is a constant in our simulations, but we include the predicted scaling with $C$ as an indication of how bubble properties are expected to change if the conductivity were reduced by the presence of a magnetic field. 

Finally, an estimate for the mass flux from the shell into the bubble can be calculated by combining Equation~(\ref{eq:T_sb}) with the cooling-modified version of Equation~(\ref{eq:Enfracs}).  We note that because the temperature is not constant throughout the interior, the outer value $T_{\rm sb}$ should only be considered an approximate mean value.  With this, we have:
\begin{align}
    \label{eq:eth_equal_Nkt}
    \frac{3}{2}\frac{\dot{m}}{\mu m_{p}}kT_{{\rm sb}}\sim \frac{5}{11}\left(1-\theta\right)\dot{E}_{{\rm in}}.
\end{align}
Adopting the scalings in Equation ~(\ref{eq:T_sb}), we solve to obtain an expression for $\dot{m}$:
\begin{multline}
    \label{eq:mdot_scaling}
    \dot{m}=\dot{m}_{0}\frac{\left(1-\theta\right)^{37/35}}{\theta^{2/7}}\left(\frac{\mu}{0.62}\right)\left(\frac{\Delta t_{{\rm SNe}}}{0.1\,{\rm Myr}}\right)^{-27/35} \times \\ \left(\frac{\rho_{0}}{1.4\,m_{p}\,{\rm cm}^{-3}}\right)^{-2/35}\left(\frac{t}{1\,{\rm Myr}}\right)^{6/35}\left(\frac{C}{6\times10^{-7}\,{\rm cgs}}\right)^{2/7}.
\end{multline}
Calibrating to the simulations, we find $\dot{m}_0\approx 2900\,M_{\odot}\,\rm Myr^{-1}$. Integrating Equation~(\ref{eq:mdot_scaling}) yields the total mass of hot gas in the SB. The hot gas mass produced per SNe is:
\begin{multline}
    \label{eq:mhot_scaling}
    \hat{M}_{{\rm hot}}\approx250\,M_{\odot}\frac{\left(1-\theta\right)^{37/35}}{\theta^{2/7}}\left(\frac{\mu}{0.62}\right)\left(\frac{\Delta t_{{\rm SNe}}}{0.1\,{\rm Myr}}\right)^{8/35}\times \\ \left(\frac{\rho_{0}}{1.4\,m_{p}\,{\rm cm}^{-3}}\right)^{-2/35}\left(\frac{t}{1\,{\rm Myr}}\right)^{6/35}\left(\frac{C}{6\times10^{-7}\,{\rm cgs}}\right)^{2/7}.
\end{multline}
This can be combined with the radial evolution from Equation~(\ref{eq:weaver_R_t_cooling}) to predict the mean number density of the interior, $n_{{\rm sb}}=3M_{{\rm hot}}/(4\pi R^{3}\mu m_{p})$. We find 
\begin{multline}
    \label{eq:nsb_scaling}
    n_{{\rm sb}}=0.068\,{\rm cm^{-3}}\frac{\left(1-\theta\right)^{16/35}}{\theta^{2/7}}\left(\frac{\Delta t_{{\rm SNe}}}{0.1\,{\rm Myr}}\right)^{-6/35}\times \\ \left(\frac{\rho_{0}}{1.4\,m_{p}\,{\rm cm}^{-3}}\right)^{-19/35}\left(\frac{t}{1\,{\rm Myr}}\right)^{-22/35}\left(\frac{C}{6\times10^{-7}\,{\rm cgs}}\right)^{2/7}.
\end{multline}

We note that our analytic predictions for $T_{\rm SB}$ and $\dot{m}$, and thus, for $\hat{M}_{\rm hot}$ and $n_{\rm SB}$, break down at small $\theta$ ($\theta \lesssim 0.05$, corresponding to $n_{\rm H, 0}\lesssim 0.001\,\rm cm^{-3}$), where they predict $\dot{m} \to \infty $ and $T_{\rm sb} \to 0$, rather than recovering the scaling predicted by the Weaver solution. This occurs because in the limit of small $\theta$, the cooling term in Equation~(\ref{eq:energy_balance_dimensionless}) no longer dominates the left-hand side, leading to a different asymptotic form of $\tau(\xi)$ and no dependence of $\tau$ on $\theta$.

Figure~\ref{fig:sne_rate_and_density} compares the results of a suite of simulations with different $n_{\rm H,0}$ and $\lambda  \delta v$ to the scaling relations we predict for $\mathcal{L}_{{\rm int}}/\dot{E}_{{\rm th}}$ (top panel; Equation~\ref{eq:cooling_eth_ratio_norm}), 
$\theta$ (second panel; Equation~\ref{eq:theta}), $\dot{m}$ (3rd panel; Equation~\ref{eq:mdot_scaling}), and $T_{\rm sb}$ (4th panel; Equation~\ref{eq:T_sb}). The bottom panel shows the ratio of the total energy flux out of the interface from advection and cooling to the heat flux into the interface; this is predicted to be $\approx 1$. All the simulations have $\Delta t_{\rm SNe} = 0.1$\,Myr. For context, we also show the predictions of the Weaver solution in the bottom three panels. 

We measure all the quantities plotted at $t\approx 10$\,Myr. At late times, shocks due to individual SNe have usually subsided near the outer regions of the bubble, causing the energy fluxes to reach quasi-equilibrium. However, as we discuss in Section~\ref{sec:discrete_shocks}, this does not occur in simulations with very efficient cooling (large $n_{\rm H,0}$ and $\lambda \delta v$), because the mass of hot gas in the bubble interior is reduced, causing SNe exploding within the bubble to sweep up less mass as they expand and decelerate more slowly than they would without cooling. In such cases, the quantities plotted in Figure~\ref{fig:sne_rate_and_density} can fluctuate significantly between the arrival of blast waves. We minimize the effects of these fluctuations by averaging all quantities over a timescale of $\Delta t_{\rm SNe}$.

In calibrating our analytic predictions for the cooling efficiency (Equation \ref{eq:cooling_eth_ratio_norm}), interior temperature (Equation \ref{eq:T_sb}), and evaporation rate (Equation \ref{eq:mdot_scaling}), we choose coefficients such that the prediction exactly matches the simulation with $n_{\rm H, 0}= 1\,\rm cm^{-3}$ and $\lambda \delta v = 3\,\rm pc\,km\,s^{-1}$. All other analytic curves shown in Figure \ref{fig:sne_rate_and_density} for differing $n_{\rm H, 0}$ and $\lambda \delta v$ adopt the same value of the  respective  coefficient. 

Comparison of the simulated  and predicted values of $\dot m$ and $T_{\rm sb}$ with the Weaver solution shows that cooling significantly reduces the evaporation rate, and modestly changes the temperature.  More cooling (larger $n_{\rm H,0}$ and/or larger $\lambda \delta v$) is associated with greater reductions in evaporation, and greater increases in the temperature of the hot gas. We further note that a decrease in the Spitzer conductivity coefficient $C$ would decrease $\dot m$ and increase $T_{\rm SB}$.

\begin{figure}
    \includegraphics[width=\columnwidth]{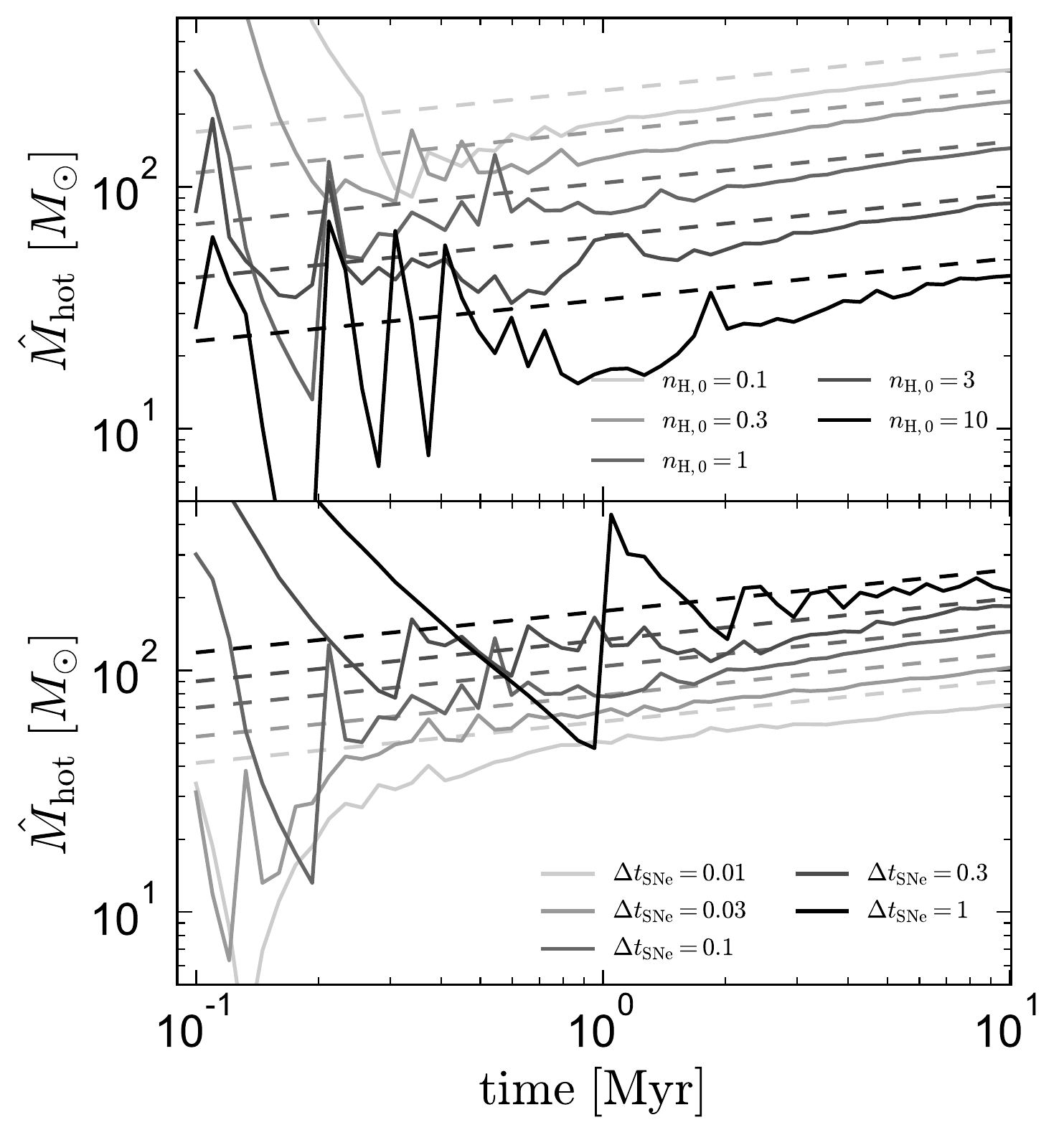}
    \caption{Mass of hot gas in the SB interior, normalized by the number of SNe that have exploded so far. We vary the ambient density in the top panel and the SNe rate in the bottom panel. Dashed lines show the prediction of Equation~(\ref{eq:mhot_scaling}), which assumes that $M_{\rm hot}$ is the integral of all the mass flux into the SB since $t=0$, with the mass flux following the equilibrium prediction (Equation~\ref{eq:mdot_scaling}) at all times. The simulations are in reasonably good agreement with the predicted scaling relation at late times, but generally have slightly lower $M_{\rm hot}$ than predicted by the analytic model because the mass flux into the SB is lower than the equilibrium prediction at early times, when shocks from discrete SNe cannot be neglected.}
    \label{fig:Mhot}
\end{figure}

We compare the prediction of Equation~(\ref{eq:mhot_scaling}) to the simulations in Figure~\ref{fig:Mhot}, where we show simulations with a range of ambient density (top) and supernovae rate (bottom). There is significant disagreement with the simulation prediction at early times, both because $M_{\rm hot}$ is enhanced prior to shell formation, and because the evaporative mass flux is reduced at early times when shocks from discrete SNe propagate into the shell (Figure~\ref{fig:mass_fluxes}). 

The bottom panel of Figure~\ref{fig:Mhot} shows that the simulation with the highest SN rate, $\Delta t_{\rm SNe} = 0.01$\,Myr produces a somewhat lower $\hat{M}_{\rm hot}$ than predicted by Equation~(\ref{eq:mhot_scaling}), even though shocks from individual SNe become negligible quite early. The primary reason for this is that the effects of saturation of conduction become more significant when the SB interior is hotter, and $T_{\rm SB}$ increases with increasing SNe rate. Saturation suppresses evaporation into the SB and is not included in our analytic models, leading to a modest overprediction of $\hat{M}_{\rm hot}$ at smaller $\Delta t_{\rm SNe}$.

Overall, we regard the agreement between the simulations and analytic predictions in Figures \ref{fig:radius_evolution} - \ref{fig:Mhot} as quite satisfactory. We emphasize that while a normalization coefficient is calibrated using the simulations, the scaling relations with ambient density $n_{\rm H,0}$, SN interval $\Delta t_{\rm SNe}$, and mixing parameter $\lambda \delta v$ are not. Nevertheless, Figures \ref{fig:radius_evolution} and \ref{fig:Mhot} show that the predicted analytic scaling behaviors  are only reached after the SB has evolved for some time.  We return to the issue of early vs. late evolution in Section \ref{sec:discrete_shocks}.

Some deviations from the analytic prediction for $\mathcal{L}_{{\rm int}}/\dot{E}_{{\rm th}}$ are expected due to issues such as (a) contributions of the photoelectric heating term to $\mathcal{L}_{\rm int}$, particularly at high $n_{\rm H,0}$, (b) cooling occurring at a range of temperatures rather than at a single temperature $T_{\rm pk}$, and (c) some cooling occurring outside the nonlinear mixing region.

Deviations from the analytic predictions for $\dot{m}$ and $T_{\rm sb}$ are expected to arise due to deviations from the predicted $\mathcal{L}_{{\rm int}}/\dot{E}_{{\rm th}}$ as well as from saturation of conduction and the discreteness of individual SNe. The largest deviation of  $\dot{m}$ and $T_{\rm sb}$ in Figure \ref{fig:sne_rate_and_density} from the analytically predicted scaling relation occurs at large $\lambda \delta v$ and high ambient densities. This is the regime where cooling is most efficient and deviations from the classical Weaver solution are largest. The effects of the saturation of conduction are also largest when cooling is very efficient, because blast waves from individual SNe remain supersonic when they reach the shell until late times.

\begin{figure*}
    \includegraphics[width=\textwidth]{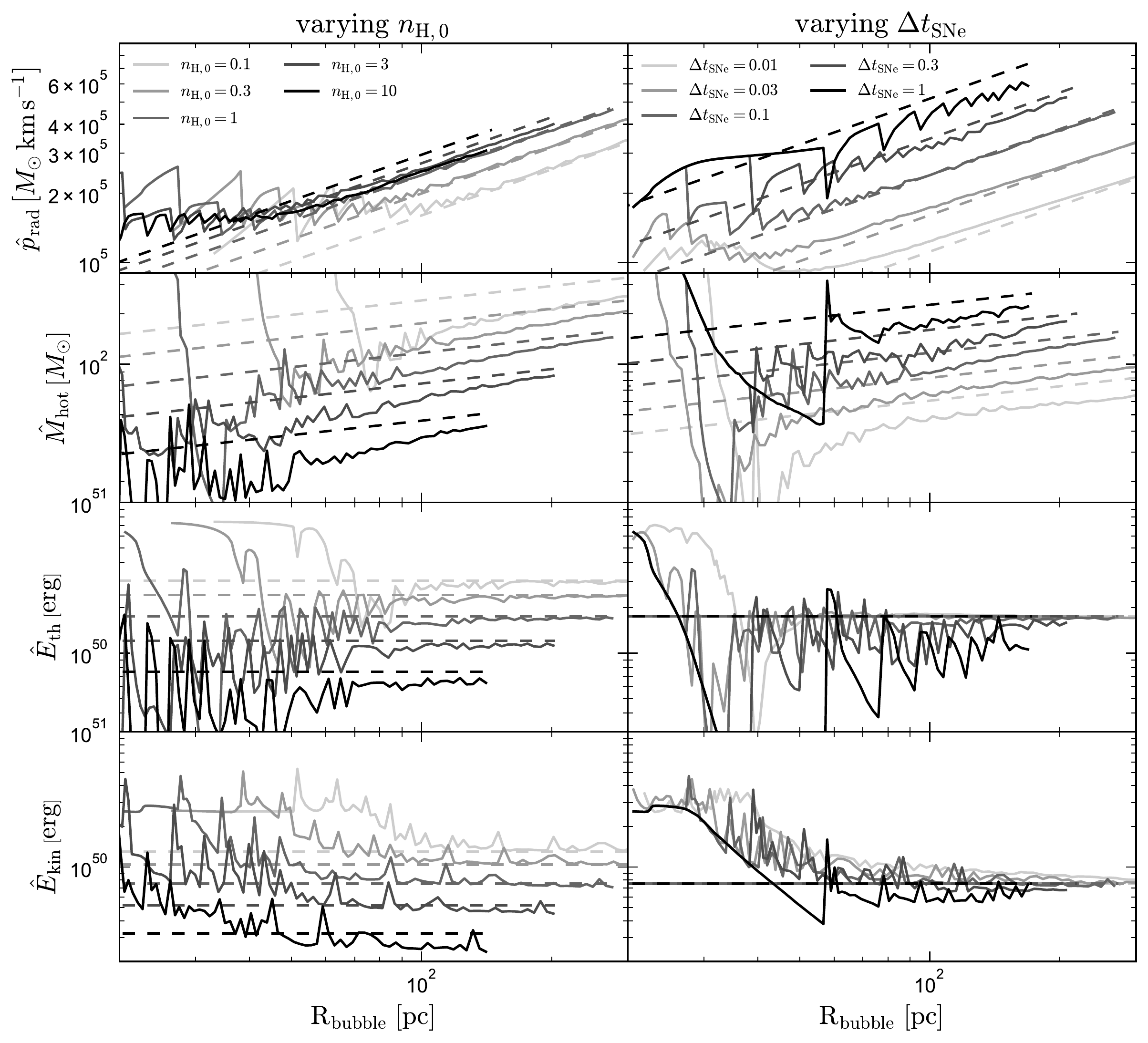}
    \caption{SB radial momentum (top row), hot gas mass (2nd row), thermal energy (3rd row), and kinetic energy (bottom row). All quantities are per SN and are plotted as a function of SB radius. We vary the ambient density (left, with $\Delta t_{\rm SNe} = 0.1$\,Myr held fixed) and the SN rate (right, with $n_{\rm H,0}=1\,\rm cm^{-3}$ held fixed). Dashed lines show the predictions of our (late-time)  analytic model. Both the simulations and analytic predictions terminate at $t=10$\,Myr. Our idealized problem setup with a homogeneous, unstratified ambient medium is expected to break down when $R_{\rm bubble }$ is of order the scale height of the host galaxy's warm-cold gas disk, typically $\sim$100-300\,pc. }
    \label{fig:vs_radius}
\end{figure*}

\subsection{Evolution as a function of radius}
\label{sec:evol_with_R}

The assumption of an unstratified homogeneous medium breaks down when the SB is sufficiently large, for $R_{\rm bubble}$ comparable to the scale height of the host galaxy's ISM disk, $H\sim 100-300$\,pc. The ``feedback yields'' delivered to the ISM are also limited by their values when $R_{\rm bubble}\sim H$. It is therefore useful to quantify the specific energy, momentum, and hot gas mass as a function of $R_{\rm bubble}$.

Figure~\ref{fig:vs_radius} shows the time evolution of the specific radial momentum, hot gas mass, and thermal and kinetic energy as a function of SB radius for simulations with a range of ambient densities and SN rates. The SB properties at $R_{\rm bubble}\sim 100-200$\,pc quantify approximately how much energy and momentum are available to drive turbulence prior to breakout. The value of $\hat M_{\rm hot}$ when the SB reaches a similar radius is indicative of the mass available to drive a hot galactic wind.

Solving Equation~(\ref{eq:weaver_R_t_cooling}) for $t$ allows us to rewrite the analytically predicted scaling relations in terms of SB radius: 
\begin{align}
    \label{eq:phat_R}
    \hat{p}_{{\rm rad}}&\propto\left(1-\theta\right)^{2/3}\rho_{0}^{1/3}\Delta t_{{\rm SNe}}^{1/3}R^{2/3}\\
    \label{eq:Mhat_R}
    \hat{M}_{{\rm hot}}&\propto\frac{\left(1-\theta\right)}{\theta^{2/7}}\Delta t_{{\rm SNe}}^{2/7}R^{2/7}\\
    \label{eq:Eth_R}
    \hat{E}_{{\rm th}}&\propto\hat{E}_{{\rm kin}}\propto\left(1-\theta\right).
\end{align}
Equations~(\ref{eq:phat_R})--(\ref{eq:Eth_R}) have the same explicit dependence on $R$, $\rho_0$, and $\Delta t_{{\rm SNe}}$ as in the classical Weaver solution but include an additional dependence on $\theta$.
In the Weaver solution, $\hat{M}_{\rm hot}(R)$, $\hat{E}_{\rm th}(R)$, and $\hat{E}_{\rm kin}(R)$ are independent of $\rho_0$. This is no longer the case when cooling is included, because $\theta$ increases with $\rho_0$. The comparison between the numerical results and analytic predictions show that the ``late-time'' solutions are generally reached by the time $R_{\rm bubble} \sim 100$\,pc, so that the predicted scalings with density and SN rate are satisfied. 

At $R_{\rm bubble}\sim 100-200$\,pc, our models predict $10^{5}\lesssim\hat{p}_{{\rm rad}}/\left(M_{\odot}\,{\rm km\,s^{-1}}\right)\lesssim 6\times10^{5}$, $40\lesssim\hat{M}_{{\rm hot}}/M_{\odot}\lesssim 300$, $6\times10^{49}\lesssim\hat{E}_{{\rm th}}/{\rm erg}\lesssim3\times10^{50}$, and  $3\times10^{49}\lesssim\hat{E}_{{\rm kin}}/{\rm erg}\lesssim1.5\times10^{50}$. After breakout, a mass of hot gas not exceeding $ \hat{M}_{{\rm hot}}$ per SN would be able to vent out of the disk.  The venting of hot gas also effectively limits both the momentum injection to the ISM and to the galactic wind, as shown in simulations with clustered star formation in stratified disks \citep[e.g.][]{Kim_2018,Fielding_2018}.

\section{Early evolution}

\subsection{Effects of discrete shocks}
\label{sec:discrete_shocks}

\begin{figure*}
\includegraphics[width=\textwidth]{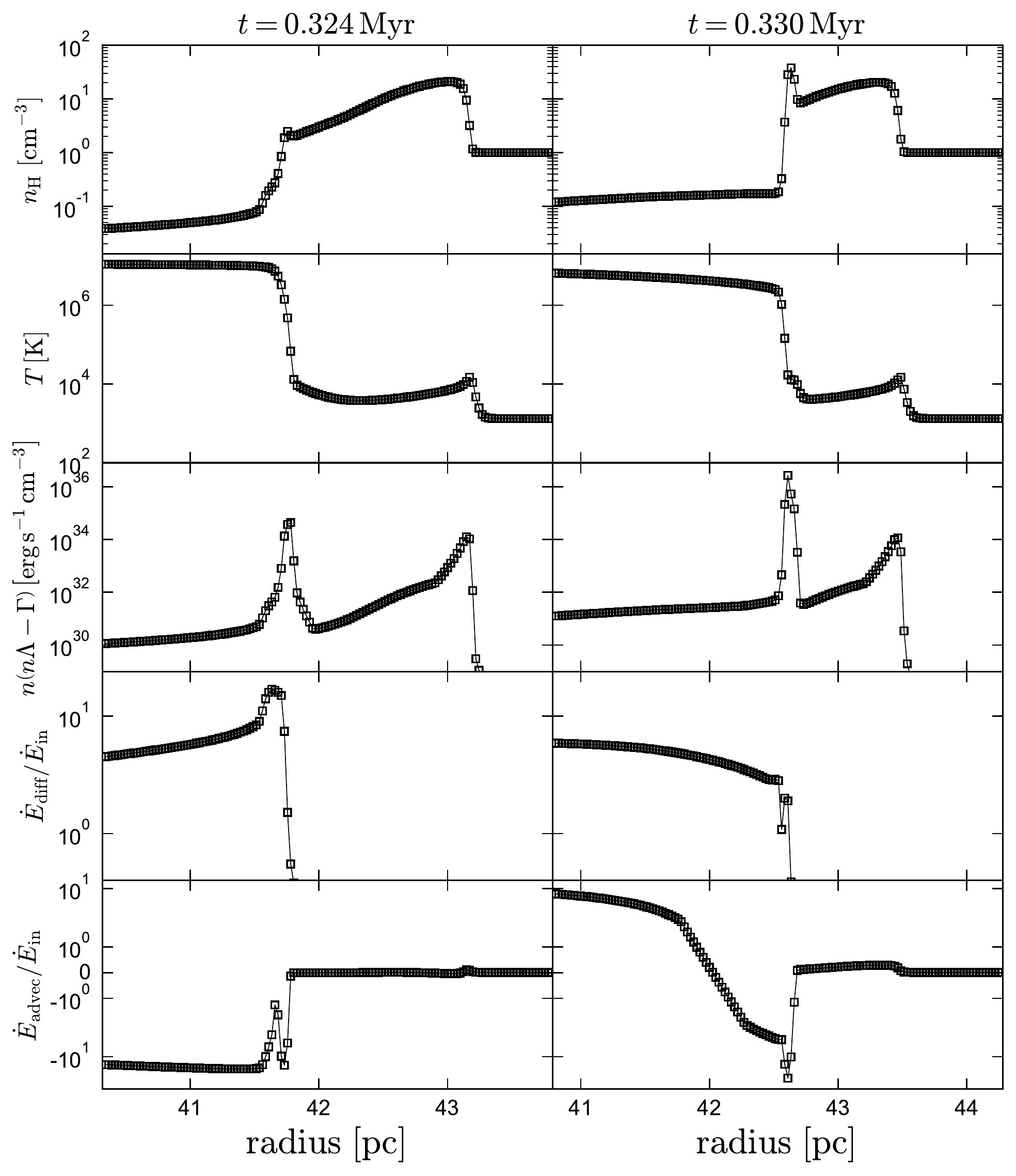}
\caption{Zoom-in on the shell/interior interface at early times for the fiducial model, during (left) and shortly after (right) the arrival of a shock from a new SN at the interface. The bottom three panels show profiles of the cooling loss, the diffusive energy flux {\it into} the shell (i.e. conduction plus mixing), and the advective energy flux {\it out of} the shell. As the shock arrives (left panels), it pushes hot gas from the bubble interior into the shell, resulting in a strong negative advective energy flux (into the shell). After the shock hits the shell and compresses it (right panels, where the shell is noticeably thinner than in the left panels), cooling losses increase, and there is a period of high mass flux into the shell. }
\label{fig:interface_early}
\end{figure*}

The analysis in Section~\ref{sec:pred_cooling_loss} primarily considered the limit in which repeated SNe can be approximated as a continuous energy source with luminosity $\dot{E}_{\rm in}=E_{\rm SN}/\Delta t_{\rm SNe}$. This approximation is not valid at early times. We now consider the effect of shocks from individual SNe.

Figure~\ref{fig:interface_early} shows the structure of the shell/bubble interface at an early time ($t\approx 0.3$\,Myr), during and shortly after the arrival of a blast wave from a new SN at the shell. The left panels show the shell/interface region just as the shock arrives and begins to compress the interface. There is a strongly negative advective flux into the bubble interior (i.e., hot gas from the interior is being pushed into the interface), as well as a strong heat flux into the shell. The right panels show the same fluxes a short period later, after the shock has compressed the dense shell and the direction of the advective fluxes has reversed, with both mass and energy flowing from the interface into the bubble interior. The cooling rate in the interface is also significantly elevated over its value before the shock's arrival. The energy fluxes into and out of the interface are more than an order of magnitude larger during and shortly after the arrival of the shock at the interface than their equilibrium values (see Figure~\ref{fig:interface}). As we discuss below, this leads to rapid increases in $M_{\rm hot}$ followed by a period of enhanced cooling. 

\begin{figure}
\includegraphics[width=\columnwidth]{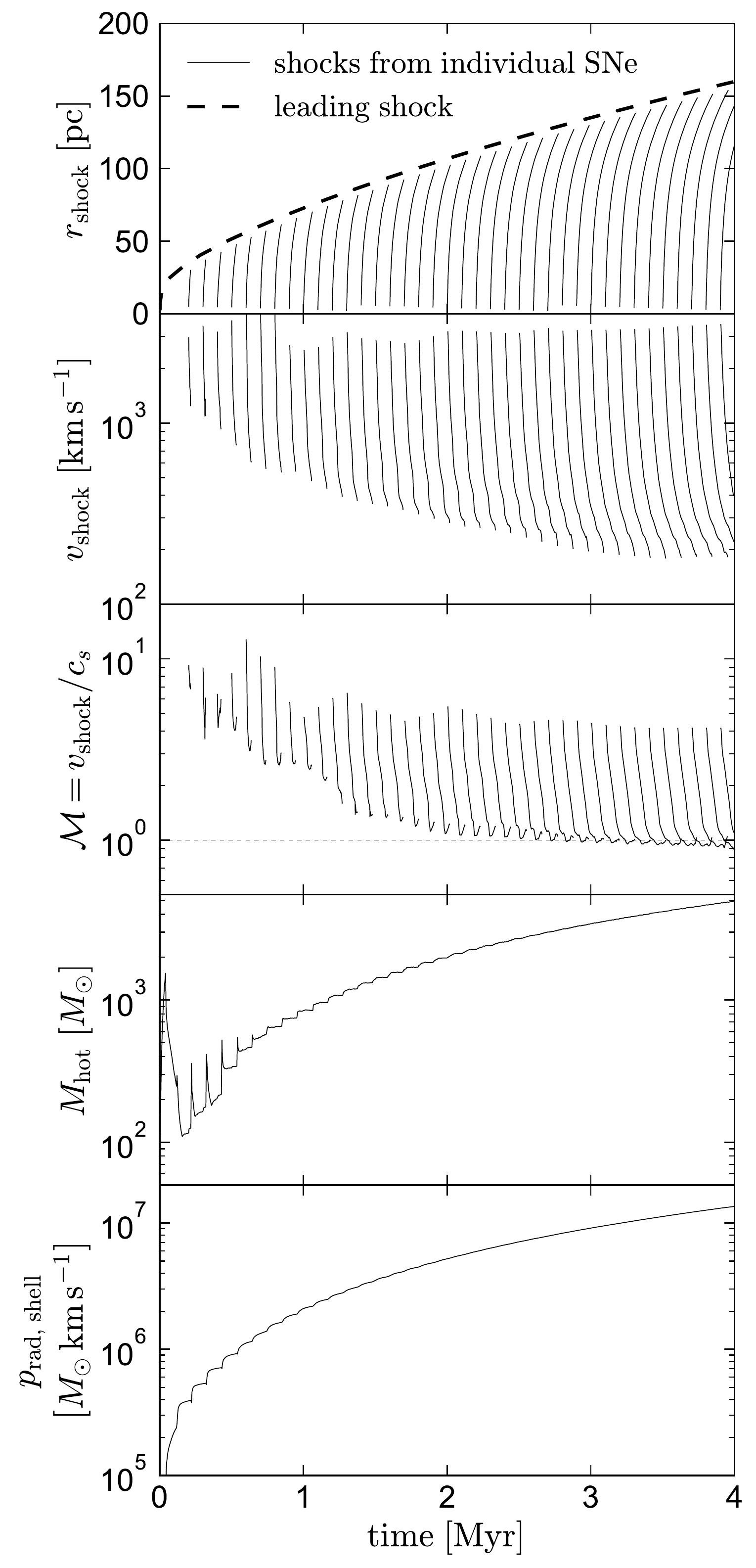}
\caption{Top three panels show radius, velocity, and Mach number of blast waves from individual SNe as they travel through the hot bubble in the fiducial model. For the first $\sim$3 Myr, these remain supersonic until they collide with the cool shell. The resulting shocks thus carry significant hot gas mass into the shell and also directly shock-heat shell gas, leading to a period of strong conduction. During this period, cooling cannot remove all the energy conducted into the interface. At later times, blast waves from individual SNe become subsonic before reaching the shell. Panels 4 and 5 show the total mass of hot ($T > 10^5$\,K) gas in the SB interior and the radial momentum of the shell. At early times, when shocks reach the shell supersonically, $M_{\rm hot}$ and $p_{\rm rad,shell}$ increase stepwise, with rapid increases each time a shock propagates into the shell. At later times, ($t\gtrsim 3$\,Myr, when SNe shocks become subsonic before arriving at the shell), the SB reaches quasi-equilibrium and the growth of $M_{\rm hot}$ and $p_{\rm rad,\,shell}$ becomes continuous.}
\label{fig:individual_shocks}
\end{figure}

To further assess the effects of shocks from SNe within the SB, we trace the blast waves from individual SNe as they travel through the bubble interior. In each simulation snapshot following a new SNe explosion, we identify the shock from the most recent SN as the cell with maximum pressure. We then track the blast wave as it expands by identifying the local maximum in $P(r)$ in each snapshot that is nearest to the predicted location of the shock based on its position and velocity in the previous snapshot. We follow each blast wave until it reaches the shell or fades (i.e., no suitable local maximum can be identified in $P(r)$ that is associated with the location of the shock in the previous snapshot). 

Figure~\ref{fig:individual_shocks} (top panel) shows the radial expansion of blast waves from individual SNe in the first 4 Myr of the fiducial simulation. We calculate the velocity of each blast wave (2nd panel) as $v_{{\rm shock}}={\rm d}r_{{\rm shock}}/{\rm d}t$. We compute the Mach number, $\mathcal{M}$, of each shock as a function of time (3rd panel) from the adiabatic sound speed, $c_{s}=\sqrt{\gamma P/\rho}$. Because gas at the pressure local maximum has already been heated by the shock, we calculate $c_s$ in front of the shock; i.e., in the medium into which it is propagating. In practice, we use the median sound speed in the region $(2 - 5)$\,pc in front of the pressure local maximum to calculate $\mathcal{M}$.

Wiggles in the blast wave velocity curves at late times  correspond to periods when the blast waves collide and interfere with reflected waves from previous SNe in the bubble interior. The sound speed in the bubble generally decreases with radius (see Figure~\ref{fig:profiles}), so the shock velocity drops faster than the Mach number. In some cases, the Mach number actually increases as the blast wave approaches the shell and the temperature falls; in such cases, the sound speed falls more steeply with radius than the shock velocity.

Shocks from individual SNe remain supersonic when they reach the shell at early times, only becoming subsonic after $\sim$3\,Myr. This transition coincides with the time when $M_{\rm hot}$ ceases to increase stepwise, and SNe blast waves stop driving significant hot gas mass into the shell (Figure~\ref{fig:mass_fluxes}). It is important to distinguish when individual blast waves are supersonic because this determines (a) whether they sweep up significant mass from the interior, and (b) whether shocks propagate into and directly heat gas in the shell.

As long as the blast waves remain supersonic, they slam into the shell impulsively. This can result in (a) a brief period when a large amount of hot gas is carried in and/or kinetic energy is partly converted to thermal energy as the shock hits the shell, leading to enhanced conduction, and (b) a sudden increase in the kinetic energy and momentum of the shell, as well as $M_{\rm hot}$.  It is also possible for swept-up hot gas deposited in the shell to condense and cool there, leading to a net loss of energy and hot gas mass from the bubble. Models with continuous energy injection are less applicable at early times, when blast waves within the bubble reach the shell supersonically. At later times, once blast waves become subsonic before reaching the shell, most of their kinetic energy has been thermalized, and the continuous luminosity approximation is appropriate. 

We derive a scaling relation for the time after which blast waves from individual SNe become subsonic before reaching the shell as follows. We suppose that a strong shock propagating through the interior of the bubble can be represented as a front with velocity $v_s$ containing a fraction of the mass in the SB interior, such that $E_{\rm SN} = f M_{\rm hot} v_s^2$. Here the constant $f$ encapsulates uncertainty in the fraction of mass that is swept up and the fraction of the blast wave's energy that is kinetic.
We also have $M_{\rm hot}c_s^2 \sim (50/99)(1-\theta)E_{\rm SN}(t/\Delta t_{\rm SNe})$ from integrating over the life of the bubble.  We thus expect that shocks will become subsonic at a time $t_{\rm subsonic}\sim (2/f) \Delta t_{\rm SNe}/(1-\theta)$. In practice, we find
\begin{align}
    \label{eq:scaling_subsonic}
    t_{{\rm subsonic}}\approx \frac{10}{1-\theta}\Delta t_{\rm SNe},
\end{align}
where the constant ``10'' is found by calibrating to the simulations.
Equation~(\ref{eq:scaling_subsonic}) can also be cast in terms of the total number of SNe that have exploded, as $N_{{\rm SNe,\,subsonic}}\approx 10/\left(1-\theta\right)$. 

\begin{figure}
    \includegraphics[width=\columnwidth]{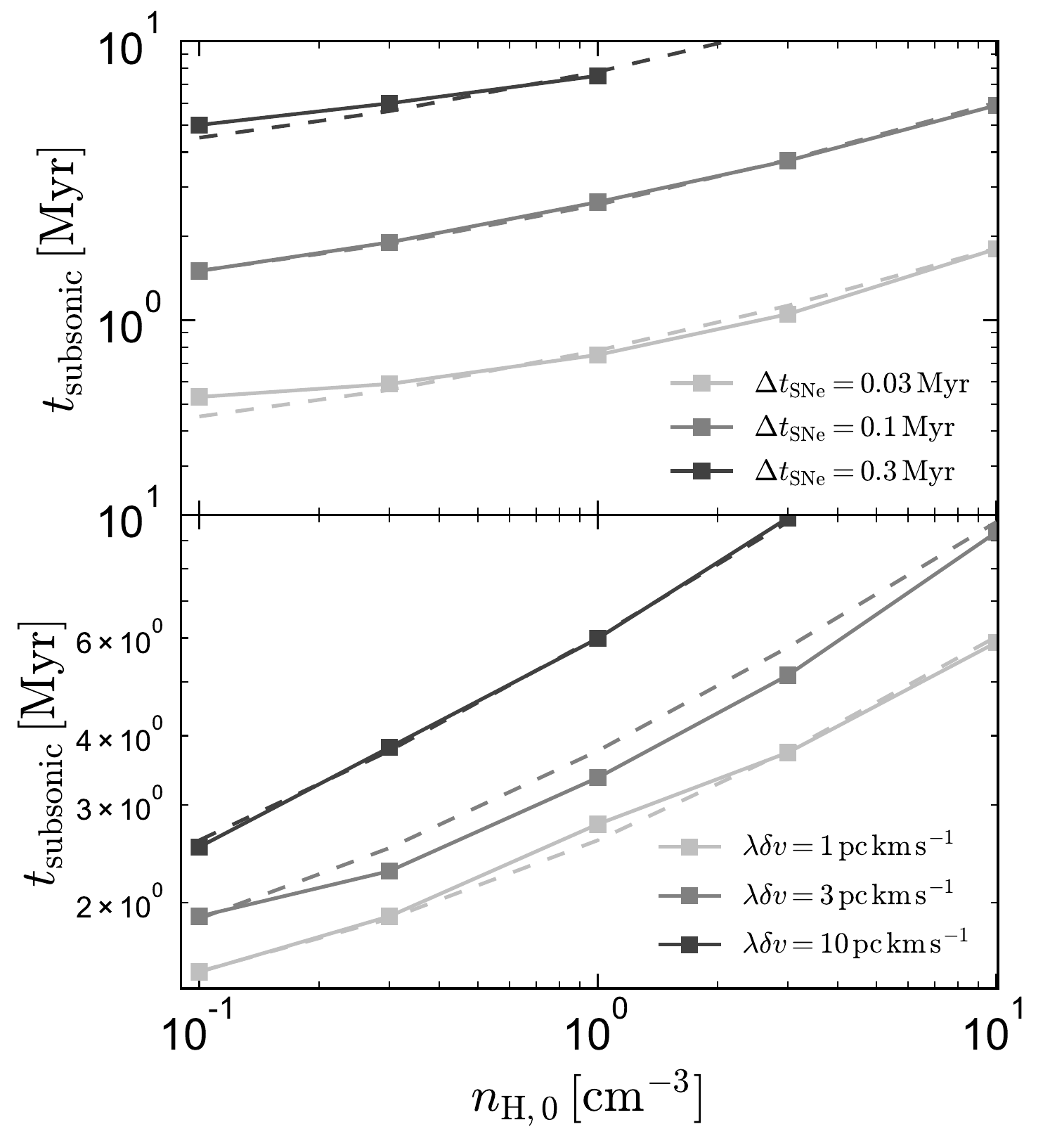}
    \caption{Time when blast waves from individual SNe become subsonic before reaching the shell.  
    SNe expanding within the SB no longer collide with the shell as strong shocks after this time, making the approximation of continuous energy injection more applicable. In the top panel, we vary the SNe rate and ambient density, leaving $\lambda \delta v = 1 \rm\,pc\,km\,s^{-1}$ fixed; in the bottom panel, we fix $\Delta t_{\rm SNe} = 0.1$\,Myr and vary $\lambda \delta v$. Dashed lines show the prediction of Equation~(\ref{eq:scaling_subsonic}).}
    \label{fig:t_subsonic}
\end{figure}

We compare this prediction to the simulations in Figure~\ref{fig:t_subsonic}, where we plot the time after which SNe become subsonic for simulations with a range of $n_{\rm H,0}$, $\lambda \delta v$, and $\Delta t_{\rm SNe}$. We measure $t_{\rm subsonic}$ in the simulations as the first time at which the Mach number of an blast wave falls below $\mathcal{M} = 1$ prior to reaching the shell. As in the previous sections, we compute $\theta$ for the analytic prediction using Equations~(\ref{eq:cooling_eth_ratio_norm}) and~(\ref{eq:theta}). The overall agreement with Equation~(\ref{eq:scaling_subsonic}) is quite good. For a typical ISM density of $n_{\rm H,0}=1\,\rm cm^{-3}$, we thus predict that a relatively massive cluster with $M_{\rm star}\sim 4\times 10^4\,M_{\odot}$ (corresponding to $\Delta t_{\rm SNe} = 0.1$\,Myr) will transition from the discrete-SNe regime to the regime of continuous energy injection after $\approx 3$\,Myr. Small clusters with $M_{\rm star}\lesssim 2\times10^3 M_{\odot}$ will never enter the continuous energy injection regime, because SNe will cease before $t_{\rm subsonic}$. 

Given the radial evolution predicted by Equation~(\ref{eq:weaver_R_t_cooling}), we can also formulate Equation~(\ref{eq:scaling_subsonic}) in terms of the radius of the SB:
\begin{align}
    \label{eq:R_subsonic}
    R_{{\rm subsonic}}=\frac{R_{{\rm 0}}}{\left(1-\theta\right)^{2/5}}\left(\frac{\Delta t_{{\rm SNe}}}{0.1\,{\rm Myr}}\right)^{2/5}\left(\frac{\rho_{0}}{1.4\,m_{p}\,{\rm cm}^{-3}}\right)^{-1/5},
\end{align}
with $R_0\approx 83$\,pc. Implicit in Equation~(\ref{eq:R_subsonic}) is the assumption that SNe do not cease before $t_{\rm subsonic}$. For the fiducial bubble parameters considered in Section~\ref{sec:results}, this yields $R_{\rm subsonic}\approx 120$\,pc, which is comparable to the radius at which the bubble is expected to begin breaking out of the galactic disk 
(see Section \ref{sec:evol_with_R}).

\citet{Maclow_1988} also considered whether shock waves from individual SNe within a SB should become subsonic. They assumed that the bubble interior was well-described by the \citet{Weaver_1977} similarity solution (with no cooling) and found, contrary to the result of our simulations, that blast waves from individual SNe should essentially {\it always} become subsonic before reaching the shell.  We can understand the differences in these conclusions as follows:  
The deceleration of blast waves within the SB is due to the sweeping up of mass in the bubble interior. Cooling reduces the mass in the interior and thus increases the velocity and Mach number of individual SNe blast waves. The scaling relations from Figure~\ref{fig:t_subsonic} predict the Mach number to scale as $\mathcal{M}\sim\left[\Delta t_{{\rm SNe}}/\left(\left(1-\theta\right)t\right)\right]^{1/2}$, so $\mathcal{M}$ is larger when $\theta$ is nonzero. This is the primary reason for the discrepancy between the prediction from \citet{Maclow_1988} and our simulation results.\footnote{Equation~(\ref{eq:scaling_subsonic}) appears to predict that blast waves within the SB should remain supersonic for the first $\sim$\,10 SNe when $\theta =0$. As discussed in Section~\ref{sec:pred_cooling_loss}, the solution for the interior structure of the SB that we derive is not expected to hold in the limit of $\theta \to 0$. Using the SB properties predicted by the Weaver solution and reproducing the calculations from \citet{Maclow_1988} leads to the same scaling relation with $\Delta t_{\rm SNe}$ that is predicted by Equation~(\ref{eq:scaling_subsonic}). However, in this case the analytically predicted coefficient to $\Delta t_{\rm SNe}$ is $\approx 1.3$, while we found it to be $\approx 10$ when we calibrated to the simulation results with cooling.}

\subsection{Effects of stellar winds}
\label{sec:winds}
\noindent Our calculations thus far have only included energy injected by SNe, neglecting the effects of stellar winds and radiation from young stars. Integrating over the whole lifetime of a young star cluster, the energy injected into the ISM by SNe dominates over that injected by stellar winds by an order of magnitude  \citep[e.g.][]{Leitherer_1999}, and the dynamical impact from expansion of \ion{H}{II} regions (driven by both radiation pressure and ionized gas pressure) is also an order of magnitude or more lower than that of SNe \citep[e.g.][]{JGKim_2018}.  However, stellar winds and radiation are dominant for the first few Myr of a cluster's life, before SNe begin \citep[e.g.][]{Geen_2015, Agertz_2013}. We therefore now consider the effects of stellar winds on the early-time evolution of SBs.

For a solar-metallicity cluster with a \citet{Kroupa_2001} IMF, the mechanical wind luminosity is roughly constant, $L_{{\rm winds}}\approx9\times10^{33}\,{\rm erg\,s^{-1}\times\left(M_{{\rm star}}/M_{\odot}\right)}$, over the first $\sim$3\,Myr, after which it declines steeply \citep[e.g.][]{Leitherer_1999}. This is, coincidentally, quite similar to the mean mechanical SNe luminosity at later times, which is $L_{{\rm SNe}}\approx 8\times10^{33}\,{\rm erg\,s^{-1}\times\left(M_{{\rm star}}/M_{\odot}\right)}$. This means that, for any SN rate, the effects of winds can be approximated by adding a $3\,\rm Myr$ period of continuous energy injection prior to the first SNe.

We have carried out a set of simulations with $n_{\rm H,0}=(0.1-10)\,\rm cm^{-3}$ and $\Delta t_{\rm SNe}=(0.03-0.3)\,\rm Myr$ that include a 3-Myr long period of continuous energy injection prior to the explosion of the first SN. The time-averaged wind luminosity is taken to be $\dot{E}_{\rm in}=10^{51}\,\rm erg/\Delta t_{\rm SNe}$, equal to the energy input rate once SNe commence, and winds cease as soon as SNe begin. We neglect any additional feedback due to ionization or radiation pressure in these calculations, since its character is quite different from that of mechanical feedback due to winds and SNe. Following \citet{Leitherer_1999}, the injected mass flux associated with the wind is a factor of 3.7 higher than the value for SNe; i.e., $37 M_{\odot}/\Delta t_{\rm SNe}$ instead of $10 M_{\odot}/\Delta t_{\rm SNe}$.

\begin{figure}
    \includegraphics[width=\columnwidth]{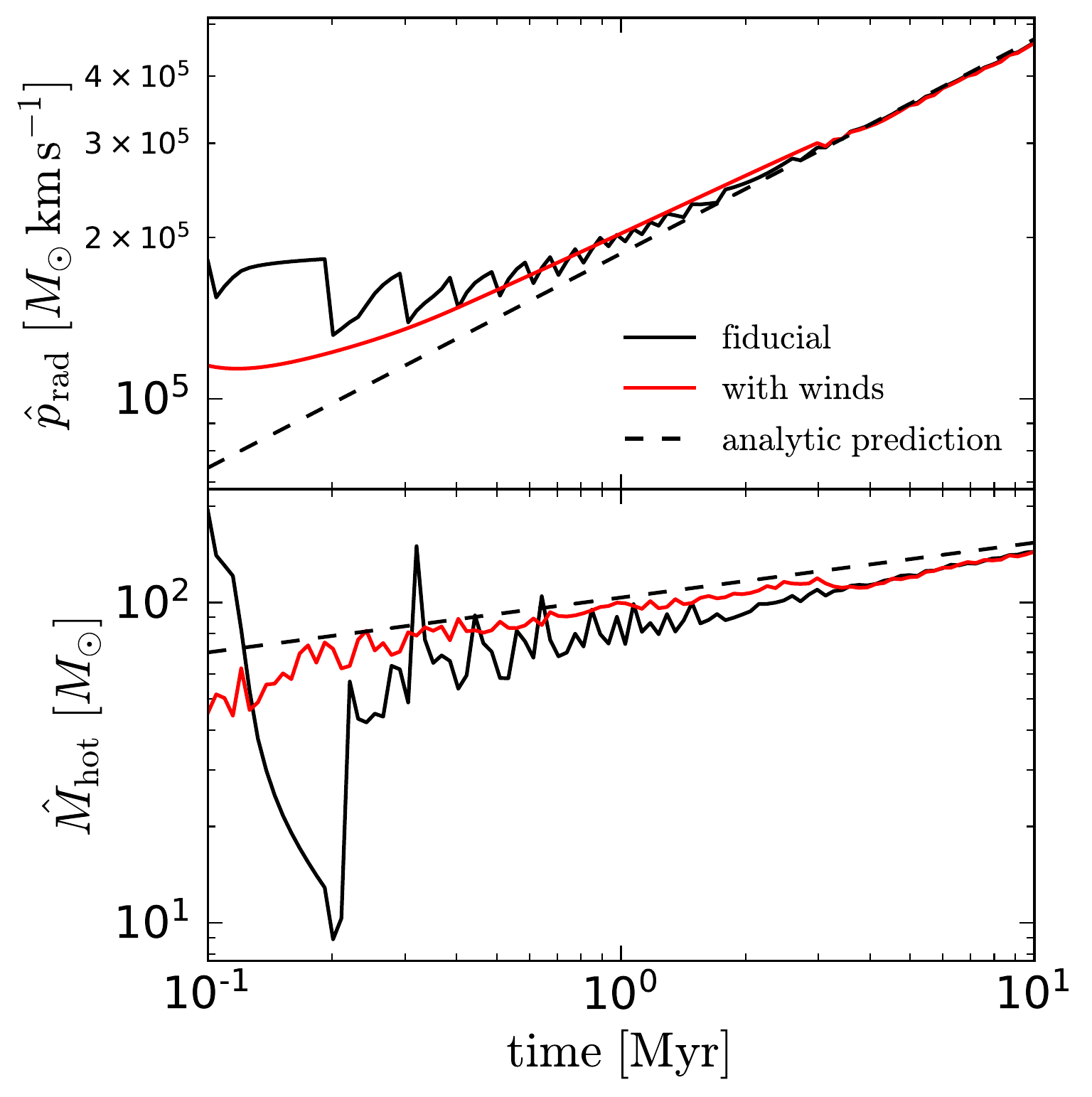}
    \caption{Comparison of simulations with and without the inclusion of stellar winds. Black line shows the fiducial simulation, with $n_{{\rm H,0}}=1\,{\rm cm^{-3}}$ and $\lambda\delta v=1\,{\rm pc\,km\,s^{-1}}$, in which SNe explode every 0.1\,Myr beginning at $t=0$. Red line shows a simulation in which energy injection is continuous between $t=0$ and $t=3$\,Myr (representing stellar winds) and switches to discrete SNe after $t=3$\,Myr; the mean energy input rate is constant. At $t>3$\,Myr, the evolution of the two simulations is essentially identical.  
    At earlier times, $\hat{p}_{\rm rad}$ and $\hat{M}_{\rm hot}$ are slightly lower and higher, respectively, in the simulation with winds, but the two simulations still agree within a factor $\sim 2$.}
    \label{fig:winds}
\end{figure}

Figure~\ref{fig:winds} compares the evolution of $\hat{p}_{\rm rad}$ and $\hat{M}_{\rm hot}$ in the fiducial simulation to one in which energy injection is through winds for the first 3\,Myr. For both simulations, the SB's age is measured from the time when energy injection begins. For the simulation with winds, normalized quantities at $t < 3$ Myr are obtained by dividing by $(t/\Delta t_{\rm SNe})$ as opposed to $\left\lfloor t/\Delta t_{{\rm SNe}}\right\rfloor +1$. That is, for both simulations, we normalize $\hat{p}_{\rm rad}$ and $\hat{M}_{\rm hot}$ relative to the amount of energy injected so far.

Overall, the global evolution the SB is quite similar in the two simulations, with $\hat{p}_{\rm rad}$ and $\hat{M}_{\rm hot}$ generally agreeing within a factor of two. At $t > 3$\,Myr, once SNe have commenced in the simulation with winds, the two simulations agree within a few percent. This is perhaps unsurprising, because the total energy injected at fixed time is the same after $t=3$\,Myr, with or without winds. Once several SNe have exploded, the SB's global properties are well described by the continuous energy injection approximation even in the case of discrete SNe explosions (e.g. Figure~\ref{fig:radius_evolution}). We find good agreement in global SB properties between simulations with and without winds over the full range of SNe rates and ambient densities we tested. Differences between simulations with and without winds are generally larger for SBs with lower SNe rates or higher ambient densities. 

Although the SB's global properties are relatively insensitive to whether energy is injected continuously or through discrete SNe, inclusion of winds can diminish the importance of shocks at early times, because there is more time for hot gas to evaporate into the wind-driven bubble before SNe begin. If the SB's age is measured from the time of the first SN, then including winds effectively shifts the SB's evolution in time, as if SNe had begun 3 Myr earlier. This additional hot gas mass decelerates shocks when individual SNe begin, such that they can become subsonic at earlier times relative to the onset of SNe.

We find that for cases where $t_{\rm subsonic}$ predicted by Equation~(\ref{eq:scaling_subsonic}) is less than 3 Myr, inclusion of winds causes all SN blastwaves to become subsonic before reaching the shell. For cases where $t_{\rm subsonic}$ predicted by Equation~(\ref{eq:scaling_subsonic}) exceeds 3 Myr, we find that when winds are included (and $t$ is measured from the time when winds begin), the time at which SN blasts become subsonic always agrees with the prediction of Equation~(\ref{eq:scaling_subsonic}) within 20\%. With the inclusion of winds, shocks from early SNe still reach the shell supersonically when the ambient density is high ($n_{{\rm H},0}\geq1\,{\rm cm}^{-3}$ for $\Delta t_{\rm SNe} = 0.1\rm \,Myr$) or the SN rate is low ($\Delta t_{{\rm SNe}}\geq0.1\,{\rm Myr}$ for $n_{{\rm H},0}=1\,{\rm cm}^{-3}$), but not when the SNe rate is high or the ambient density is low.

The above results suggest that winds can diminish the importance of shocks from discrete SNe at early times.  However, in practice ambient ISM densities are high enough and  most cluster masses are low enough that the early SNe will still have supersonic blasts.  Moreover, the stage when winds are important coincides with the time clusters are still embedded within dense ($n_{\rm H,0}>100\,{\rm cm}^{-3}$) molecular clouds.  With the expected $\theta \gtrsim 0.9$ for gas at these densities (see Equations~
\ref{eq:cooling_eth_ratio_norm} and \ref{eq:theta}), Equation~(\ref{eq:weaver_R_t_cooling}) suggests that SB evolution would in many cases transition from wind-dominated to supernova-dominated even before breaking out of the natal cloud. In such situations, SBs within the low-density diffuse ISM will be driven primarily by SNe.

Additionally, we note that observations of young clusters find feedback from stellar winds to be weaker than naively expected. That is, the X-ray luminosities observed from the regions surrounding young clusters ($\rm age \lesssim 3\,\rm Myr$, when no SNe are expected to have occurred yet) are much lower (typically by a factor of $20-50$) than predicted by models that take the wind luminosity from stellar population synthesis and assume the kinetic energy of winds is efficiently thermalized through shocks \citep[e.g.][]{Chu_2003, Harper_Clark_2009, Rosen_2014, Leroy_2018}. These observations remain  imperfectly understood theoretically: potential explanations include slower-than-predicted winds from massive stars, efficient turbulent mixing of shocked winds with the cool ISM, and escape of hot gas through holes in the shell at early times. Indeed, at the densities of molecular clouds, the high values of $\theta$ would be consistent (through Equation~\ref{eq:Eth_R}) with an  order  of magnitude or more reduction in the hot-gas energy that powers X-rays.  In any case, the fact that the observable signatures of winds in real SBs are weaker than naively expected suggests a corresponding reduction  of the dynamical effects of winds in real SBs, motivating the primary focus on the SN-driven stage of SB evolution in this paper.  

\section{Summary and Discussion}
\label{sec:discussion}
We have carried out idealized 1D simulations of superbubbles (SBs) driven by repeated supernovae (SNe), focusing in particular on the effects of thermal conduction and radiative cooling near the interface between the cool shell and the hot bubble interior. In order to account for the effects of interface instabilities that occur in real SBs but are not captured in 1D simulations, we also include a diffusion term in the energy equation to represent  nonlinear mixing. 

Unlike most previous studies, which either neglected conduction or assumed that conduction is balanced exactly by evaporation into the shell while neglecting cooling, we model conduction, cooling, and evaporation self-consistently. By modifying the model of \citet{Weaver_1977} to include cooling losses, we have derived an analytic model to predict the thermal properties and dynamical evolution of SBs.

Our primary results are as follows:
\begin{enumerate}
\item {\it Bubble expansion and momentum}: With cooling, the bubble expansion rate is reduced compared to the classical Weaver solution, and the momentum injection per supernova is correspondingly reduced (Figure \ref{fig:integrated_bubble_properties} and Equations \ref{eq:weaver_R_t_cooling}-\ref{eq:prad_cooling}). For our fiducial simulation with one SN every 0.1 Myr and an ambient density of $n_{\rm H,0} = \rm 1\,cm^{-3}$, the radial momentum per SN, $\hat p_{\rm rad}$, is reduced by a factor $\sim2$ compared to the no-cooling case. For higher ambient densities, there is a monotonic increase in the cooling efficiency, and a decrease in the expansion rate and momentum.  However, $\hat p_{\rm rad}$ of a few times $10^5 M_{\odot}\, {\rm km \,s^{-1}}$ is still typical for the full parameter range we have considered  (Figure~\ref{fig:radius_evolution} and \ref{fig:vs_radius}). 

\item {\it The bubble interior}: Thermal conduction substantially increases the mass of hot gas in SB interiors (Figures~\ref{fig:profiles} and~\ref{fig:integrated_bubble_properties}) relative to models without conduction, although this is still reduced below that of the classical Weaver solution due to cooling in the shell. For our fiducial simulation parameters, $\hat M_{\rm hot} \sim$150\,$M_{\odot}$ of gas evaporates from the cool shell into the hot interior per SN.  Values of $\hat M_{\rm hot}$ are somewhat lower for higher density and SN rate (see Figures~\ref{fig:Mhot},\ref{fig:vs_radius}).  {\it Without} conduction, only $\sim$10\,$M_{\odot}$ of hot gas is produced per SN; this is attributable entirely to the ejecta and circumstellar material. The temperature of the bubble interior is much cooler ($T\sim 10^{6-7}$\,K) with conduction than without it ($T > 10^{8}$\,K).

\item {\it Shocks from individual SNe}: At early times ($t\lesssim 3$\,Myr and $R \lesssim 120\,\rm pc$ for our fiducial simulation parameters), blast waves from individual SNe remain supersonic throughout their traversal of the bubble interior (Figure~\ref{fig:individual_shocks}). This is not predicted for the classical SB solution \citep[e.g][]{McCray_1987, Maclow_1988}; it occurs because cooling losses reduce the gas mass swept up by expanding shocks in the interior, causing blast waves to decelerate more slowly than in the classical solution. The duration of the period dominated by discrete supersonic blasts depends on the time interval between SNe and the cooling efficiency (Figure~\ref{fig:t_subsonic} and  Equations~\ref{eq:scaling_subsonic}-\ref{eq:R_subsonic}), such that SBs around lower-mass clusters and SBs in gas with higher ambient density will remain in the discrete-SNe limit longer. 

When blast waves from individual SNe reach the shell before becoming subsonic, they remain strong shocks that carry energy and mass into the shell, accompanied by an enhancement in  the conductive heat flux.  This quickly heats the inner layer of the shell and precipitates an outflux of hot gas  into the bubble (Figures~\ref{fig:interface_early} and~\ref{fig:individual_shocks}). In all models, there is a brief period following the arrival of a shock when the energy flux into the shell exceeds what can be carried away by cooling, leading to an enhanced mass flux into the bubble. Nevertheless, the time-averaged energy advection is into rather than out of the shell at early  times, and the time-averaged cooling is higher at early  times compared to late times (Figure \ref{fig:energy_fluxes}).

We have tested the effects of including a 3-Myr period where energy is injected continuously due to stellar winds before SNe begin (Section~\ref{sec:winds}). This has little effect on the global evolution of the SB because the mechanical luminosity of winds is similar to that of SNe. However, because including winds allows more hot gas to accumulate in the SB interior before SNe begin to explode, winds tend to decrease the effects of shocks from individual SNe. 
Nevertheless,except for models with very low ambient density and high SN rate, the first SNe to occur will still produce shocks when they collide with the SB shell.

\item {\it Cooling in the shell/bubble interface}: 
Once the SB reaches a state in which transient blast waves can be neglected, energy is lost from the interior only by conduction to, and pressure work on, the shell. The conductive heat flux from the hot bubble into the shell is balanced by the combination of (a) mass evaporation from the shell into the bubble and (b) cooling at a rate ${\cal L}_{\rm int}$ in the shell/bubble interface (Figures~\ref{fig:interface}, \ref{fig:energy_fluxes}). For a typical ambient density of $n_{\rm H,0}=1\,\rm cm^{-3}$ and our fiducial mixing efficiency, cooling losses are dominant: $\mathcal{L}_{{\rm int}}/\dot{E}_{{\rm cond}}\approx 0.8$ and ${\cal L}_{\rm int}/\dot{E}_{\rm in} \approx 0.6$. In other words, roughly 80\% of the energy conducted into the interface is cooled away, and $\sim$\,60\% of the total bubble energy is lost to cooling in the interface.  As a result, the mass flux from evaporation of the shell into the interior is only $\sim$20\% of what it would be without cooling losses (Figure~\ref{fig:mass_fluxes}). Once cooling losses are subtracted, the mass flux can be fully accounted for by what is needed to balance the conductive heat flux. The classical similarity solution of \citet{Weaver_1977} did not account for cooling in the interface, so we find significantly less mass in the bubble interior than was predicted by this earlier model. 
\item {\it New analytic models}:
We construct analytic models to predict the dynamical evolution of SBs in the presence of cooling (Section~\ref{sec:sb_dyn}), the dependence of the cooling efficiency on SB parameters (Section~\ref{sec:cooling_eff}), and the evaporation rate and interior temperature profile of SBs with cooling (Section~\ref{sec:mhot_modified_pred}). Our models build on the classical solution for bubbles driven by continuous energy injection \citep{Weaver_1977, McCray_1987}, but self-consistently account for cooling losses as well as effects of thermal conduction.  

The effects of cooling in our analytic model are encapsulated in a dimensionless, time-independent cooling efficiency, $\theta\equiv {\cal L}_{\rm int}/\dot E_{\rm in}$, which depends only on the ambient density and the efficiency of turbulent mixing at the interface (Equations~\ref{eq:cooling_eth_ratio_norm}-\ref{eq:theta}). For $n_{\rm H,0}=1\,\rm cm^{-3}$ and our fiducial mixing parameter $\lambda \delta v = 1 {\rm pc\ km s^{-1}}$, $\theta \approx 0.61$. For a given $\theta$, we provide scaling relations for the evolution of a SB's radius, internal pressure, radial momentum, and total energy (see Equations~\ref{eq:weaver_R_t_cooling}-\ref{eq:prad_cooling}  and related discussion in Section \ref{sec:sb_dyn}). 
In addition, we provide predictions for the interior temperature, hot gas mass, and density (Equations~\ref{eq:T_sb}-\ref{eq:nsb_scaling}). We also predict the shape of the temperature profile at the edge of the bubble, which differs from the no-cooling case (Equation~\ref{eq:tau_lim}). 
Our simulations are in good agreement with the analytically predicted scaling relations (Figures~\ref{fig:radius_evolution} - \ref{fig:vs_radius}
and~\ref{fig:t_subsonic}).

\item {\it Role of mixing and conduction in setting SB properties}:  Cooling losses are set primarily by turbulent mixing, not conduction (Figure~\ref{fig:integrated_bubble_properties}). The dynamics and many global properties of the SB (size, momentum, pressure, energy) are uniquely specified by the cooling efficiency $\theta$. 
In turn, $\theta$ depends on the ambient ISM density and on the efficiency of nonlinear mixing (Figure~\ref{fig:sne_rate_and_density}).
Thermal conduction sets the evaporation rate of the shell, and thus, the temperature and density of the bubble interior. 
Because the evaporation rate also depends on the amount of cooling losses in the interface, the interior properties of the SB depend on both the cooling efficiency and thermal conductivity (Section~\ref{sec:mhot_modified_pred}).  However, for any nonzero level of conduction, the interior temperature is fairly insensitive to the values of parameters (Equation~\ref{eq:T_sb}).

\item {\it Numerical convergence}: Without conduction or nonlinear mixing, it is very difficult to achieve convergence in 1D simulations. Most cooling losses occur in the thin interface between the hot bubble interior and cool shell, and the thickness of this interface becomes arbitrarily small at high resolution in simulations without conduction or mixing, leading to negligible cooling losses. Including thermal conduction alone broadens the interface and somewhat increases cooling losses but does not lead to full convergence at the resolution level we can achieve. Inclusion of diffusive mixing broadens the interface further and leads to more rapid convergence (Figure~\ref{fig:cooling_res}).

\item {\it Saturation of conduction}: Particularly at early times, conduction is often saturated (i.e., the classical conductive flux exceeds what can actually be transported by electrons). Failure to account for saturation of conduction leads to a modest but non-negligible ($\sim$\,20\%) overestimate of the hot gas mass in the bubble (Figure~\ref{fig:saturation}). 
\end{enumerate}

\subsection{Limitations of 1D models}
\label{sec:limitations}
Because our simulations are 1D and highly idealized, assuming that SNe explode at uniform time intervals in an unstratified homogeneous medium, several processes that affect real superbubbles are imperfectly modeled.

Most significantly, 1D simulations cannot explicitly model instabilities that lead to mixing of gas of different temperature phases. We represent the effects of turbulent mixing and heat diffusion  using an effective conductivity (Equation~\ref{eq:kappa_mix}).  This model is motivated by shear-driven instabilities (Section~\ref{sec:mix}) and has one free parameter $\lambda \delta v$; i.e, the wavelength of unstable modes times the turbulent velocity, which quantifies the efficiency of mixing. As shown in Section~\ref{sec:cooling_eff} and Appendix~\ref{sec:subgrid}, this parameter determines the cooling rate for a given ambient density and thus enters as a scaling factor in many of the SB properties predicted by our model.    

Since cooling depends on $\lambda \delta v$, one possible approach is to calibrate $\lambda \delta v$ based on the cooling rates in 3D SB  simulations.  Presently available results are generally consistent with the range of $\lambda \delta v$ that we have considered (See Appendix \ref{sec:subgrid}), but were not designed to enable this calibration.  Some quantitative uncertainty remains regarding interface cooling.  
Previous works that included an inhomogeneous and/or turbulent ambient medium \citep{Kim_2017, Fielding_2018} found converged cooling and momentum injection rates at moderate resolution in 3D. However, instabilities in such simulations may  not be have been fully resolved.  Other simulations in which the ambient medium is uniform and instabilities grow from the resolution scale have not found convergence \citep[e.g.][]{Gentry_2019}.  Moreover, the amount numerical diffusion decreases rather slowly as the resolution improves \citep{Agertz_2007, Hopkins_2015}. 

Cooling losses in SBs also affect the injection of the momentum that drives ISM turbulence (and regulates star formation), and the mass of hot gas that can escape as a wind.  Thus, present limitations in understanding of interface mixing remains a persistent source of uncertainty in theoretical research in several areas.  As such, obtaining more precise constraints on the ``true'' value of $\lambda \delta v$ represents an important open problem, which we will explore in future work. This will require very high resolution 3D simulations of SB expansion in models of  the ISM with realistic ambient inhomogenous structure. Repeating the simulations presented here in 3D at the same resolution is currently prohibitively expensive computationally, but it may be possible to resolve the interface instabilities using adaptive mesh refinement. Given a calibration of mixing, it may be possible to implement it as a subgrid diffusion model in galactic and CGM simulations where resolution is inadequate to directly capture the relevant turbulent mixing.

Our simulations also do not include magnetic fields. Because electrons move along magnetic field lines, magnetic fields tend to suppress conduction perpendicular to the magnetic field \citep[e.g.][]{Braginskii_1965}. 
It is thus possible that our simulations somewhat overestimate the true conductive flux from the bubble to shell. However, the expulsion of gas from the shell into the bubble is expected to  comb the magnetic field in the radial direction, such that any magnetic field frozen into the shell may have little effect on the conductive energy flux \citep[e.g.][]{Cowie_1977}. The RM instability also tends to produce radially aligned magnetic fields \citep{Inoue_2013}. 3D simulations with conduction are required to quantify the effect of magnetic fields on the conductive heat flux. Previous work has also suggested that magnetic fields suppress the development of instabilities in the shell/interior interface \citep[e.g.][]{Ntormousi_2017}. In the context of the 1D models studied in this work, suppression of conduction due to magnetic fields can be approximated as decreasing the constant $C$ in the thermal conductivity. The scaling relations we provide (Equations~\ref{eq:T_sb}-\ref{eq:nsb_scaling}) show that the effects of re-scaling thermal conductivity are relatively weak: for example, lowering $C$ by a factor of 10 only reduces $M_{\rm hot}$ by a factor of 1.9. 

\subsection{Comparison to previous work}
\label{sec:comparison}
Compared to the classical SB similarity solution \citep[e.g.][]{Weaver_1977, McCray_1987, Maclow_1988}, our simulations predict SBs to be smaller, have lower energy and momentum, and contain less hot gas, usually with slightly higher temperature. The magnitude of the difference between our predictions and the classical solution depends on the ambient density and efficiency of mixing in the interface (Figure~\ref{fig:sne_rate_and_density}). For our fiducial problem setup, the momentum per SN $\hat p_{\rm rad}$ is a factor of $2-3$ lower than in the classical solution; in SBs with high ambient density and efficient mixing, cooling reduces $\hat p_{\rm rad}$ by up to a factor of $\sim$\,10. 

Although they did not account for cooling in the interface in their similarity solution, \citet{Weaver_1977} and \citet{McCray_1987} did note that at late times, cooling would eventually become important in the SB interior. They predicted that at this time, a SB's expansion would cease to be pressure-driven, causing the momentum equation and predicted radial evolution to change. In particular, \citet[][their Equation 11]{McCray_1987} predicted that a SB's radial evolution would depart from the $R\sim t^{3/5}$ behavior predicted for a pressure-driven SB after a cooling time of
\begin{align}
    \label{eq:tc}
    t_{c}\sim 24\,{\rm Myr}\left(\frac{\Delta t_{{\rm SNe}}}{0.1\,{\rm Myr}}\right)^{-0.3}\left(\frac{n_{{\rm H,}0}}{1\,{\rm cm^{-3}}}\right)^{-0.7},
\end{align}
so that, for example, the interior of a SB with $\Delta t_{\rm SNe} = 0.1$\, and $n_{\rm H,0}=10\,\rm cm^{-3}$ would cool after $t_c\sim 5$\,Myr.

None of our simulations reach such a regime within the first 10\,Myr: Figure~\ref{fig:radius_evolution} shows that they all follow $R\sim t^{3/5}$. We also experimented with running the simulation with $n_{\rm H,0}=10\,\rm cm^{-3}$ (which is the most likely to cool according to Equation~\ref{eq:tc}) for 40\,Myr. Even in this case, we do not find any significant departure from the predicted $R\sim t^{3/5}$ behavior. 

The reason for this discrepancy is that cooling becomes important in the interface long before it becomes important in the hot interior. As we show in Section~\ref{sec:mhot_modified_pred}, cooling in the interface (a) increases the temperature and (b) decreases the density in the interior. Both effects delay interior cooling. This delay is more severe at high ambient density and large $\theta$ (Equation~\ref{eq:nsb_scaling}), such that we do not expect cooling in the interior to become important for any of the range of SB parameters studied in this work. 

\citet{Kim_2017} recently carried out 3D simulations of SBs in an inhomogeneous medium consisting of cold clouds formed via thermal instability embedded in a warm intercloud medium. Their simulations included cooling and allowed for development of instabilities at the hot/cold interface, but did not include thermal conduction. The hot gas mass in the interior of their simulated SBs thus came primarily from the ablation of cold clouds within the SB, rather than from shell evaporation as it does in our simulations. 
Despite contributions to $M_{\rm hot}$ from ablation of cold clouds, they found relatively low values of $M_{\rm hot}$, with $\hat{M}_{\rm hot} =(10-100) M_{\odot}$ in most cases. This owed in part to the fact that clouds in the interior often cooled quickly even after being surrounded by the hot SB interior. \citet{Fielding_2018} also found, from 3D simulations of SBs evolving in an inhomogeneous, turbulent medium, that $\hat{M}_{\rm hot} =(10-100) M_{\odot}$ prior to breakout from the ISM. 

In our simulations and analytic model, the fraction of the total injected energy that is retained in the SB approaches a constant value (Figure~\ref{fig:integrated_bubble_properties}). In contrast, \citet{Kim_2017} found that the bubble energy per SN declined steadily at late times. This discrepancy is very likely due to the fact that the ambient medium in the simulations of \citet{Kim_2017} was cloudy, while the ambient medium in our simulations is uniform. When comparing 3D simulations of SBs in a uniform medium to otherwise identical simulations in an inhomogeneous, turbulent medium, \citet{Fielding_2018} found that those in a uniform medium approached a constant energy per SNe, while the energy per SNe decreased with time in the case of a turbulent medium. In the context of the models we develop, a decrease in $\hat{E}_{\rm SB}$ at late times is expected if the efficiency of mixing, parameterized by $\lambda \delta v$, increases with time. 

\citet{Yadav_2017} also found an approximately constant asymptotic energy per SNe in 3D simulations with a uniform medium. Which behavior is more realistic is not yet clear: the ISM is indeed inhomogeneous and turbulent, but numerical diffusion, which can lead to excess cooling losses, is also enhanced in an inhomogeneous medium. \citet{Yadav_2017} also included conduction in one of their 3D simulations of SBs driven by young star clusters. In agreement with our results, they found that although conduction increases $M_{\rm hot}$, it does not significantly change the dynamics or total energy of the SB because the degree of cooling loss is primarily set by turbulent mixing. One question of interest for future work is whether this conclusion holds true in a turbulent medium, where conduction might be expected to accelerate the ablation of cold clouds within the SB interior. 

While $\hat M_{\rm hot}$ in our models with conduction is significantly enhanced relative to simulations without conduction, the values of the radial momentum per SN at $t\lesssim 3$Myr are generally similar to recent results from 3D simulations.  In particular, \citet{Kim_2017} found $\hat p_{\rm rad} \sim (0.8-3)\times 10^5\,{\rm M_\odot\,km\, s^{-1}}$, relatively insensitive to ambient density but increasing with $\Delta t_{\rm SNe}$, comparable to the results shown in Figure~\ref{fig:radius_evolution} at similar times. This is expected, as the early times in our simulations are dominated by shocks from individual SN, similar to the results in 3D simulations without conduction.  At late times, however, the pressure in the bubble interior becomes more important in driving expansion than individual SNe shocks in our simulations, and the values of $\hat p_{\rm rad}$ rise above those in 3D simulations without conduction. In the \citet{Kim_2017} simulations, the pressure within SBs at late times was reduced by efficient mixing and cooling. An interesting question for future research is whether, during the period before SB breakout, $\hat p_{\rm rad}$ can be enhanced by conduction, when considering multiphase ISM disks with realistic scale heights and cluster masses (e.g. Figure \ref{fig:vs_radius}).  

\subsection{Implications and prospects}
\label{sec:prospects}

One of the important conclusions of this work is that, compared to the levels produced by the combination of supernova ejecta and ablation by KH instabilities of embedded clouds by high-velocity hot flows, conduction can significantly enhance the mass of hot gas created in SBs. The enhancement of $\hat M_{\rm hot}$ by conduction has potentially important implications for phase balance in the ISM and circumgalactic medium, as well as for galactic winds.  

In 3D simulations of SBs expanding in a two-phase ISM, \citet{Kim_2017} found that both warm and hot gas were accelerated by SBs, but only the hot gas had high enough velocity that it might be expected to escape from an intermediate- or high-mass galaxy's gravitational potential and drive a galactic wind. \citet[][]{Fielding_2018} similarly found little warm gas with $v>100 {\rm km/s}$ in their simulations of SBs in a turbulent medium.  Assuming that one SN explodes for every $\sim 100\,M_{\odot}$ of stars formed (and that only the hot gas escapes in a wind) the \citet{Kim_2017} results correspond to a relatively low wind mass loading factor, $\eta=\dot{M}_{{\rm outflow}}/{\rm SFR}$, with $0.1 \lesssim \eta_{\rm hot} \lesssim 1$. The results for mass loading from \citet{Fielding_2018} were similar, even after breakout from the disk. \citet{Kim_2017} thus concluded that SBs are only likely to launch winds with $\eta_{\rm hot} > 1$ if the SNe rate is sufficiently high that SBs break out of the disk before shell formation, when cooling is not yet important and $\hat{M}_{\rm hot}$ is much higher. This is only expected to occur if the SNe rate is very high, as from a massive cluster \citep[see also][]{Fielding_2018}, and in particular if there is a burst of star formation above typical local values (this rate determines the disk scale height in the first place).  

Due to the conduction-driven evaporation in our simulations, we predict values of $\hat{M}_{\rm hot}$ that are roughly an order of magnitude higher than those found by \citet{Kim_2017} (see Equation~\ref{eq:mhot_scaling} and Figure~\ref{fig:Mhot}). This implies mass loading factors $\eta_{\rm hot}$ that are usually larger than 1. Such values of $\eta$ are more compatible with the values needed to explain the low cosmic star formation efficiency, though they are still lower than the values typically found in current cosmological simulations \citep[e.g.][]{Muratov_2015, Angles_2017}. We note, however, that there is not necessarily a simple correspondence between $\hat  M_{\rm hot}$ and the mass that is actually removed from galaxies in winds: hot outflowing gas can both entrain additional gas in the galaxy or CGM after leaving a SB, increasing $\eta$, or be sufficiently decelerated by the material it sweeps up that it fails to escape, decreasing $\eta$ \citep[e.g.][]{Muratov_2015, ElBadry_2018, Fielding_2018}.
Nevertheless, \citet{Kim_2018} found, using self-consistent simulations of star-forming three-phase ISM disks with uniformly high resolution in both midplane and outflow/fountain regions, that the value $\eta_{\rm hot}\sim 0.1$ in hot winds was comparable to that found in idealized SB simulations without conduction.\footnote{In an extended suite of simulations for different galactic conditions covering gas surface density up to $100\,M_\odot\,{\rm pc}^{-2}$ (Kim et al., in prep), the mass loading factors of hot winds also do not vary much, with typical $\eta_{\rm hot}\sim 0.1$.} The interaction of hot, outflowing gas with warm fountain gas did not significantly accelerate the latter, nor reduce the energy flux of the former.  

The possibility of a large increase of $\hat M_{\rm hot}$ in SBs due to thermal conduction has important implications for galactic winds from normal galaxies. In addition to increasing the hot-gas mass outflow loading to $\eta_{\rm hot}> 1$, a greater momentum flux in the hot medium could boost the acceleration of warm gas out of galaxies, possibly increasing the wind loading factor in galaxies with deep potential wells.

Finally, we remark that our analytic solution for SBs with cooling can be easily implemented in any application that currently uses the \citet{Weaver_1977} model. In addition to use in interpretation of observed supershells \citep[e.g.][]{Oey_1995, Jaskot_2011}, other potential uses include setting the inputs to models of starburst-driven winds for varying environments \citep[e.g.][]{Thompson_2016}, comparing to outflows driven by nascent super-star clusters within dense molecular clouds \citep[e.g.][]{Leroy_2018}, and implementing subgrid models for galaxy formation simulations \citep[e.g.][]{Keller_2014}. 

\section*{Acknowledgements}
We thank the referee for a thoughtful report. We are grateful to Chris Mckee, Yuan Li, and Ken Shen for helpful discussions, and to Drummond Fielding for providing an implementation of radiative cooling in \texttt{Athena++}.
KE was supported by an NSF graduate research fellowship. The work of ECO was supported in part by ATP grant NNX17AG26G from NASA and Investigator grant 510940 from the Simons Foundation. CGK was supported in part by grant  528307 from the Simons Foundation.   
EQ and KE are supported by a Simons Investigator Award from the Simons Foundation and by NSF grant AST-1715070.
DRW is supported by a fellowship from the Alfred P. Sloan Foundation and acknowledges support from the Alexander von Humboldt Foundation.
This work was initiated as a project for the Kavli Summer Program in Astrophysics held at the Center for Computational Astrophysics of the Flatiron Institute in 2018. The program was co-funded by the Kavli Foundation and the Simons Foundation. We thank them for their generous support.



\bibliographystyle{mnras}

\begin{thebibliography}{}
\makeatletter
\relax
\def\mn@urlcharsother{\let\do\@makeother \do\$\do\&\do\#\do\^\do\_\do\%\do\~}
\def\mn@doi{\begingroup\mn@urlcharsother \@ifnextchar [ {\mn@doi@}
  {\mn@doi@[]}}
\def\mn@doi@[#1]#2{\def\@tempa{#1}\ifx\@tempa\@empty \href
  {http://dx.doi.org/#2} {doi:#2}\else \href {http://dx.doi.org/#2} {#1}\fi
  \endgroup}
\def\mn@eprint#1#2{\mn@eprint@#1:#2::\@nil}
\def\mn@eprint@arXiv#1{\href {http://arxiv.org/abs/#1} {{\tt arXiv:#1}}}
\def\mn@eprint@dblp#1{\href {http://dblp.uni-trier.de/rec/bibtex/#1.xml}
  {dblp:#1}}
\def\mn@eprint@#1:#2:#3:#4\@nil{\def\@tempa {#1}\def\@tempb {#2}\def\@tempc
  {#3}\ifx \@tempc \@empty \let \@tempc \@tempb \let \@tempb \@tempa \fi \ifx
  \@tempb \@empty \def\@tempb {arXiv}\fi \@ifundefined
  {mn@eprint@\@tempb}{\@tempb:\@tempc}{\expandafter \expandafter \csname
  mn@eprint@\@tempb\endcsname \expandafter{\@tempc}}}

\bibitem[\protect\citeauthoryear{{Agertz} et~al.,}{{Agertz}
  et~al.}{2007}]{Agertz_2007}
{Agertz} O.,  et~al., 2007, \mn@doi [\mnras]
  {10.1111/j.1365-2966.2007.12183.x}, \href
  {http://adsabs.harvard.edu/abs/2007MNRAS.380..963A} {380, 963}

\bibitem[\protect\citeauthoryear{{Agertz}, {Kravtsov}, {Leitner}  \&
  {Gnedin}}{{Agertz} et~al.}{2013}]{Agertz_2013}
{Agertz} O.,  {Kravtsov} A.~V.,  {Leitner} S.~N.,   {Gnedin} N.~Y.,  2013,
  \mn@doi [\apj] {10.1088/0004-637X/770/1/25}, \href
  {http://adsabs.harvard.edu/abs/2013ApJ...770...25A} {770, 25}

\bibitem[\protect\citeauthoryear{{Angl{\'e}s-Alc{\'a}zar},
  {Faucher-Gigu{\`e}re}, {Kere{\v s}}, {Hopkins}, {Quataert}  \&
  {Murray}}{{Angl{\'e}s-Alc{\'a}zar} et~al.}{2017}]{Angles_2017}
{Angl{\'e}s-Alc{\'a}zar} D.,  {Faucher-Gigu{\`e}re} C.-A.,  {Kere{\v s}} D.,
  {Hopkins} P.~F.,  {Quataert} E.,   {Murray} N.,  2017, \mn@doi [\mnras]
  {10.1093/mnras/stx1517}, \href
  {http://adsabs.harvard.edu/abs/2017MNRAS.470.4698A} {470, 4698}

\bibitem[\protect\citeauthoryear{{Badjin}, {Glazyrin}, {Manukovskiy}  \&
  {Blinnikov}}{{Badjin} et~al.}{2016}]{Badjin_2016}
{Badjin} D.~A.,  {Glazyrin} S.~I.,  {Manukovskiy} K.~V.,   {Blinnikov} S.~I.,
  2016, \mn@doi [\mnras] {10.1093/mnras/stw790}, \href
  {http://adsabs.harvard.edu/abs/2016MNRAS.459.2188B} {459, 2188}

\bibitem[\protect\citeauthoryear{{Bagetakos}, {Brinks}, {Walter}, {de Blok},
  {Usero}, {Leroy}, {Rich}  \& {Kennicutt}}{{Bagetakos}
  et~al.}{2011}]{Bagetakos_2011}
{Bagetakos} I.,  {Brinks} E.,  {Walter} F.,  {de Blok} W.~J.~G.,  {Usero} A.,
  {Leroy} A.~K.,  {Rich} J.~W.,   {Kennicutt} Jr. R.~C.,  2011, \mn@doi [\aj]
  {10.1088/0004-6256/141/1/23}, \href
  {http://adsabs.harvard.edu/abs/2011AJ....141...23B} {141, 23}

\bibitem[\protect\citeauthoryear{{Balbus} \& {McKee}}{{Balbus} \&
  {McKee}}{1982}]{Balbus_1982}
{Balbus} S.~A.,  {McKee} C.~F.,  1982, \mn@doi [\apj] {10.1086/159581}, \href
  {http://adsabs.harvard.edu/abs/1982ApJ...252..529B} {252, 529}

\bibitem[\protect\citeauthoryear{{Bale}, {Pulupa}, {Salem}, {Chen}  \&
  {Quataert}}{{Bale} et~al.}{2013}]{Bale_2013}
{Bale} S.~D.,  {Pulupa} M.,  {Salem} C.,  {Chen} C.~H.~K.,   {Quataert} E.,
  2013, \mn@doi [\apjl] {10.1088/2041-8205/769/2/L22}, \href
  {http://adsabs.harvard.edu/abs/2013ApJ...769L..22B} {769, L22}

\bibitem[\protect\citeauthoryear{{Begelman} \& {McKee}}{{Begelman} \&
  {McKee}}{1990}]{Begelman_1990}
{Begelman} M.~C.,  {McKee} C.~F.,  1990, \mn@doi [\apj] {10.1086/168994}, \href
  {http://adsabs.harvard.edu/abs/1990ApJ...358..375B} {358, 375}

\bibitem[\protect\citeauthoryear{{Blondin}, {Wright}, {Borkowski}  \&
  {Reynolds}}{{Blondin} et~al.}{1998}]{Blondin_1998}
{Blondin} J.~M.,  {Wright} E.~B.,  {Borkowski} K.~J.,   {Reynolds} S.~P.,
  1998, \mn@doi [\apj] {10.1086/305708}, \href
  {http://adsabs.harvard.edu/abs/1998ApJ...500..342B} {500, 342}

\bibitem[\protect\citeauthoryear{{Bowen} et~al.,}{{Bowen}
  et~al.}{2008}]{Bowen_2008}
{Bowen} D.~V.,  et~al., 2008, \mn@doi [\apjs] {10.1086/524773}, \href
  {http://adsabs.harvard.edu/abs/2008ApJS..176...59B} {176, 59}

\bibitem[\protect\citeauthoryear{{Braginskii}}{{Braginskii}}{1965}]{Braginskii_1965}
{Braginskii} S.~I.,  1965, Reviews of Plasma Physics, \href
  {http://adsabs.harvard.edu/abs/1965RvPP....1..205B} {1, 205}

\bibitem[\protect\citeauthoryear{{Brinks} \& {Bajaja}}{{Brinks} \&
  {Bajaja}}{1986}]{Brinks_1986}
{Brinks} E.,  {Bajaja} E.,  1986, \aap, \href
  {http://adsabs.harvard.edu/abs/1986A%26A...169...14B} {169, 14}

\bibitem[\protect\citeauthoryear{{Bucciantini}, {Amato}, {Bandiera}, {Blondin}
  \& {Del Zanna}}{{Bucciantini} et~al.}{2004}]{Bucciantini_2004}
{Bucciantini} N.,  {Amato} E.,  {Bandiera} R.,  {Blondin} J.~M.,   {Del Zanna}
  L.,  2004, \mn@doi [\aap] {10.1051/0004-6361:20040360}, \href
  {https://ui.adsabs.harvard.edu/\#abs/2004A&A...423..253B} {423, 253}

\bibitem[\protect\citeauthoryear{{Cash}, {Charles}, {Bowyer}, {Walter},
  {Garmire}  \& {Riegler}}{{Cash} et~al.}{1980}]{Cash_1980}
{Cash} W.,  {Charles} P.,  {Bowyer} S.,  {Walter} F.,  {Garmire} G.,
  {Riegler} G.,  1980, \mn@doi [\apjl] {10.1086/183261}, \href
  {http://adsabs.harvard.edu/abs/1980ApJ...238L..71C} {238, L71}

\bibitem[\protect\citeauthoryear{{Castor}, {McCray}  \& {Weaver}}{{Castor}
  et~al.}{1975}]{Castor_1975}
{Castor} J.,  {McCray} R.,   {Weaver} R.,  1975, \mn@doi [\apjl]
  {10.1086/181908}, \href {http://adsabs.harvard.edu/abs/1975ApJ...200L.107C}
  {200, L107}

\bibitem[\protect\citeauthoryear{{Cecil}, {Bland-Hawthorn}  \&
  {Veilleux}}{{Cecil} et~al.}{2002}]{Cecil_2002}
{Cecil} G.,  {Bland-Hawthorn} J.,   {Veilleux} S.,  2002, \mn@doi [\apj]
  {10.1086/341861}, \href {http://adsabs.harvard.edu/abs/2002ApJ...576..745C}
  {576, 745}

\bibitem[\protect\citeauthoryear{{Chevalier} \& {Clegg}}{{Chevalier} \&
  {Clegg}}{1985}]{Chevalier_1985}
{Chevalier} R.~A.,  {Clegg} A.~W.,  1985, \mn@doi [\nat] {10.1038/317044a0},
  \href {http://adsabs.harvard.edu/abs/1985Natur.317...44C} {317, 44}

\bibitem[\protect\citeauthoryear{{Chu}, {Gruendl}  \& {Guerrero}}{{Chu}
  et~al.}{2003}]{Chu_2003}
{Chu} Y.-H.,  {Gruendl} R.~A.,   {Guerrero} M.~A.,  2003, in {Arthur} J.,
  {Henney} W.~J.,  eds,  Revista Mexicana de Astronomia y Astrofisica, vol.~27
  Vol. 15, Revista Mexicana de Astronomia y Astrofisica Conference Series. pp
  62--67 (\mn@eprint {} {astro-ph/0212152})

\bibitem[\protect\citeauthoryear{{Cioffi}, {McKee}  \& {Bertschinger}}{{Cioffi}
  et~al.}{1988}]{Cioffi_1988}
{Cioffi} D.~F.,  {McKee} C.~F.,   {Bertschinger} E.,  1988, \mn@doi [\apj]
  {10.1086/166834}, \href {http://adsabs.harvard.edu/abs/1988ApJ...334..252C}
  {334, 252}

\bibitem[\protect\citeauthoryear{{Courant}, {Friedrichs}  \& {Lewy}}{{Courant}
  et~al.}{1928}]{Courant_1928}
{Courant} R.,  {Friedrichs} K.,   {Lewy} H.,  1928, \mn@doi [Mathematische
  Annalen] {10.1007/BF01448839}, \href
  {http://adsabs.harvard.edu/abs/1928MatAn.100...32C} {100, 32}

\bibitem[\protect\citeauthoryear{{Cowie} \& {McKee}}{{Cowie} \&
  {McKee}}{1977}]{Cowie_1977}
{Cowie} L.~L.,  {McKee} C.~F.,  1977, \mn@doi [\apj] {10.1086/154911}, \href
  {http://adsabs.harvard.edu/abs/1977ApJ...211..135C} {211, 135}

\bibitem[\protect\citeauthoryear{{Duffell}}{{Duffell}}{2016}]{Duffell_2016}
{Duffell} P.~C.,  2016, \mn@doi [\apj] {10.3847/0004-637X/821/2/76}, \href
  {http://adsabs.harvard.edu/abs/2016ApJ...821...76D} {821, 76}

\bibitem[\protect\citeauthoryear{{Dunne}, {Points}  \& {Chu}}{{Dunne}
  et~al.}{2001}]{Dunne_2001}
{Dunne} B.~C.,  {Points} S.~D.,   {Chu} Y.-H.,  2001, \mn@doi [\apjs]
  {10.1086/321794}, \href {http://adsabs.harvard.edu/abs/2001ApJS..136..119D}
  {136, 119}

\bibitem[\protect\citeauthoryear{{El-Badry} et~al.,}{{El-Badry}
  et~al.}{2018}]{ElBadry_2018}
{El-Badry} K.,  et~al., 2018, \mn@doi [\mnras] {10.1093/mnras/stx2482}, \href
  {http://adsabs.harvard.edu/abs/2018MNRAS.473.1930E} {473, 1930}

\bibitem[\protect\citeauthoryear{{Elmegreen} \& {Scalo}}{{Elmegreen} \&
  {Scalo}}{2004}]{Elmegreen_2004}
{Elmegreen} B.~G.,  {Scalo} J.,  2004, \mn@doi [\araa]
  {10.1146/annurev.astro.41.011802.094859}, \href
  {http://adsabs.harvard.edu/abs/2004ARA%26A..42..211E} {42, 211}

\bibitem[\protect\citeauthoryear{{Fielding}, {Quataert}, {Martizzi}  \&
  {Faucher-Gigu{\`e}re}}{{Fielding} et~al.}{2017}]{Fielding_2017}
{Fielding} D.,  {Quataert} E.,  {Martizzi} D.,   {Faucher-Gigu{\`e}re} C.-A.,
  2017, \mn@doi [\mnras] {10.1093/mnrasl/slx072}, \href
  {http://adsabs.harvard.edu/abs/2017MNRAS.470L..39F} {470, L39}

\bibitem[\protect\citeauthoryear{{Fielding}, {Quataert}  \&
  {Martizzi}}{{Fielding} et~al.}{2018}]{Fielding_2018}
{Fielding} D.,  {Quataert} E.,   {Martizzi} D.,  2018, \mn@doi [\mnras]
  {10.1093/mnras/sty2466}, \href
  {http://adsabs.harvard.edu/abs/2018MNRAS.tmp.2355F} {}

\bibitem[\protect\citeauthoryear{{Fierlinger}, {Burkert}, {Ntormousi},
  {Fierlinger}, {Schartmann}, {Ballone}, {Krause}  \& {Diehl}}{{Fierlinger}
  et~al.}{2016}]{Fierlinger_2016}
{Fierlinger} K.~M.,  {Burkert} A.,  {Ntormousi} E.,  {Fierlinger} P.,
  {Schartmann} M.,  {Ballone} A.,  {Krause} M.~G.~H.,   {Diehl} R.,  2016,
  \mn@doi [\mnras] {10.1093/mnras/stv2699}, \href
  {http://adsabs.harvard.edu/abs/2016MNRAS.456..710F} {456, 710}

\bibitem[\protect\citeauthoryear{{Folini} \& {Walder}}{{Folini} \&
  {Walder}}{2006}]{Folini_2006}
{Folini} D.,  {Walder} R.,  2006, \mn@doi [\aap] {10.1051/0004-6361:20053898},
  \href {http://adsabs.harvard.edu/abs/2006A%26A...459....1F} {459, 1}

\bibitem[\protect\citeauthoryear{{Geen}, {Hennebelle}, {Tremblin}  \&
  {Rosdahl}}{{Geen} et~al.}{2015}]{Geen_2015}
{Geen} S.,  {Hennebelle} P.,  {Tremblin} P.,   {Rosdahl} J.,  2015, \mn@doi
  [\mnras] {10.1093/mnras/stv2272}, \href
  {https://ui.adsabs.harvard.edu/abs/2015MNRAS.454.4484G} {454, 4484}

\bibitem[\protect\citeauthoryear{{Gentry}, {Krumholz}, {Dekel}  \&
  {Madau}}{{Gentry} et~al.}{2017}]{Gentry_2017}
{Gentry} E.~S.,  {Krumholz} M.~R.,  {Dekel} A.,   {Madau} P.,  2017, \mn@doi
  [\mnras] {10.1093/mnras/stw2746}, \href
  {http://adsabs.harvard.edu/abs/2017MNRAS.465.2471G} {465, 2471}

\bibitem[\protect\citeauthoryear{{Gentry}, {Krumholz}, {Madau}  \&
  {Lupi}}{{Gentry} et~al.}{2019}]{Gentry_2019}
{Gentry} E.~S.,  {Krumholz} M.~R.,  {Madau} P.,   {Lupi} A.,  2019, \mn@doi
  [\mnras] {10.1093/mnras/sty3319}, \href
  {https://ui.adsabs.harvard.edu/\#abs/2019MNRAS.483.3647G} {483, 3647}

\bibitem[\protect\citeauthoryear{{Gupta}, {Nath}, {Sharma}  \&
  {Shchekinov}}{{Gupta} et~al.}{2016}]{Gupta_2016}
{Gupta} S.,  {Nath} B.~B.,  {Sharma} P.,   {Shchekinov} Y.,  2016, \mn@doi
  [\mnras] {10.1093/mnras/stw1920}, \href
  {http://adsabs.harvard.edu/abs/2016MNRAS.462.4532G} {462, 4532}

\bibitem[\protect\citeauthoryear{{Harper-Clark} \& {Murray}}{{Harper-Clark} \&
  {Murray}}{2009}]{Harper_Clark_2009}
{Harper-Clark} E.,  {Murray} N.,  2009, \mn@doi [\apj]
  {10.1088/0004-637X/693/2/1696}, \href
  {https://ui.adsabs.harvard.edu/abs/2009ApJ...693.1696H} {693, 1696}

\bibitem[\protect\citeauthoryear{{Heiles}}{{Heiles}}{1979}]{Heiles_1979}
{Heiles} C.,  1979, \mn@doi [\apj] {10.1086/156986}, \href
  {http://adsabs.harvard.edu/abs/1979ApJ...229..533H} {229, 533}

\bibitem[\protect\citeauthoryear{{Hopkins}}{{Hopkins}}{2015}]{Hopkins_2015}
{Hopkins} P.~F.,  2015, \mn@doi [\mnras] {10.1093/mnras/stv195}, \href
  {http://adsabs.harvard.edu/abs/2015MNRAS.450...53H} {450, 53}

\bibitem[\protect\citeauthoryear{{Iffrig} \& {Hennebelle}}{{Iffrig} \&
  {Hennebelle}}{2015}]{Iffrig_2015}
{Iffrig} O.,  {Hennebelle} P.,  2015, \mn@doi [\aap]
  {10.1051/0004-6361/201424556}, \href
  {http://adsabs.harvard.edu/abs/2015A%26A...576A..95I} {576, A95}

\bibitem[\protect\citeauthoryear{{Inoue}, {Shimoda}, {Ohira}  \&
  {Yamazaki}}{{Inoue} et~al.}{2013}]{Inoue_2013}
{Inoue} T.,  {Shimoda} J.,  {Ohira} Y.,   {Yamazaki} R.,  2013, \mn@doi [\apjl]
  {10.1088/2041-8205/772/2/L20}, \href
  {http://adsabs.harvard.edu/abs/2013ApJ...772L..20I} {772, L20}

\bibitem[\protect\citeauthoryear{{Jaskot}, {Strickland}, {Oey}, {Chu}  \&
  {Garc{\'{\i}}a-Segura}}{{Jaskot} et~al.}{2011}]{Jaskot_2011}
{Jaskot} A.~E.,  {Strickland} D.~K.,  {Oey} M.~S.,  {Chu} Y.-H.,
  {Garc{\'{\i}}a-Segura} G.,  2011, \mn@doi [\apj]
  {10.1088/0004-637X/729/1/28}, \href
  {http://adsabs.harvard.edu/abs/2011ApJ...729...28J} {729, 28}

\bibitem[\protect\citeauthoryear{{Keller}, {Wadsley}, {Benincasa}  \&
  {Couchman}}{{Keller} et~al.}{2014}]{Keller_2014}
{Keller} B.~W.,  {Wadsley} J.,  {Benincasa} S.~M.,   {Couchman} H.~M.~P.,
  2014, \mn@doi [\mnras] {10.1093/mnras/stu1058}, \href
  {http://adsabs.harvard.edu/abs/2014MNRAS.442.3013K} {442, 3013}

\bibitem[\protect\citeauthoryear{{Kim} \& {Ostriker}}{{Kim} \&
  {Ostriker}}{2015}]{Kim_2015}
{Kim} C.-G.,  {Ostriker} E.~C.,  2015, \mn@doi [\apj]
  {10.1088/0004-637X/802/2/99}, \href
  {http://adsabs.harvard.edu/abs/2015ApJ...802...99K} {802, 99}

\bibitem[\protect\citeauthoryear{{Kim} \& {Ostriker}}{{Kim} \&
  {Ostriker}}{2017}]{Kim_Ostriker_2017}
{Kim} C.-G.,  {Ostriker} E.~C.,  2017, \mn@doi [\apj]
  {10.3847/1538-4357/aa8599}, \href
  {http://adsabs.harvard.edu/abs/2017ApJ...846..133K} {846, 133}

\bibitem[\protect\citeauthoryear{{Kim} \& {Ostriker}}{{Kim} \&
  {Ostriker}}{2018}]{Kim_2018}
{Kim} C.-G.,  {Ostriker} E.~C.,  2018, \mn@doi [\apj]
  {10.3847/1538-4357/aaa5ff}, \href
  {http://adsabs.harvard.edu/abs/2018ApJ...853..173K} {853, 173}

\bibitem[\protect\citeauthoryear{{Kim}, {Ostriker}  \& {Kim}}{{Kim}
  et~al.}{2013}]{Kim_2013}
{Kim} C.-G.,  {Ostriker} E.~C.,   {Kim} W.-T.,  2013, \mn@doi [\apj]
  {10.1088/0004-637X/776/1/1}, \href
  {http://adsabs.harvard.edu/abs/2013ApJ...776....1K} {776, 1}

\bibitem[\protect\citeauthoryear{{Kim}, {Ostriker}  \& {Raileanu}}{{Kim}
  et~al.}{2017}]{Kim_2017}
{Kim} C.-G.,  {Ostriker} E.~C.,   {Raileanu} R.,  2017, \mn@doi [\apj]
  {10.3847/1538-4357/834/1/25}, \href
  {http://adsabs.harvard.edu/abs/2017ApJ...834...25K} {834, 25}

\bibitem[\protect\citeauthoryear{{Kim}, {Kim}  \& {Ostriker}}{{Kim}
  et~al.}{2018}]{JGKim_2018}
{Kim} J.-G.,  {Kim} W.-T.,   {Ostriker} E.~C.,  2018, \mn@doi [\apj]
  {10.3847/1538-4357/aabe27}, \href
  {https://ui.adsabs.harvard.edu/abs/2018ApJ...859...68K} {859, 68}

\bibitem[\protect\citeauthoryear{{Koo} \& {McKee}}{{Koo} \&
  {McKee}}{1992}]{Koo_1992}
{Koo} B.-C.,  {McKee} C.~F.,  1992, \mn@doi [\apj] {10.1086/171132}, \href
  {http://adsabs.harvard.edu/abs/1992ApJ...388...93K} {388, 93}

\bibitem[\protect\citeauthoryear{{Koyama} \& {Inutsuka}}{{Koyama} \&
  {Inutsuka}}{2002}]{Koyama_2002}
{Koyama} H.,  {Inutsuka} S.-i.,  2002, \mn@doi [\apjl] {10.1086/338978}, \href
  {http://adsabs.harvard.edu/abs/2002ApJ...564L..97K} {564, L97}

\bibitem[\protect\citeauthoryear{{Krause}, {Fierlinger}, {Diehl}, {Burkert},
  {Voss}  \& {Ziegler}}{{Krause} et~al.}{2013}]{Krause_2013}
{Krause} M.,  {Fierlinger} K.,  {Diehl} R.,  {Burkert} A.,  {Voss} R.,
  {Ziegler} U.,  2013, \mn@doi [\aap] {10.1051/0004-6361/201220060}, \href
  {https://ui.adsabs.harvard.edu/\#abs/2013A&A...550A..49K} {550, A49}

\bibitem[\protect\citeauthoryear{{Kregenow} et~al.,}{{Kregenow}
  et~al.}{2006}]{Kregenow_2006}
{Kregenow} J.,  et~al., 2006, \mn@doi [\apjl] {10.1086/505196}, \href
  {http://adsabs.harvard.edu/abs/2006ApJ...644L.167K} {644, L167}

\bibitem[\protect\citeauthoryear{{Kroupa}}{{Kroupa}}{2001}]{Kroupa_2001}
{Kroupa} P.,  2001, \mn@doi [\mnras] {10.1046/j.1365-8711.2001.04022.x}, \href
  {http://adsabs.harvard.edu/abs/2001MNRAS.322..231K} {322, 231}

\bibitem[\protect\citeauthoryear{{Leitherer} et~al.,}{{Leitherer}
  et~al.}{1999}]{Leitherer_1999}
{Leitherer} C.,  et~al., 1999, \mn@doi [\apjs] {10.1086/313233}, \href
  {http://adsabs.harvard.edu/abs/1999ApJS..123....3L} {123, 3}

\bibitem[\protect\citeauthoryear{{Leroy} et~al.,}{{Leroy}
  et~al.}{2018}]{Leroy_2018}
{Leroy} A.~K.,  et~al., 2018, \mn@doi [\apj] {10.3847/1538-4357/aaecd1}, \href
  {http://esoads.eso.org/abs/2018ApJ...869..126L} {869, 126}

\bibitem[\protect\citeauthoryear{{Li}, {Bryan}  \& {Ostriker}}{{Li}
  et~al.}{2017}]{Li_2017}
{Li} M.,  {Bryan} G.~L.,   {Ostriker} J.~P.,  2017, \mn@doi [\apjl]
  {10.3847/2041-8213/835/1/L10}, \href
  {http://adsabs.harvard.edu/abs/2017ApJ...835L..10L} {835, L10}

\bibitem[\protect\citeauthoryear{{Mac Low} \& {McCray}}{{Mac Low} \&
  {McCray}}{1988}]{Maclow_1988}
{Mac Low} M.-M.,  {McCray} R.,  1988, \mn@doi [\apj] {10.1086/165936}, \href
  {http://adsabs.harvard.edu/abs/1988ApJ...324..776M} {324, 776}

\bibitem[\protect\citeauthoryear{{Mac Low}, {McCray}  \& {Norman}}{{Mac Low}
  et~al.}{1989}]{Maclow_1989}
{Mac Low} M.-M.,  {McCray} R.,   {Norman} M.~L.,  1989, \mn@doi [\apj]
  {10.1086/167094}, \href {http://adsabs.harvard.edu/abs/1989ApJ...337..141M}
  {337, 141}

\bibitem[\protect\citeauthoryear{{Martizzi}, {Faucher-Gigu{\`e}re}  \&
  {Quataert}}{{Martizzi} et~al.}{2015}]{Martizzi_2015}
{Martizzi} D.,  {Faucher-Gigu{\`e}re} C.-A.,   {Quataert} E.,  2015, \mn@doi
  [\mnras] {10.1093/mnras/stv562}, \href
  {http://adsabs.harvard.edu/abs/2015MNRAS.450..504M} {450, 504}

\bibitem[\protect\citeauthoryear{Max, McKee  \& Mead}{Max
  et~al.}{1980}]{Max_1980}
Max C.~E.,  McKee C.~F.,   Mead W.~C.,  1980, \mn@doi [The Physics of Fluids]
  {10.1063/1.863183}, 23, 1620

\bibitem[\protect\citeauthoryear{{McCray} \& {Kafatos}}{{McCray} \&
  {Kafatos}}{1987}]{McCray_1987}
{McCray} R.,  {Kafatos} M.,  1987, \mn@doi [\apj] {10.1086/165267}, \href
  {http://adsabs.harvard.edu/abs/1987ApJ...317..190M} {317, 190}

\bibitem[\protect\citeauthoryear{{McKee} \& {Cowie}}{{McKee} \&
  {Cowie}}{1977}]{McKee_1977}
{McKee} C.~F.,  {Cowie} L.~L.,  1977, \mn@doi [\apj] {10.1086/155350}, \href
  {http://adsabs.harvard.edu/abs/1977ApJ...215..213M} {215, 213}

\bibitem[\protect\citeauthoryear{{McKee} \& {Ostriker}}{{McKee} \&
  {Ostriker}}{2007}]{McKee_2007}
{McKee} C.~F.,  {Ostriker} E.~C.,  2007, \mn@doi [\araa]
  {10.1146/annurev.astro.45.051806.110602}, \href
  {http://adsabs.harvard.edu/abs/2007ARA%26A..45..565M} {45, 565}

\bibitem[\protect\citeauthoryear{{Michaut}, {Cavet}, {Bouquet}, {Roy}  \&
  {Nguyen}}{{Michaut} et~al.}{2012}]{Michaut_2012}
{Michaut} C.,  {Cavet} C.,  {Bouquet} S.~E.,  {Roy} F.,   {Nguyen} H.~C.,
  2012, \mn@doi [\apj] {10.1088/0004-637X/759/2/78}, \href
  {http://adsabs.harvard.edu/abs/2012ApJ...759...78M} {759, 78}

\bibitem[\protect\citeauthoryear{{Miles}}{{Miles}}{2009}]{Miles_2009}
{Miles} A.~R.,  2009, \mn@doi [\apj] {10.1088/0004-637X/696/1/498}, \href
  {http://adsabs.harvard.edu/abs/2009ApJ...696..498M} {696, 498}

\bibitem[\protect\citeauthoryear{{Muratov}, {Kere{\v s}},
  {Faucher-Gigu{\`e}re}, {Hopkins}, {Quataert}  \& {Murray}}{{Muratov}
  et~al.}{2015}]{Muratov_2015}
{Muratov} A.~L.,  {Kere{\v s}} D.,  {Faucher-Gigu{\`e}re} C.-A.,  {Hopkins}
  P.~F.,  {Quataert} E.,   {Murray} N.,  2015, \mn@doi [\mnras]
  {10.1093/mnras/stv2126}, \href
  {http://adsabs.harvard.edu/abs/2015MNRAS.454.2691M} {454, 2691}

\bibitem[\protect\citeauthoryear{{Nath} \& {Shchekinov}}{{Nath} \&
  {Shchekinov}}{2013}]{Nath_2013}
{Nath} B.~B.,  {Shchekinov} Y.,  2013, \mn@doi [\apjl]
  {10.1088/2041-8205/777/1/L12}, \href
  {http://adsabs.harvard.edu/abs/2013ApJ...777L..12N} {777, L12}

\bibitem[\protect\citeauthoryear{{Norman} \& {Ferrara}}{{Norman} \&
  {Ferrara}}{1996}]{Norman_1996}
{Norman} C.~A.,  {Ferrara} A.,  1996, \mn@doi [\apj] {10.1086/177603}, \href
  {http://adsabs.harvard.edu/abs/1996ApJ...467..280N} {467, 280}

\bibitem[\protect\citeauthoryear{{Norman} \& {Ikeuchi}}{{Norman} \&
  {Ikeuchi}}{1989}]{Norman_1989}
{Norman} C.~A.,  {Ikeuchi} S.,  1989, \mn@doi [\apj] {10.1086/167912}, \href
  {http://adsabs.harvard.edu/abs/1989ApJ...345..372N} {345, 372}

\bibitem[\protect\citeauthoryear{{Ntormousi}, {Burkert}, {Fierlinger}  \&
  {Heitsch}}{{Ntormousi} et~al.}{2011}]{Ntormousi_2011}
{Ntormousi} E.,  {Burkert} A.,  {Fierlinger} K.,   {Heitsch} F.,  2011, \mn@doi
  [\apj] {10.1088/0004-637X/731/1/13}, \href
  {http://adsabs.harvard.edu/abs/2011ApJ...731...13N} {731, 13}

\bibitem[\protect\citeauthoryear{{Ntormousi}, {Dawson}, {Hennebelle}  \&
  {Fierlinger}}{{Ntormousi} et~al.}{2017}]{Ntormousi_2017}
{Ntormousi} E.,  {Dawson} J.~R.,  {Hennebelle} P.,   {Fierlinger} K.,  2017,
  \mn@doi [\aap] {10.1051/0004-6361/201629268}, \href
  {https://ui.adsabs.harvard.edu/abs/2017A&A...599A..94N} {599, A94}

\bibitem[\protect\citeauthoryear{{Oey} \& {Massey}}{{Oey} \&
  {Massey}}{1995}]{Oey_1995}
{Oey} M.~S.,  {Massey} P.,  1995, \mn@doi [\apj] {10.1086/176292}, \href
  {http://adsabs.harvard.edu/abs/1995ApJ...452..210O} {452, 210}

\bibitem[\protect\citeauthoryear{{Oppenheimer} \& {Schaye}}{{Oppenheimer} \&
  {Schaye}}{2013}]{Oppenheimer_2013}
{Oppenheimer} B.~D.,  {Schaye} J.,  2013, \mn@doi [\mnras]
  {10.1093/mnras/stt1043}, \href
  {http://adsabs.harvard.edu/abs/2013MNRAS.434.1043O} {434, 1043}

\bibitem[\protect\citeauthoryear{{Ostriker} \& {Shetty}}{{Ostriker} \&
  {Shetty}}{2011}]{Ostriker_2011}
{Ostriker} E.~C.,  {Shetty} R.,  2011, \mn@doi [\apj]
  {10.1088/0004-637X/731/1/41}, \href
  {http://adsabs.harvard.edu/abs/2011ApJ...731...41O} {731, 41}

\bibitem[\protect\citeauthoryear{{Parker}}{{Parker}}{1953}]{Parker_1953}
{Parker} E.~N.,  1953, \mn@doi [\apj] {10.1086/145707}, \href
  {http://adsabs.harvard.edu/abs/1953ApJ...117..431P} {117, 431}

\bibitem[\protect\citeauthoryear{{Parker}}{{Parker}}{1963}]{Parker_1963}
{Parker} E.~N.,  1963, {Interplanetary dynamical processes.}

\bibitem[\protect\citeauthoryear{{Pittard}}{{Pittard}}{2013}]{Pittard_2013}
{Pittard} J.~M.,  2013, \mn@doi [\mnras] {10.1093/mnras/stt1552}, \href
  {http://adsabs.harvard.edu/abs/2013MNRAS.435.3600P} {435, 3600}

\bibitem[\protect\citeauthoryear{{Puche}, {Westpfahl}, {Brinks}  \&
  {Roy}}{{Puche} et~al.}{1992}]{Puche_1992}
{Puche} D.,  {Westpfahl} D.,  {Brinks} E.,   {Roy} J.-R.,  1992, \mn@doi [\aj]
  {10.1086/116199}, \href {http://adsabs.harvard.edu/abs/1992AJ....103.1841P}
  {103, 1841}

\bibitem[\protect\citeauthoryear{{Rosen}, {Lopez}, {Krumholz}  \&
  {Ramirez-Ruiz}}{{Rosen} et~al.}{2014}]{Rosen_2014}
{Rosen} A.~L.,  {Lopez} L.~A.,  {Krumholz} M.~R.,   {Ramirez-Ruiz} E.,  2014,
  \mn@doi [\mnras] {10.1093/mnras/stu1037}, \href
  {https://ui.adsabs.harvard.edu/abs/2014MNRAS.442.2701R} {442, 2701}

\bibitem[\protect\citeauthoryear{{Sankrit} \& {Dixon}}{{Sankrit} \&
  {Dixon}}{2007}]{Sankrit_2007}
{Sankrit} R.,  {Dixon} W.~V.~D.,  2007, \mn@doi [\pasp] {10.1086/513288}, \href
  {http://adsabs.harvard.edu/abs/2007PASP..119..284S} {119, 284}

\bibitem[\protect\citeauthoryear{{Sano}, {Nishihara}, {Matsuoka}  \&
  {Inoue}}{{Sano} et~al.}{2012}]{Sano_2012}
{Sano} T.,  {Nishihara} K.,  {Matsuoka} C.,   {Inoue} T.,  2012, \mn@doi [\apj]
  {10.1088/0004-637X/758/2/126}, \href
  {http://adsabs.harvard.edu/abs/2012ApJ...758..126S} {758, 126}

\bibitem[\protect\citeauthoryear{{Savage} \& {Lehner}}{{Savage} \&
  {Lehner}}{2006}]{Savage_2006}
{Savage} B.~D.,  {Lehner} N.,  2006, \mn@doi [\apjs] {10.1086/497915}, \href
  {http://adsabs.harvard.edu/abs/2006ApJS..162..134S} {162, 134}

\bibitem[\protect\citeauthoryear{{Sharma}, {Roy}, {Nath}  \&
  {Shchekinov}}{{Sharma} et~al.}{2014}]{Sharma_2014}
{Sharma} P.,  {Roy} A.,  {Nath} B.~B.,   {Shchekinov} Y.,  2014, \mn@doi
  [\mnras] {10.1093/mnras/stu1307}, \href
  {http://adsabs.harvard.edu/abs/2014MNRAS.443.3463S} {443, 3463}

\bibitem[\protect\citeauthoryear{{Silich}, {Franco}, {Palous}  \&
  {Tenorio-Tagle}}{{Silich} et~al.}{1996}]{Silich_1996}
{Silich} S.~A.,  {Franco} J.,  {Palous} J.,   {Tenorio-Tagle} G.,  1996,
  \mn@doi [\apj] {10.1086/177728}, \href
  {http://adsabs.harvard.edu/abs/1996ApJ...468..722S} {468, 722}

\bibitem[\protect\citeauthoryear{{Spitzer}}{{Spitzer}}{1962}]{Spitzer_1962}
{Spitzer} L.,  1962, {Physics of Fully Ionized Gases}

\bibitem[\protect\citeauthoryear{{Stone} \& {Gardiner}}{{Stone} \&
  {Gardiner}}{2010}]{Stone_2010}
{Stone} J.~M.,  {Gardiner} T.~A.,  2010, \mn@doi [\apjs]
  {10.1088/0067-0049/189/1/142}, \href
  {http://adsabs.harvard.edu/abs/2010ApJS..189..142S} {189, 142}

\bibitem[\protect\citeauthoryear{{Stone}, {Gardiner}, {Teuben}, {Hawley}  \&
  {Simon}}{{Stone} et~al.}{2008}]{Stone_2008}
{Stone} J.~M.,  {Gardiner} T.~A.,  {Teuben} P.,  {Hawley} J.~F.,   {Simon}
  J.~B.,  2008, \mn@doi [\apjs] {10.1086/588755}, \href
  {http://adsabs.harvard.edu/abs/2008ApJS..178..137S} {178, 137}

\bibitem[\protect\citeauthoryear{{Stone}, {Tomida}, {White}  \&
  {Felker}}{{Stone} et~al.}{2019}]{Stone_2019}
{Stone} J.,  {Tomida} K.,  {White} C.,   {Felker} K.,  2019, ApJS, submitted.

\bibitem[\protect\citeauthoryear{{Strickland}, {Heckman}, {Colbert}, {Hoopes}
  \& {Weaver}}{{Strickland} et~al.}{2004}]{Strickland_2004}
{Strickland} D.~K.,  {Heckman} T.~M.,  {Colbert} E.~J.~M.,  {Hoopes} C.~G.,
  {Weaver} K.~A.,  2004, \mn@doi [\apjs] {10.1086/382214}, \href
  {http://adsabs.harvard.edu/abs/2004ApJS..151..193S} {151, 193}

\bibitem[\protect\citeauthoryear{{Thompson}, {Quataert}, {Zhang}  \&
  {Weinberg}}{{Thompson} et~al.}{2016}]{Thompson_2016}
{Thompson} T.~A.,  {Quataert} E.,  {Zhang} D.,   {Weinberg} D.~H.,  2016,
  \mn@doi [\mnras] {10.1093/mnras/stv2428}, \href
  {http://esoads.eso.org/abs/2016MNRAS.455.1830T} {455, 1830}

\bibitem[\protect\citeauthoryear{{Thornton}, {Gaudlitz}, {Janka}  \&
  {Steinmetz}}{{Thornton} et~al.}{1998}]{Thornton_1998}
{Thornton} K.,  {Gaudlitz} M.,  {Janka} H.-T.,   {Steinmetz} M.,  1998, \mn@doi
  [\apj] {10.1086/305704}, \href
  {http://esoads.eso.org/abs/1998ApJ...500...95T} {500, 95}

\bibitem[\protect\citeauthoryear{{Tomisaka} \& {Ikeuchi}}{{Tomisaka} \&
  {Ikeuchi}}{1986}]{Tomisaka_1986}
{Tomisaka} K.,  {Ikeuchi} S.,  1986, \pasj, \href
  {http://adsabs.harvard.edu/abs/1986PASJ...38..697T} {38, 697}

\bibitem[\protect\citeauthoryear{{Tomisaka}, {Habe}  \& {Ikeuchi}}{{Tomisaka}
  et~al.}{1981}]{Tomisaka_1981}
{Tomisaka} K.,  {Habe} A.,   {Ikeuchi} S.,  1981, \mn@doi [\apss]
  {10.1007/BF00648941}, \href
  {http://adsabs.harvard.edu/abs/1981Ap%26SS..78..273T} {78, 273}

\bibitem[\protect\citeauthoryear{{Townsley}, {Broos}, {Feigelson}, {Brandl},
  {Chu}, {Garmire}  \& {Pavlov}}{{Townsley} et~al.}{2006}]{Townsley_2006}
{Townsley} L.~K.,  {Broos} P.~S.,  {Feigelson} E.~D.,  {Brandl} B.~R.,  {Chu}
  Y.-H.,  {Garmire} G.~P.,   {Pavlov} G.~G.,  2006, \mn@doi [\aj]
  {10.1086/500532}, \href {http://adsabs.harvard.edu/abs/2006AJ....131.2140T}
  {131, 2140}

\bibitem[\protect\citeauthoryear{{Vasiliev}, {Nath}  \&
  {Shchekinov}}{{Vasiliev} et~al.}{2015}]{Vasiliev_2015}
{Vasiliev} E.~O.,  {Nath} B.~B.,   {Shchekinov} Y.,  2015, \mn@doi [\mnras]
  {10.1093/mnras/stu2133}, \href
  {http://adsabs.harvard.edu/abs/2015MNRAS.446.1703V} {446, 1703}

\bibitem[\protect\citeauthoryear{{Vasiliev}, {Shchekinov}  \&
  {Nath}}{{Vasiliev} et~al.}{2017}]{Vasiliev_2017}
{Vasiliev} E.~O.,  {Shchekinov} Y.~A.,   {Nath} B.~B.,  2017, \mn@doi [\mnras]
  {10.1093/mnras/stx719}, \href {http://esoads.eso.org/abs/2017MNRAS.468.2757V}
  {468, 2757}

\bibitem[\protect\citeauthoryear{{Vishniac} \& {Ryu}}{{Vishniac} \&
  {Ryu}}{1989}]{Vishniac_1989}
{Vishniac} E.~T.,  {Ryu} D.,  1989, \mn@doi [\apj] {10.1086/167161}, \href
  {http://adsabs.harvard.edu/abs/1989ApJ...337..917V} {337, 917}

\bibitem[\protect\citeauthoryear{{Walch} \& {Naab}}{{Walch} \&
  {Naab}}{2015}]{Walch_2015}
{Walch} S.,  {Naab} T.,  2015, \mn@doi [\mnras] {10.1093/mnras/stv1155}, \href
  {http://adsabs.harvard.edu/abs/2015MNRAS.451.2757W} {451, 2757}

\bibitem[\protect\citeauthoryear{{Weaver}, {McCray}, {Castor}, {Shapiro}  \&
  {Moore}}{{Weaver} et~al.}{1977}]{Weaver_1977}
{Weaver} R.,  {McCray} R.,  {Castor} J.,  {Shapiro} P.,   {Moore} R.,  1977,
  \mn@doi [\apj] {10.1086/155692}, \href
  {http://adsabs.harvard.edu/abs/1977ApJ...218..377W} {218, 377}

\bibitem[\protect\citeauthoryear{{Weisz}, {Skillman}, {Cannon}, {Dolphin},
  {Kennicutt}, {Lee}  \& {Walter}}{{Weisz} et~al.}{2009}]{Weisz_2009}
{Weisz} D.~R.,  {Skillman} E.~D.,  {Cannon} J.~M.,  {Dolphin} A.~E.,
  {Kennicutt} Jr. R.~C.,  {Lee} J.,   {Walter} F.,  2009, \mn@doi [\apj]
  {10.1088/0004-637X/704/2/1538}, \href
  {http://adsabs.harvard.edu/abs/2009ApJ...704.1538W} {704, 1538}

\bibitem[\protect\citeauthoryear{{Yadav}, {Mukherjee}, {Sharma}  \&
  {Nath}}{{Yadav} et~al.}{2017}]{Yadav_2017}
{Yadav} N.,  {Mukherjee} D.,  {Sharma} P.,   {Nath} B.~B.,  2017, \mn@doi
  [\mnras] {10.1093/mnras/stw2522}, \href
  {http://adsabs.harvard.edu/abs/2017MNRAS.465.1720Y} {465, 1720}

\makeatother
\end{thebibliography}


\appendix

\section{Numerical convergence}
\label{sec:res_test}

\begin{figure*}
\includegraphics[width=\textwidth]{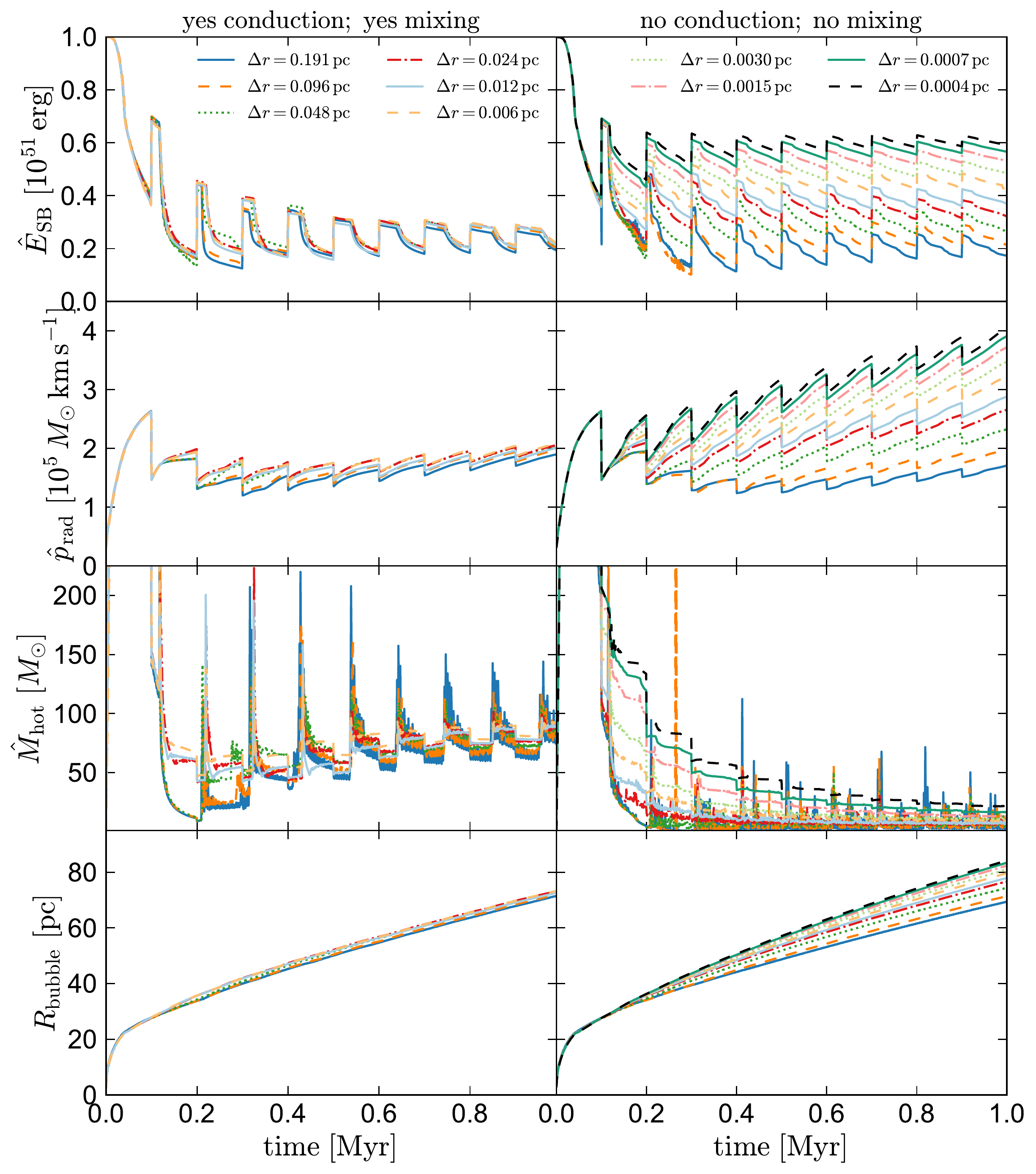}
\caption{Resolution convergence study for simulations with conduction and diffusive mixing (left) and without conduction or mixing (right). The bubble energy, radial momentum, and hot gas mass are all normalized by the total number of SNe that have exploded so far. The colors and line styles for the first six resolution levels are the same for the left and right panels; the four highest-resolution simulations shown in the right panels do not have corresponding simulations in the left panel because simulations with conduction are more expensive. Without conduction, convergence is very slow, because the shell/interior interface becomes increasingly thin at high resolution, leading to less and less cooling. Conduction and mixing prevent the shell/interior interface from becoming arbitrarily thin, enabling convergence. }
\label{fig:res_test}
\end{figure*}

In Figure~\ref{fig:res_test}, we show the effects of varying the simulation resolution on integrated SB properties, for the fiducial model. Simulations with (without) conduction and diffusive mixing (adopting $\lambda \delta v = 1 {\rm pc\ km \ s^{-1}}$) are shown in the left (right) panels. Because the simulations with conduction are computationally more expensive than those without it, the four highest resolution levels ($\Delta r \leq  0.003$\,pc) are only reached in the no-conduction, no-mixing simulations.

The right panels show that the simulations without conduction or diffusive mixing do not fully converge, even at the highest resolution level: increasing the resolution leads to higher SB energy and radial momentum, and a larger bubble radius. Radiative losses in the shell/interior interface are progressively reduced at higher resolution. Indeed, the highest-resolution simulation approaches the $\hat{E}_{\rm SB}/E_{\rm SN} \approx 0.65$ value predicted by the Weaver solution if there is {\it no} cooling within the bubble or interface. The simulations with conduction and mixing (left panels) converge much more rapidly: although there are some minor differences between resolution levels at early times, none of the integrated SB properties show systematic trends with resolution at late times. 
In addition, the simulations without conduction show that the hot gas mass is lower than with conduction, and does not converge with resolution.  

\begin{figure}
\includegraphics[width=\columnwidth]{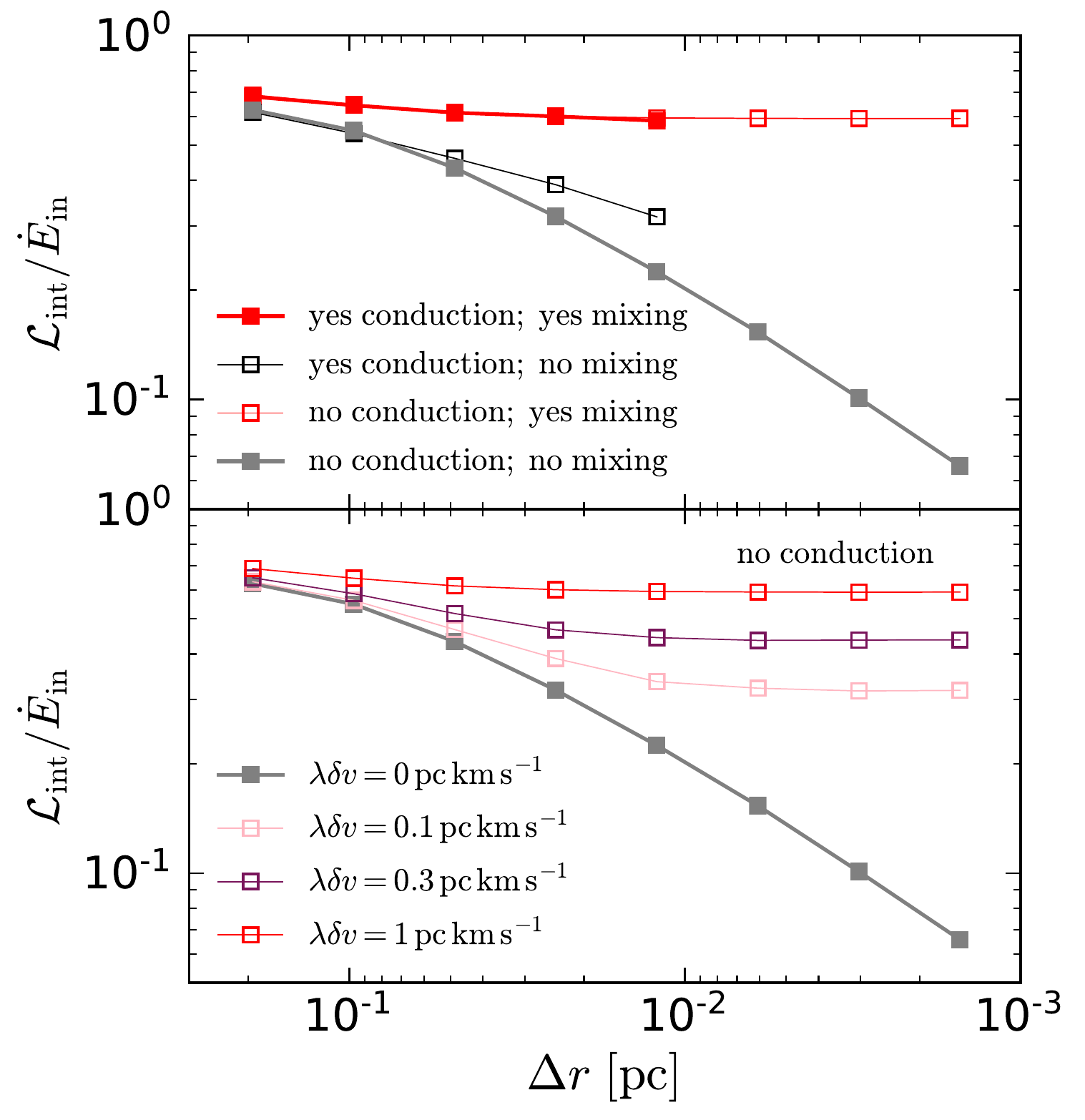}
\caption{Tests of cooling rate in the shell/bubble interface as a function of resolution for the fiducial problem setup with $n_{\rm H,0} = 1\,{\rm cm^{-3}}$ and $\Delta t_{\rm SNe} = 0.1\,\rm Myr$, with varying levels of diffusive mixing and conduction. {\bf Top}: Without conduction or mixing (solid gray), the interface becomes increasingly thin and the cooling rate drops as the resolution improves.
Including conduction but no mixing (open black) reduces variation with resolution but does not lead to convergence at the resolution we can achieve. Including diffusion (representing nonlinear mixing with $\lambda \delta v=1\,\rm pc\,km\,s^{-1}$) broadens the interface and 
leads to convergence at relatively low resolution, both with (solid red) and without (open red) conduction.
{\bf Bottom}: Convergence for different nonlinear mixing efficiencies. Simulations with more efficient mixing (larger $\lambda \delta v$) converge at lower resolution.}
\label{fig:cooling_res}
\end{figure}

The effects of conduction and mixing on the rate of cooling in the interface are investigated further in Figure~\ref{fig:cooling_res}, which shows the cooling rate within the interface (excluding cooling of shocked ISM on the shell exterior; see Section~\ref{sec:fluxes}) as a function of resolution. This comparison is made at late time ($t\approx 10$\,Myr), such that effects from individual SN shocks are no longer important. In the top panel, we compare simulations without conduction or diffusive mixing (solid gray), those with normal Spitzer conduction but no diffusive mixing (open black), those with mixing but no other conduction (open red), and those with both conduction and mixing (solid red). 

Without conduction or mixing, the cooling rate decreases with increasing resolution, without any apparent limit. At late times, shocks do not propagate into the dense gas in the interface, and nothing else in the problem sets a physical scale for the transition between hot bubble and cool shell. The volume of the SB with $T\sim 10^{5}\,\rm K$ and efficient cooling thus 
drops with increasing numerical resolution, such that cooling losses becomes negligible at arbitrarily high resolution. At infinite resolution, $\mathcal{L}_{\rm int}$ is thus expected to go to zero \citep[see also][]{Gentry_2017}.

Including conduction makes the trend with resolution less steep, but does not lead to full convergence over the range of $\Delta r$ where our simulations with conduction are computationally tractable. This likely occurs because (temperature-dependent) thermal conductivity becomes very small at the cool outer edge of the interface bordering the shell. Conduction alone {\it is} expected to lead to convergence once $\Delta r$ is small compared to the minimum Field length in the problem \citep{McKee_1977, Begelman_1990}, but the Field length varies substantially throughout the interface. At $n_{\rm H}= 1\,\rm cm^{-3}$ and $T=2\times 10^4\,{\rm K}$ (representative of the value at the outer edge of the interface; see Figure~\ref{fig:interface}), $\lambda_{\rm Field} \approx 10^{-3}$\,pc, which is smaller than we can achieve in simulations with conduction. Our results are consistent with those of \citet{Fierlinger_2016}, who found that thermal conduction alone did not lead to convergence in 1D simulations of SNe-driven bubbles, except in an extreme case in which thermal conductivity was artificially enhanced by 14 orders of magnitude.

Irrespective of whether conduction is included, including diffusive mixing with $\lambda \delta v = 1\,\rm pc\,km\,s^{-1}$ (red lines in the top panel of Figure~\ref{fig:cooling_res}) leads to fairly rapid convergence. This reflects the fact, already shown in Figure~\ref{fig:integrated_bubble_properties}, that cooling losses are set by mixing, not conduction. Numerically, this is because the mixing coefficient is not temperature dependent, while the conductivity coefficient is.  Thus, the thickness of the layer where there is mixing and therefore strong cooling does not continue to drop at high resolution, leading to convergence in the net cooling rate (see  also  discussion in Section \ref{sec:cooling_eff}).  

The bottom panel of Figure~\ref{fig:cooling_res} compares the convergence of simulations for different values of $\lambda \delta v$. We do not include thermal conduction in these simulations, as it increases the computational expense and does not significantly affect the cooling rate. As expected, simulations with smaller $\lambda \delta v$ (less efficient mixing) require higher resolution to be converged. At the resolution used for most of the simulations in this work ($\Delta r = (0.024-0.048)$\,pc), only simulations with $\lambda \delta v \geq 1\,\rm pc\,km\,s^{-1}$ are fully converged. Simulations with lower, nonzero values of $\lambda \delta v$ do converge at higher resolution. As we showed in Section~\ref{sec:cooling_eff}, the converged cooling rates increase with increasing $\lambda \delta v$.

\section{Efficiency of mixing} \label{sec:subgrid}

\begin{figure}
\includegraphics[width=\columnwidth]{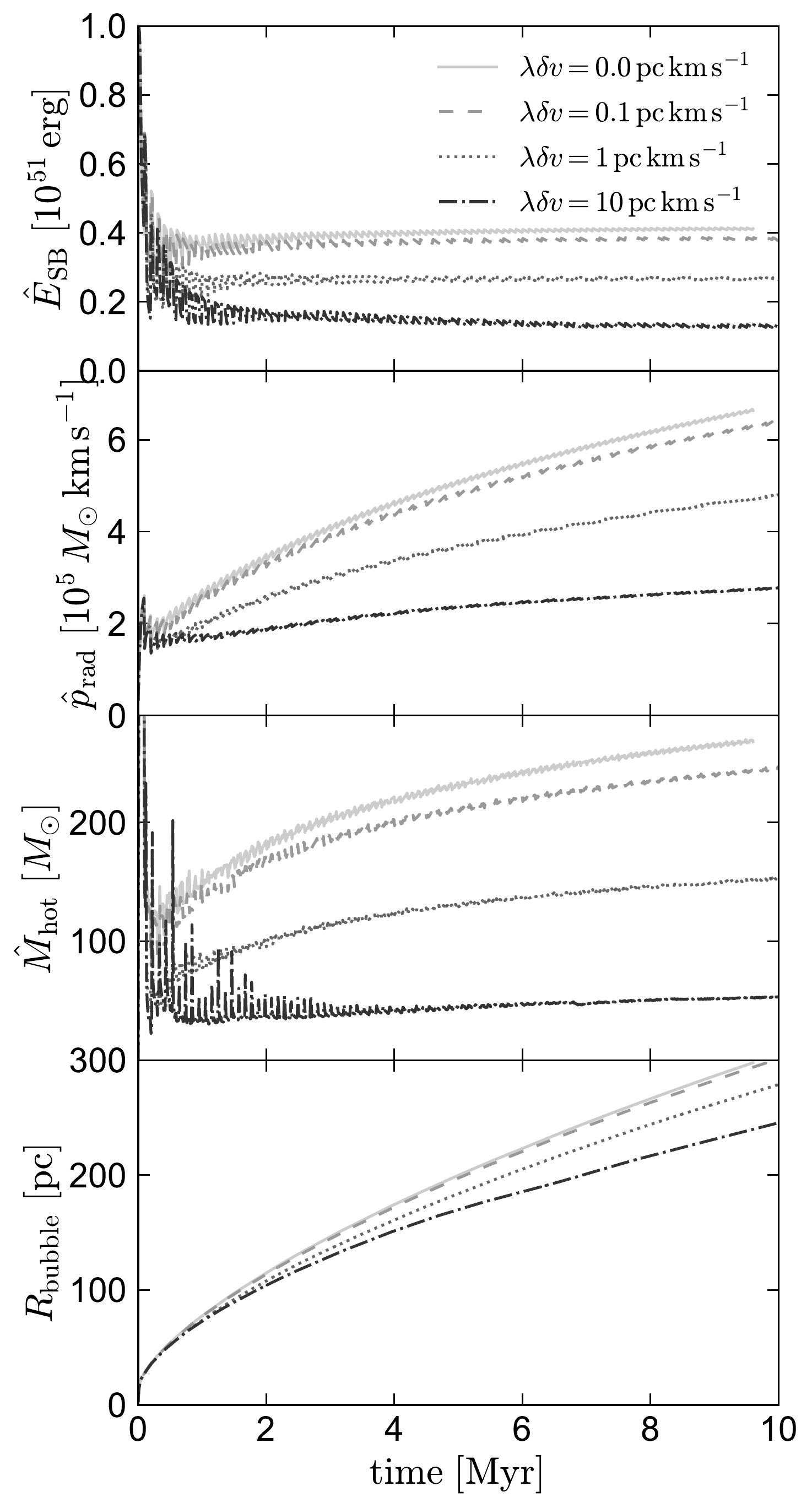}
\caption{Comparison of simulations with different values of $\lambda \delta v$, which parameterizes the efficiency of nonlinear mixing (Equation~\ref{eq:kappa_mix}) that is not captured in 1D. The resolution is fixed at $\Delta r = 0.024$\,pc and the ambient density is $n_{\rm H,0}=1\,\rm cm^{-3}$ in all cases. Increasing $\lambda \delta v$ increases the efficiency of mixing in the shell/interior interface between gas of different temperature phases. This leads to larger cooling losses, and consequently, to less mass evaporation from the shell. We use $\lambda \delta v = 1\,\rm pc\,km\,s^{-1}$ as the fiducial value in most of our simulations. 
}
\label{fig:lambda_dv_test}
\end{figure}

In Figure~\ref{fig:lambda_dv_test}, we compare the evolution of the total energy, radial momentum, hot gas mass, and bubble radius in simulations with different values of $\lambda \delta v$ at fixed resolution. 
Consistent with the predictions in Section~\ref{sec:pred_cooling_loss}, increasing $\lambda \delta v$ leads to more efficient cooling and thus lower SB energy, momentum, hot gas mass, and size. The simulation with $\lambda \delta v = 0.1\,\rm pc\,km\,s^{-1}$ appears almost identical to the one with no nonlinear mixing at all ($\lambda \delta v = 0\,\rm pc\,km\,s^{-1}$). However,  Figure~\ref{fig:cooling_res} shows that the simulations with $\lambda \delta v < 1\,\rm pc\,km\,s^{-1}$ are not fully converged at this resolution. For these two simulations,
all quantities shown in Figure~\ref{fig:lambda_dv_test} would be somewhat larger at higher resolution.

Our results thus directly demonstrate and quantify how cooling depends on the efficiency of mixing across what would otherwise be a contact discontinuity (see discussion in \citealt{Fierlinger_2016,Yadav_2017,Kim_2017, Fielding_2018,Gentry_2019}).  In our 1D simulations, the mixing efficiency is realized via the effective diffusivity $\lambda \delta v$.  In reality, this mixing is the result of nonlinear evolution of interface instabilities.  

In principle, as discussed in Section~\ref{sec:limitations}, one way to constrain the effective value of $\lambda \delta v$ from instability-driven mixing would be to calibrate to 3D simulations.  However, previous work provides only rather limited information on this. Some recent studies have been dominated by numerical diffusion at the resolution they can achieve.  Other studies at higher resolution were not designed to study late-time evolution in the limit of small $\Delta t_{\rm SNe}$. We conclude that a focused study will be required  to properly calibrate $\lambda \delta v$, as results from existing work are not sufficient to address this question. 

\section{Conductivity Ceiling}
\label{sec:kappa_max}
High values of $\kappa$ require prohibitively small timesteps for numerical stability. In particular, for Spitzer conductivity the CFL condition \citep{Courant_1928} requires that the simulation timestep, $\Delta t$, satisfies 
\begin{align}
\label{eq:delta_t}
\Delta t&\lesssim\frac{\left(\Delta r\right)^{2}\rho k_{B}}{2\kappa_S\mu m_{p}}\\&\lesssim 0.5\,{\rm yr}\times\left(\frac{\Delta r}{0.025\,{\rm pc}}\right)^{2}\left(\frac{n_{{\rm H}}}{10^{-2}\,{\rm cm}}\right)\left(\frac{T}{10^{6}\,{\rm K}}\right)^{-5/2}.
\end{align}
At high temperatures and low densities, this condition can become extremely restrictive. For example, immediately after new SN explosions, $T\sim 10^8\,{\rm K}$ and $n_{\rm H}\sim 10^{-3}$ near the center of the SB, requiring a timestep of $\sim$15 seconds at our default resolution! Because a single global timestep is used for all cells in the simulation, high values of $\kappa$ in regions with low density will grind the simulation to a halt. 

We prevent arbitrarily small timesteps by placing a density-dependent ceiling on thermal conductivity: 
\begin{align}
\label{eq:kappa_max}
\kappa_{{\rm max}}=1.8\times10^{12}\,{\rm erg\,s^{-1}cm^{-1}K^{-1}}\times \left(\frac{n_{{\rm H}}}{{\rm cm}^{-3}}\right).
\end{align}
This upper limit was chosen such that at a density of $n_{\rm H} =10^{-1}\,\rm cm^{-2}$, $\kappa_{\rm max}$ is the Spitzer value (Equation~\ref{eq:spitzer}) for a temperature of $10^7\,\rm K$. The density dependence of $\kappa_{\rm max}$ effectively sets a minimum timestep at fixed resolution: 
\begin{align}
\label{eq:min_dt}
\Delta t_{{\rm min}}\approx0.016\,{\rm yr}\times\left(\frac{\Delta r}{0.025\,{\rm pc}}\right)^{2}.
\end{align}
The actual minimum timestep in the simulations is 0.3 times $\Delta t_{\rm min}$, because we use a CFL number of 0.3. We experimented with different normalizations of $\kappa_{\rm max}$ to determine the smallest value that can safely be used without significantly changing the simulation results. Conduction is important primarily in the interface between the cool shell and hot bubble interior, where the temperature is lower and the density is higher than near the center of the SB. The ceiling on $\kappa$ does change the conductive energy flux and temperature profile near the center of the bubble, but this has a negligible effect on the integrated properties of the SB. We have verified that the adopted ceiling on $\kappa$ changes the hot gas mass (as well as the SB radius, total energy, and radial momentum) by less than 1\% from the values when no ceiling is used, while decreasing the typical computational expense by a factor of $\sim$\,100. Using a significantly smaller normalization in Equation~(\ref{eq:kappa_max}) {\it does} begin to produce deviations from simulations without a ceiling; for example, decreasing the normalization by a factor of 5 leads to a $\sim$\,10\% decrease in $M_{\rm hot}$.

\section{Saturation of conduction}
\label{sec:saturation}

\begin{figure*}
\includegraphics[width=\textwidth]{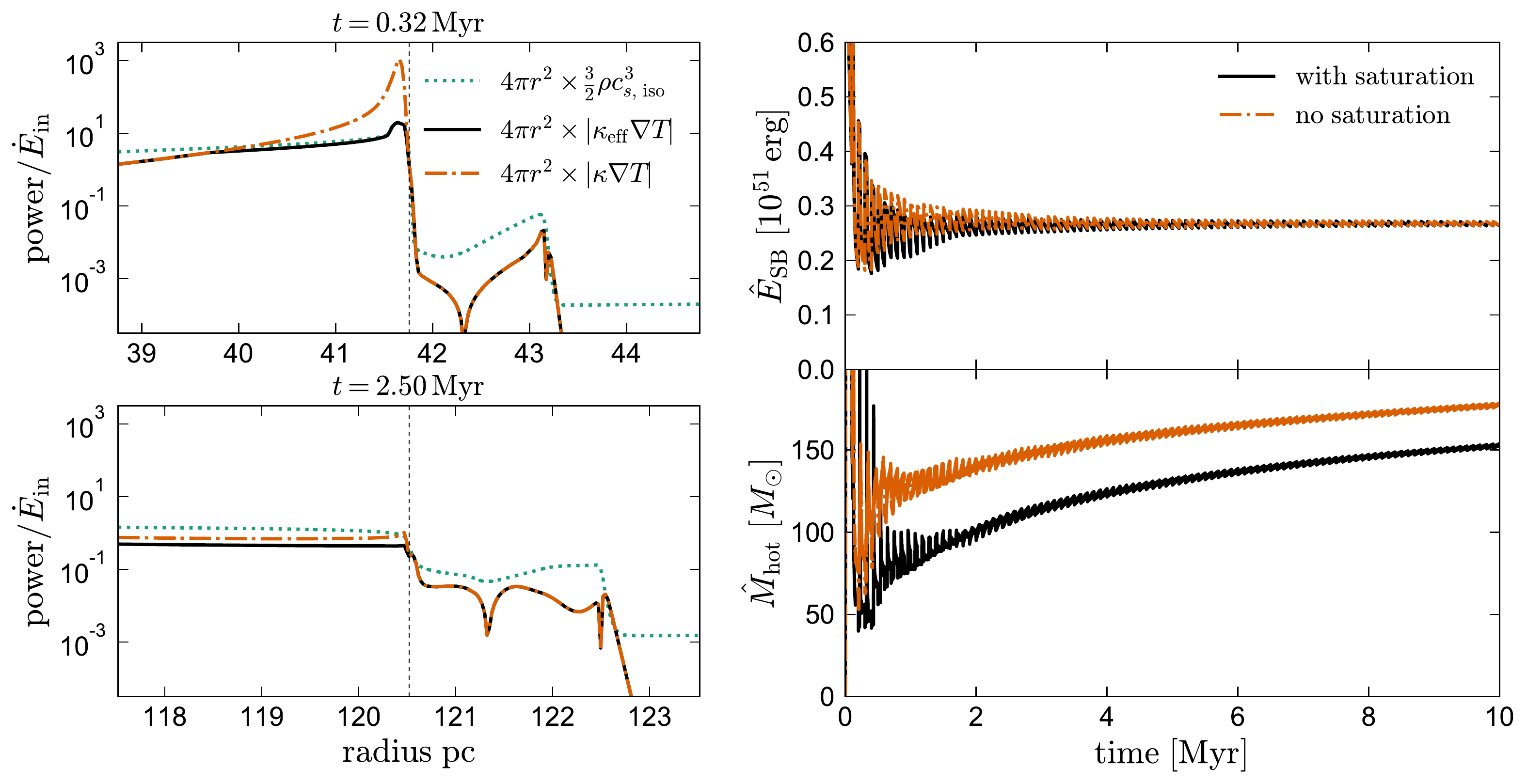}
\caption{{\bf Left}: Radial profiles of $\kappa_{s}|\nabla T|$, $\kappa_{{\rm eff}}|\nabla T|$, and $1.5\rho c_{s,\,{\rm iso}}^{3}$ from the simulation with the fiducial problem setup ($n_{\rm H,0}=1\,{\rm cm^{-3}}$ and $\Delta t_{\rm SNe}=0.1\,$Myr, with saturation implemented). We show profiles in two snapshots, one shortly after the arrival of a SN blast wave at $\sim$0.3 Myr, when the temperature gradient is very large, and one at $\sim$2.5\,Myr, when the interface has reached quasi-equilibrium. {\bf Right}: Total energy and hot gas mass per SN as a function of time for simulations which do (black) and do not (gold) account for saturation. Saturation reduces the hot gas mass produced per SNe, but the effect is modest at late times. Saturation has a negligible effect on the SB energy.}
\label{fig:saturation}
\end{figure*}

Our fiducial simulations implement saturation as described in Section~\ref{sec:numerical}, adopting an effective conductivity such that the heat flux interpolates smoothly between the classical and saturated values. 

We consider the effects of saturation in Figure~\ref{fig:saturation}. The left two panels compare the magnitude of the classical conductive heat flux (gold), the saturation limit (green), and actual flux implemented in the simulation (black), which interpolates from the classical value when $\kappa |\nabla T| \ll (3/2)\rho c_s^3$ to the saturation limit when $\kappa |\nabla T| \gg (3/2)\rho c_s^3$. We show two snapshots, one at early times, shortly after the arrival of the blast wave from a new SN, and one at later times, once the interface has reached quasi-equilibrium. Both snapshots are from the fiducial simulation, in which saturation is implemented. At early times (upper left panel), conduction is sometimes strongly saturated: after the arrival of a new SN blast wave at the interface, the classical conductive flux exceeds the saturation limit by more than an order of magnitude. At later times (lower left panel), the effects of saturation are modest, because the classical conductive flux is somewhat less than the saturation limit. The saturation prescription implemented in the simulation (Equation~\ref{eq:kappa_eff}) does slightly reduce the conductive flux relative to the classical limit, but the change is less than a factor of two. 

In the right panels of Figure~\ref{fig:saturation}, we compare the total SB energy and hot gas mass of the fiducial simulation (black) to one in which saturation has been turned off (i.e., the conductive heat flux always follows the classical prediction; gold). The differences in total energy between the two simulations are negligible at late times, implying that saturation does not significantly change the efficiency of cooling. On the other hand, there is a clear offset in $\hat{M}_{\rm hot}$ between the two simulations, in that the simulation without saturation produces somewhat more hot gas per SNe. We show in Section~\ref{sec:fluxes} that $\hat{M}_{\rm hot}$ is set by the difference between the conductive heat flux and the amount of cooling in the interface. Including saturation does not substantially change the amount of cooling, but it does reduce the conductive flux, leading to moderately less evaporation from the shell. 

The Weaver solution does not include saturation; that is, it assumes that the conductive heat flux is always given by Equation~(\ref{eq:q_cond}). Because saturation reduces the evaporative mass flux into the bubble, it heightens one of the important differences between our simulations and the Weaver solution. However, the cumulative effects of saturation are modest overall, and the change in $\hat{M}_{\rm hot}$ due to saturation is much less than the change in the evaporation due to cooling in the interface. 

\section{Cooling curve}
\label{sec:cooling_curve}
In our derivation of the expected cooling efficiency as a function of $n_{\rm H,0}$ and $\lambda \delta v$ (Section~\ref{sec:cooling_eff}), we made the approximation that most of the cooling in the interface occurs at a characteristic temperature $T_{\rm pk}$, and we assumed that $T_{\rm pk}$ does not vary much with $n_{\rm H,0}$ or $\lambda \delta v$.

\begin{figure*}
    \includegraphics[width=\textwidth]{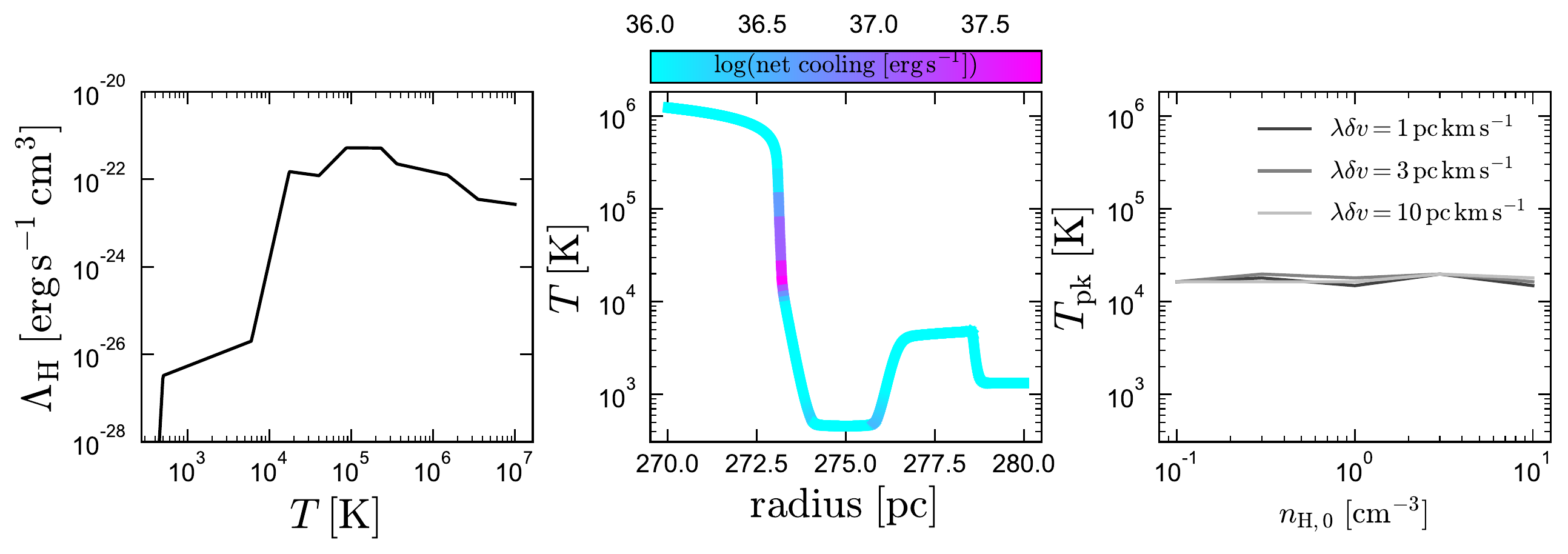}
    \caption{{\bf Left}: Cooling curve used on our simulations. {\bf Middle}: Temperature profile for the fiducial simulation at $t=10$\,Myr. Color scale shows the volume-integrated net cooling in each cell. Most of the cooling losses occur in the narrow region of the interface where the temperature is $10^4 \lesssim T/{\rm K} \lesssim 2\times 10^5$. {\bf Right}: $T_{\rm pk}$, the temperature at which most of the cooling occurs (Section~\ref{sec:cooling_eff}), for simulations with a range of $n_{\rm H,0}$ and $\lambda \delta v$. In all the simulations, $T_{\rm pk} \approx 2\times 10^4\,\rm K$.}
    \label{fig:cooling_curve}
\end{figure*}

Figure~\ref{fig:cooling_curve} examines these assumptions in more detail. The left panel shows the cooling curve used in our simulations (black line). The middle panel shows the temperature profile of a snapshot of the fiducial simulation at $t=10$\,Myr, once the interface is in flux equilibrium and transient effects from individual SNe have died down. We color the temperature profile by the cell-integrated net cooling rate, $\dot E_{\rm cool}=n(n\Lambda-\Gamma)\times 4\pi r^{2}\Delta r$, in each cell. Most of the cooling losses occur in a narrow region where the temperature is between $T \approx 10^4$\,K and $T \approx 2\times 10^5$\,K. Some additional cooling occurs at $T<10^3$\,K on both sides of the shell. In the center of the shell, the temperature reaches an equilibrium where cooling and heating exactly balance.  

Most of the cooling occurs below the peak of the cooling curve. As discussed in Section~\ref{sec:cooling_eff}, the pressure in the interface is approximately constant. Although $\Lambda(T)$ peaks at $T \sim 10^{5}\,\rm K$, the volumetric cooling rate, $\sim n^2 \Lambda(T)\sim P^2 \Lambda(T)/T^2$ peaks at $T\sim 2\times 10^4$\,K for fixed pressure $P$, and this is the temperature where most of the cooling occurs. This is shown in the right panel, which plots the temperature where cooling is most efficient, for simulation with a range of $n_{\rm H,0}$ and $\lambda \delta v$. For each simulation, we calculate $T_{\rm pk}$ as the temperature where the cooling efficiency, ${\rm d}\dot{E}_{{\rm cool}}/{\rm d}\log T$, peaks. No significant trend in $T_{\rm pk}$ is evident with $n_{\rm H,0}$ or $\lambda \delta v$, validating the assumption in Section~\ref{sec:cooling_eff}.


\bsp	
\label{lastpage}
\end{document}